\documentclass[graphics,twocolumn, usenatbib]{mn2e}

\usepackage{epsfig} 
\usepackage{graphicx} 
\usepackage{color}

\usepackage{deluxetable} 
\usepackage{aas_macros} 
\usepackage{amssymb}

\newcommand{\enzo}{\textsc{Enzo~}}
\newcommand{\cosmos}{\textsc{Cosmos~}}
\newcommand{\darwin}{\textsc{Darwin~}}

\newcommand{\kms}{\mathrm{\, km \, s^{-1}}}

\def\etal{{\it et al.}~}

\title[]{The formation of compact massive self-gravitating discs 
        in metal-free haloes with  virial temperatures of $\sim$ 13000-30000\,K}

\author[J.A. Regan \etal] 
{John A. Regan\thanks{E-mail:regan@ast.cam.ac.uk} \& Martin G. Haehnelt \\ \\
Institute of Astronomy, Madingley Road, Cambridge CB3 0HA, UK\\}

\begin{document}

\maketitle

\begin{abstract} We have used the hydrodynamical AMR code ENZO to
investigate the dynamical evolution of the gas at the centre of 
dark matter haloes  with virial velocities of  $\sim 20 - 30 \kms$ 
and virial temperatures of  $\sim 13000-30000 K$ 
at $z\sim 15$ in a cosmological context. The virial  temperature of
the  dark matter haloes  is  
above the threshold where atomic cooling by hydrogen allows the gas to
cool and collapse.  We neglect cooling by molecular  hydrogen and
metals, as may be plausible if ${\rm H}_2$ cooling is suppressed by a
meta-galactic Lyman-Werner background or an internal source of
Lyman-Werner photons, and metal  enrichment has not progressed very
far. The gas in the haloes becomes  gravitationally unstable and 
develops turbulent velocities comparable to the virial velocities 
of the dark matter haloes. Within  a few dynamical times 
it  settles into a nearly isothermal density profile over many 
decades in radius losing most of its angular momentum in the
process. About 0.1 - 1\,\% of the baryons, at the
centre of the dark matter haloes, collapse into a self-gravitating,
fat, ellipsoidal, centrifugally supported exponential disc with
scale-length of  $\sim
0.075-0.27$ pc  and rotation velocities of $25-60 \kms$. 
We are able to follow the settling of the gas 
into centrifugal support and the dynamical evolution of the compact
disc in each dark matter halo for a few dynamical
times. The dynamical evolution of the gas at the centre of the haloes
is complex. In one of the haloes the gas at the centre fragments
into a triple system leading to strong tidal perturbations and
eventually to the in-fall of a secondary smaller 
clump into the most massive primary clump. The formation of
centrifugally supported self-gravitating massive discs is likely to 
be an important intermediary stage en route to the formation of
a massive black hole seed. 
\end{abstract}


\begin{table*}
\centering
\begin{minipage}{160mm}
\begin{tabular}{ |l |l |l |l | l | l |l }
\hline \hline 
\textbf{\em Sim} & \textbf{\em Boxsize} & \textbf{\em $\bf z_{\rm init}$}
& \textbf{\em $\bf z_{\rm coll}$} & \textbf{\em DM mass}
& \textbf{\em $\bf \Delta$ R} \\ 
& \textbf{\em \rm{\bf{[Comoving $\bf h^{-1}$ Mpc]}}} & & &  \textbf{\em $\bf [M_{\odot}]$} 
& \textbf{\em \rm{\bf{[pc]}}} \\

\hline 
A & 5.0 & 200.0 & 15.0 & $9.58 \times 10^3$& $9.59 \times 10^{-3}$  \\ 
B & 2.0 & 250.0 & 14.0 & $4.42 \times 10^2$& $1.61 \times 10^{-2}$  \\ 
C & 2.0 & 250.0 & 15.0 & $4.42 \times 10^2$& $1.61 \times 10^{-2}$  \\ 

\hline 
\hline
\textbf{\em $\bf M_{\rm tot}$} & \textbf{\em $\bf R_{200}$ }  &
\textbf{\em $\bf V_{\rm c}$ }  & \textbf{\em $\bf T_{\rm vir}$} 
& \textbf{\em $\bf \rho_{\rm max}$} & \textbf{\em $\bf \lambda$} & \textbf{\em $\bf T_{\rm core}$} \\ 
\textbf{\em $\bf [M_{\odot}]$} & \textbf{\em \rm{\bf{[kpc]}}} & \textbf{\em \rm{\bf{[km $\bf s^{-1}$]}}} & \textbf{\em \rm{\bf{[K]}}} 
& \textbf{\em $\rm{\bf{[cm^{-3}]}}$} & & \textbf{\em \rm{\bf{[K]}}} \\

\hline 
$2.64 \times 10^8$  & 1.28 & 29.9 & $3.23 \times 10^4$ & $1.24 \times 10^9$ & $0.031$ &   $6.11 \times 10^3$  \\
$5.37 \times 10^7$  & 0.64 & 19.0 & $1.30 \times 10^4$ & $5.86 \times 10^8$ & $0.026$ &   $6.35 \times 10^3$  \\ 
$5.15 \times 10^7$  & 0.59 & 19.3 & $1.35 \times 10^4$ & $7.12 \times 10^8$ & $0.050$ &   $6.13 \times 10^3$  \\ 
\hline 
\hline
\end{tabular}
\end{minipage}

\caption{Basic properties of the three simulations (A, B and C): boxsize(comoving $h^{-1}$ kpc), 
starting redshift, collapse redshift, DM particle mass  ($M_{\odot}$), spatial resolution (pc), 
total mass of the halo ($M_{\odot}$), virial radius ($h^{-1}$ kpc), circular velocity (km $\mathrm {s^{-1}}$), 
virial temperature (K), maximum baryon density in the halo ($\rm cm^{-3}$), angular momentum parameter $\lambda$,  temperature
at the core of the halo (K). All units are physical units, unless
explicitly stated otherwise.}

\label{TableSims}
\end{table*}


\section{Introduction} 
Supermassive black holes (SMBH)s were
invoked  as the central engine powering  quasi-stellar objects (QSO)s soon after the first
QSOs were discovered \cite[]{Zeldovich_1964, Salpeter_1964}.  
Early predictions that supermassive black holes are ubiquitous  and maybe
found in many if not most galaxies as the remnants of dead QSOs
\cite[]{Lynden-Bell_1969} have stood the test of time. SMBHs
 are now believed to reside in most if not all  galactic bulges and
their masses correlate tightly with  the stellar velocity dispersion
and the mass of the bulges of  their host galaxies 
\cite[Gebhard et al. 2000;][]{Ferrarese_2000}. The luminosity
functions of active galactic nuclei across the  electromagnetic spectrum and its
evolution with redshift have been used to constrain the growth history
of supermassive black holes \cite[e.g.,][]{Merloni_2004, Shankar_2007, Merloni_2008}.
Little is known however, 
about how the first (super-)massive black holes came into being.
The discovery of very luminous, bright QSOs at $z \ge 6$ appears to suggest that at least
some of the most massive black holes were already in place when the Universe
was less than 1 billion years old \cite[]{Fan_2001, Fan_2006, Haiman_2006}. \\ 
\indent A wide range of generic pathways to a supermassive black hole are possible
ranging from direct collapse of gas into rather massive seed black holes,
to Eddington-limited accretion onto stellar mass black holes and the
dynamical evolution of dense  star clusters \cite[]{Rees_1978, Rees_1984}. The tight relation
between galactic bulges and their central mass suggests that they 
have  formed and grown in a connected way.  In the well-established  $
\Lambda$CDM paradigm of  structure  formation,  dark matter haloes grow by hierarchical
merging. The last decade has seen extensive and detailed modeling of
the build-up of galaxies and supermassive black holes in this context 
using semi-analytic descriptions as well as numerical simulations
\cite[]{Efstathiou_1988, Carlberg_1990, Haehnelt_1993, Kauffmann_2000, Monaco_2000, 
Cavaliere_2002, Volonteri_2003, Volonteri_2005a, Volonteri_2005b, 
Volonteri_2007b, Croton_2006, Sijacki_2008, 
DiMatteo_2008, Hopkins_2008}.
Due to the lack of observational constraints it is, however, still an
open question  at what  mass scale this hierarchical build-up starts. \\
\indent The modeling of the growth of 
supermassive black holes by  hierarchical merging of dark matter 
haloes/galaxies from the remnant  black holes of the first generation 
of stars has identified a number of 
problems. 
The shallow potentials hosting the first generation of 
stars are rather sensitive to the effect of stellar feedback 
\cite[]{Dekel_1986, Wise_2007c, OShea_2008} and may thus not provide sufficient amounts of
fuel for a rapid growth of  stellar mass black holes. The predicted frequent
merging may lead to the expulsion due to gravitational recoil 
and/or gravitational slingshot unless 
stellar seed black holes are sufficiently rare \cite[e.g.][]{Volonteri_2007a}. 
The available time for the formation of the most massive black holes requires almost
continuous accretion at the Eddington rate or super-Eddington
accretion rates. There is also circumstantial evidence that 
a certain class of X-ray sources, Ultra-luminous X-ray sources
or ULXs for short, are powered by intermediate mass black holes of
masses up to $10^{4} \, M_{\odot}$ \cite[]{Makishima_2000,
  Fabbiano_2003, Colbert_2004, Fabbiano_2006}. 
There is thus plenty of motivation to consider
seriously the possibility that the growth of supermassive black holes 
has started from massive seed black holes with masses substantially larger than
those forming as stellar remnants (see \cite{Begelman_2006} for a
recent discussion). \\
\indent The centres of high-redshift 
dark matter haloes with virial temperatures  $\ga  10^4$K, 
not yet significantly enriched with metals, have been identified as a 
promising environment to form such massive seed black holes. 
If ${\rm H}_2$ cooling is suppressed in these haloes either due to an
external UV background or more likely especially in the later stages
of the collapse due to internal sources of UV radiation \cite[cf.][]{Wise_2007b}
early fragmentation should not
occur  favouring the formation of a compact massive rotationally
supported disc \cite[]{Oh_2002, Bromm_2003, Volonteri_2003, Koushiappas_2004, Begelman_2006, 
Lodato_2006, Rees_2007, Volonteri_2008}.
This is rather different from the situation in the lower mass
haloes studied extensively as possible sites for the formation of  the
``first'' stars where ${\rm H}_2$ cooling is the dominant cooling mechanism
\cite[e.g.][]{Bromm_1999, Bromm_2002, Bromm_2004,  
Abel_1998, Abel_2000, Abel_2002, OShea_2007b, OShea_2008}. \\
\indent We use the publicly available  code \textsc{Enzo}\footnote{http://cosmos.ucsd.edu/enzo/}
 to  perform adaptive mesh
simulations  of the dynamical evolution of the gas in three
dark matter haloes  with virial velocities  between $\sim 19 \, \kms$ and  $\sim 30 \, \kms$ and
virial temperatures   between $\sim 13000 \, K$ and $\sim 30000 \, K$ at
$z \sim 15$ in a cosmological
context. The work is similar in spirit to that of \cite{Wise_2007}, hereafter W07, 
who simulated two haloes somewhat less  massive than our haloes. 
W07 find that the gas at the centre of their haloes does not reach 
rotational support and does not fragment. 
Our simulations  have a similar 
set-up but unlike  W07 we do not push for resolution 
in the central region where a very small fraction of the gas can collapse to
very high density  but rather follow the 
dynamical evolution of a significant fraction of the gas 
settling into a rotationally supported disc 
for several dynamical times.  
The paper is structured as follows. In \S \ref{Sims} we describe the
details of the numerical  simulations. In \S \ref{formulae} we
summarise some basic formulae that we use to characterise the
properties of the  haloes. In \S \ref{results} we describe the results of
our numerical simulations and in \S \ref{conclusions} we summarise our conclusions.
Throughout this paper we  assume a standard
$\Lambda$CDM cosmology with the following parameters \cite[based on the  WMAP 1st year data]{Spergel_2003},
$\Omega_{\Lambda,0}$  = 0.74, $\Omega_{\rm m,0}$ = 0.26, $\Omega_{\rm b}$ = 0.0463,
$\sigma_8$ = 0.9 and $h$ = 0.72. We further assume a spectral index of primordial 
density fluctuations of $n=1$.

\section{The Setup of the numerical simulations} 
\label{Sims} 
\subsection{ The adaptive-mesh refinement code \enzo}
We have  used the publicly available adaptive mesh refinement
(AMR) code \textsc{Enzo}.  
\enzo was originally developed  by Greg  Bryan  and
Mike  Norman at  the University of  Illinois (Bryan \& Norman 1995b,
Bryan \& Norman 1997,  Norman \& Bryan 1999, O'Shea et al. 2004).
\nocite{Bryan_1995b} \nocite{OShea_2004} \nocite{Bryan_1997}
\nocite{Norman_1999} The gravity solver in \enzo uses an N-Body
particle    mesh    technique (Efstathiou et al. 1985,  Hockney \&
Eastwood 1988). Additional finer meshes can then be added in regions
of high density to calculate the dynamics of the dark matter (DM)
particles more accurately.  \nocite{Efstathiou_1985}
\nocite{Hockney_1988}\\ The hydrodynamics solver employs the
Piecewise Parabolic  method combined with  a non-linear  Riemann
solver  for  shock  capturing.  The  Eulerian AMR  scheme  was  first
developed   by \cite{Berger_1984} and later refined by
\cite{Berger_1989}  to  solve   the  hydrodynamical equations for an
ideal gas.   \nocite{Bryan_1995b} Bryan \& Norman (1997)  adopted such
a   scheme  for cosmological   simulations.  In addition to this there
are also modules available which  compute the radiative cooling of the gas
together with a multispecies  chemical reaction network. There are two
versions of the chemistry solver available,  one with 6 species 
(${\rm H}, {\rm H}^+, {\rm He}, {\rm He}^+,  {\rm He}^{++}, {\rm e}^-$) and one  with 9 species 
(same as before plus ${\rm H}_{2}, {\rm H}_{2}^+,{\rm H}^-$). As stated previously the  
simulations conducted here do not include ${\rm H}_2$ cooling and
chemistry. 
Our simulations make
extensive use of \textsc{Enzo's} capability to employ nested  grids  which  
we describe in the next section.

\subsection{Nested grids \& initial conditions} 
Initial conditions were generated with the initial 
conditions generator supplied with the \enzo code. 
The nested grids are introduced at the initial conditions stage. 
We have first run exploratory DM only simulations with coarse resolution, 
setting the maximum refinement level to 4. These DM only simulations have a root
grid size of $128^3$ and  no nested grids.
In these exploratory simulations we have  identified the most
massive halo at a redshift of 10 and then  rerun the simulations, 
including the  hydrodynamics module. We also
introduce nested grids at this point. The nested grids  are placed
around the region of interest, as identified from the  coarse
DM simulation. We have used four levels of nested grids in our
simulations with a maximum effective resolution of $1024^3$. The
introduction of nested grids is accompanied by a corresponding 
increase in the DM resolution by increasing the number of particles 
in the region of interest. We only use the highest resolution nested 
grid to refine further thus economising  our
computational output .This corresponds to a region of $625 \, h^{-1}$ (comoving) kpc for 
simulation A and $250 \, h^{-1}$ (comoving) kpc for simulations B \& C. 
The  total number of particles in our simulation
is 4935680, with $128^3$ of these  in our highest resolution
region. The nested grids are distributed as follows in root grid
sizes, $L0 = 128^3, L1=32^3, L2=24^3, L3=16^3$.  
Table \ref{TableSims} gives the details of the simulations
discussed here. 


\begin{figure*}
\includegraphics[height=4.0cm, width=5cm]{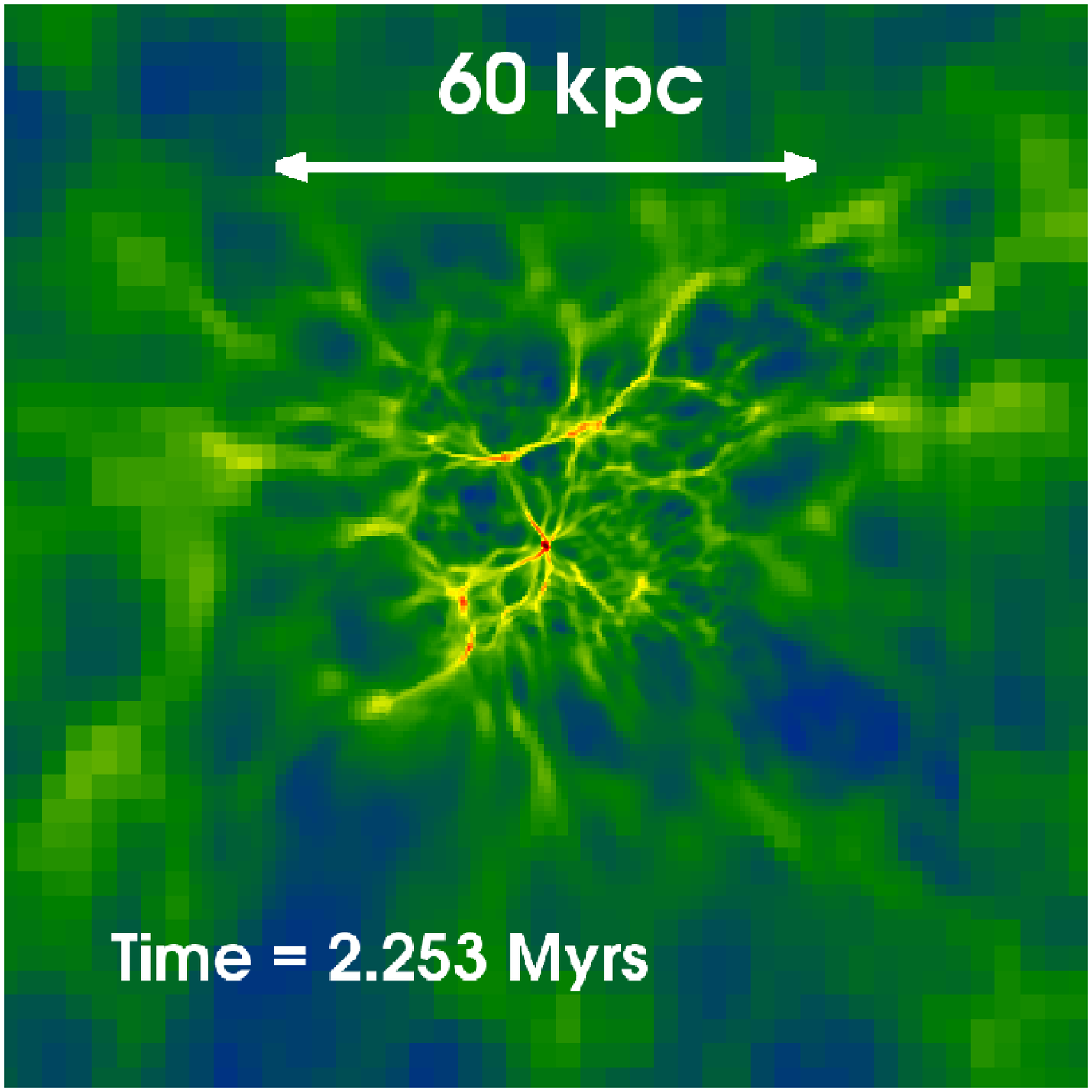} 
\includegraphics[height=4.0cm, width=5cm]{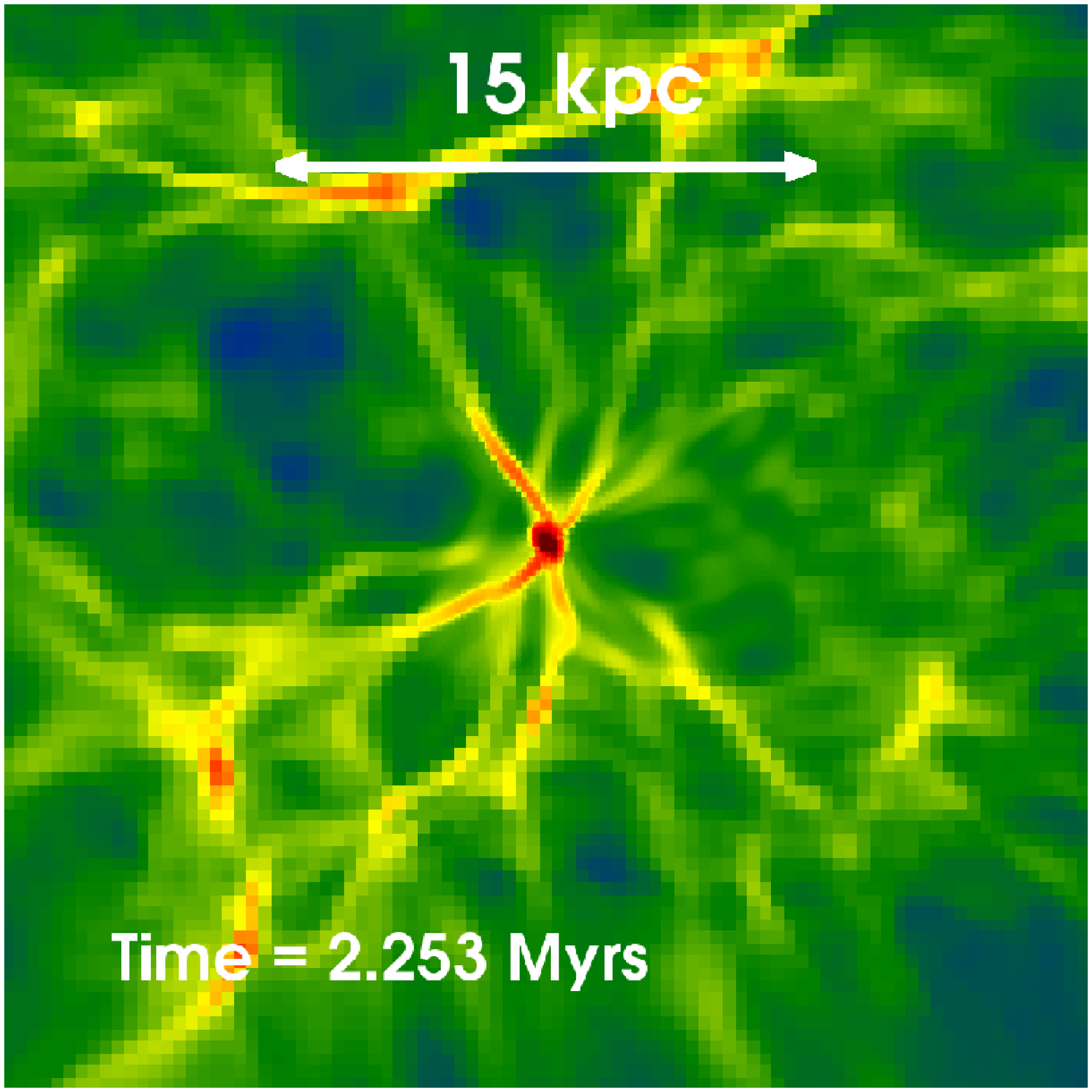} 
\includegraphics[height=4.0cm, width=5cm]{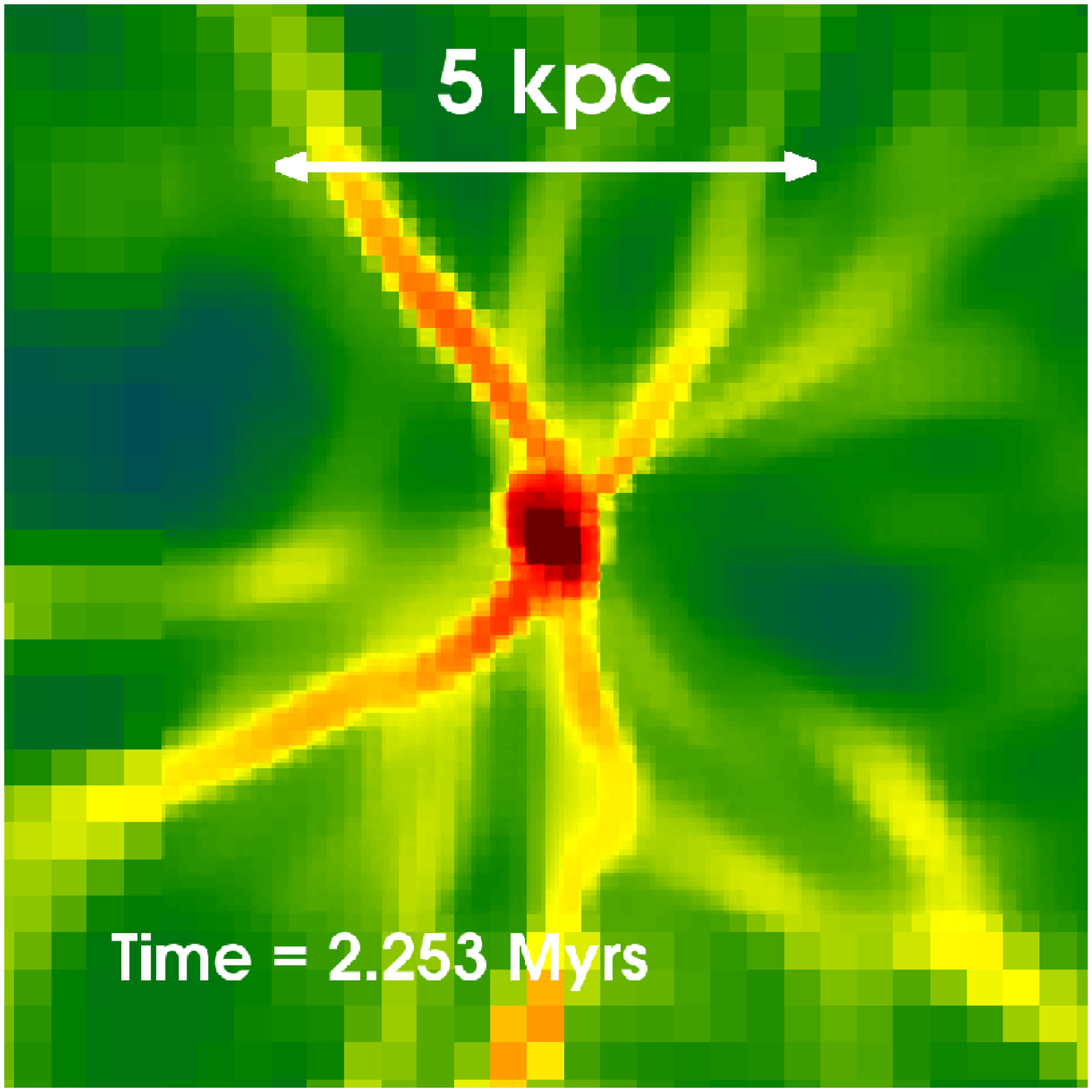} 
\includegraphics[height=4.0cm, width=5cm]{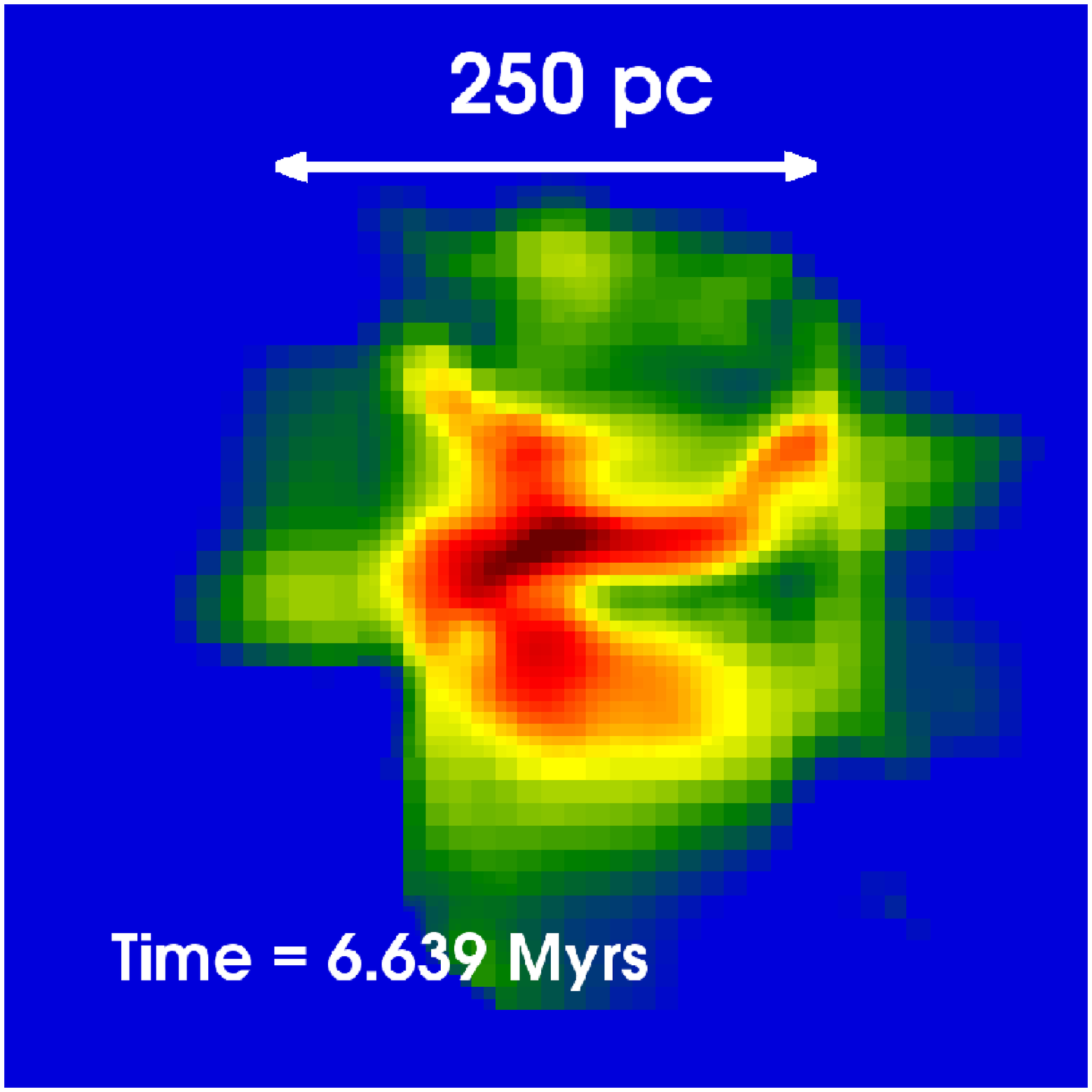} 
\includegraphics[height=4.0cm, width=5cm]{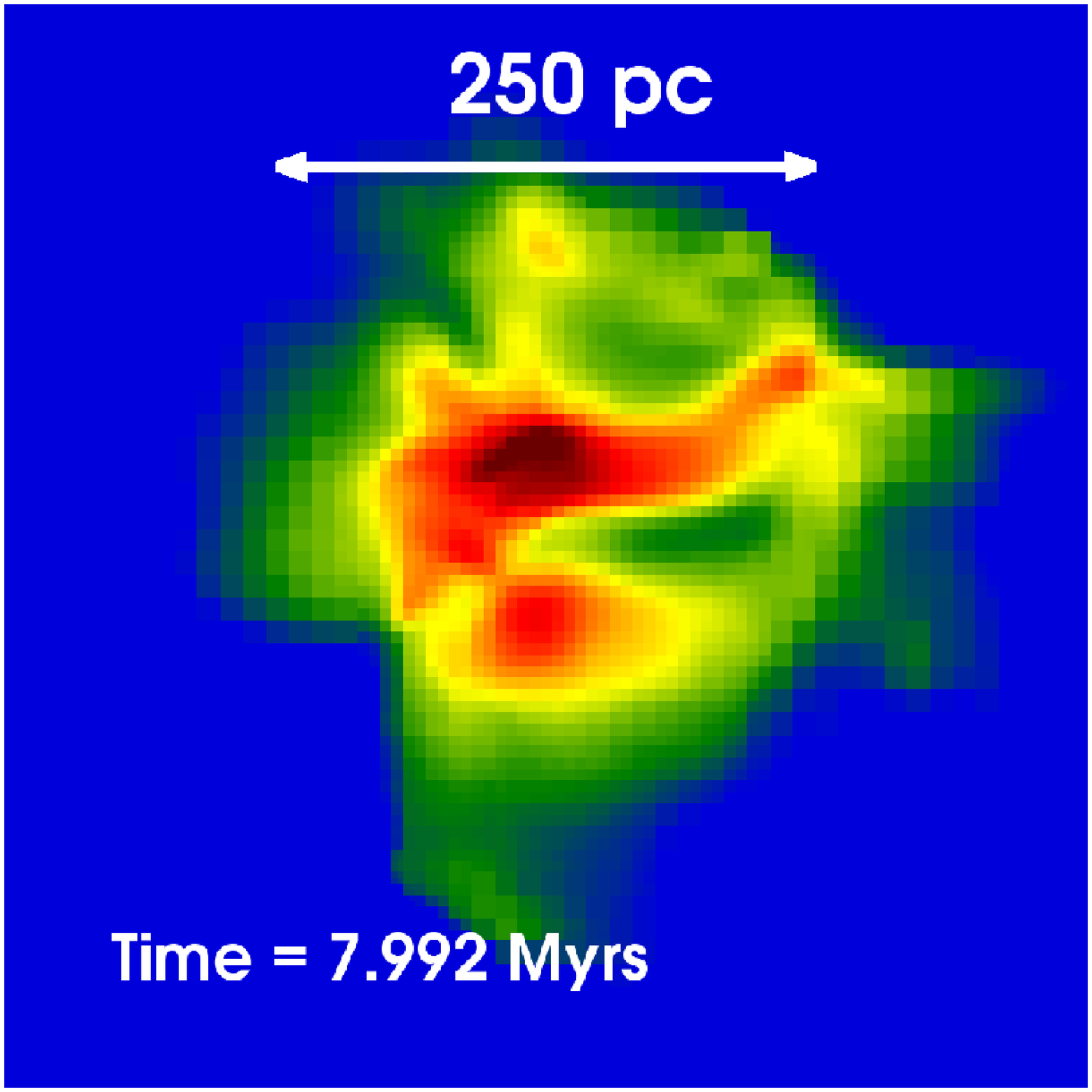} 
\includegraphics[height=4.0cm, width=5cm]{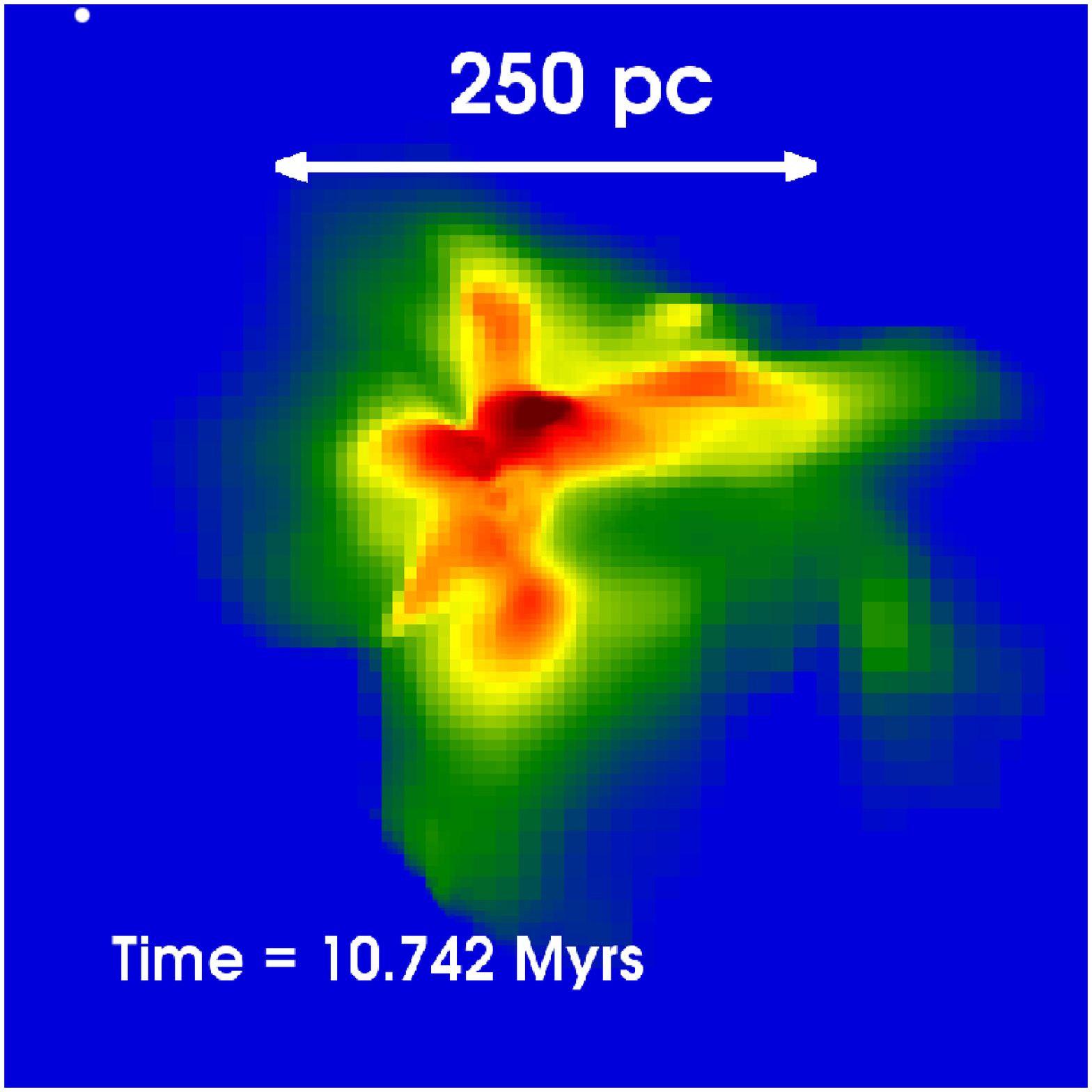} 
\includegraphics[height=4.0cm, width=5cm]{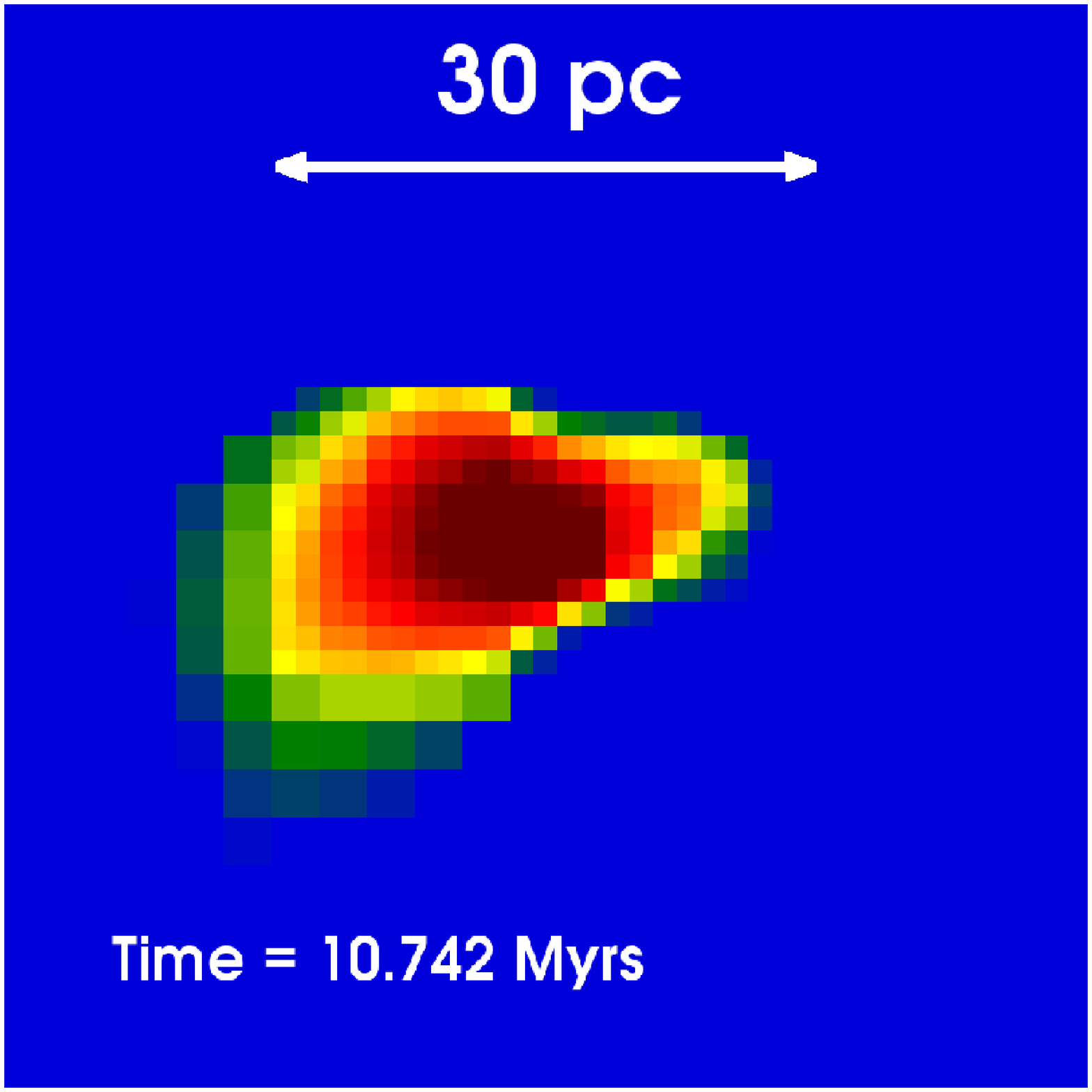} 
\includegraphics[height=4.0cm, width=5cm]{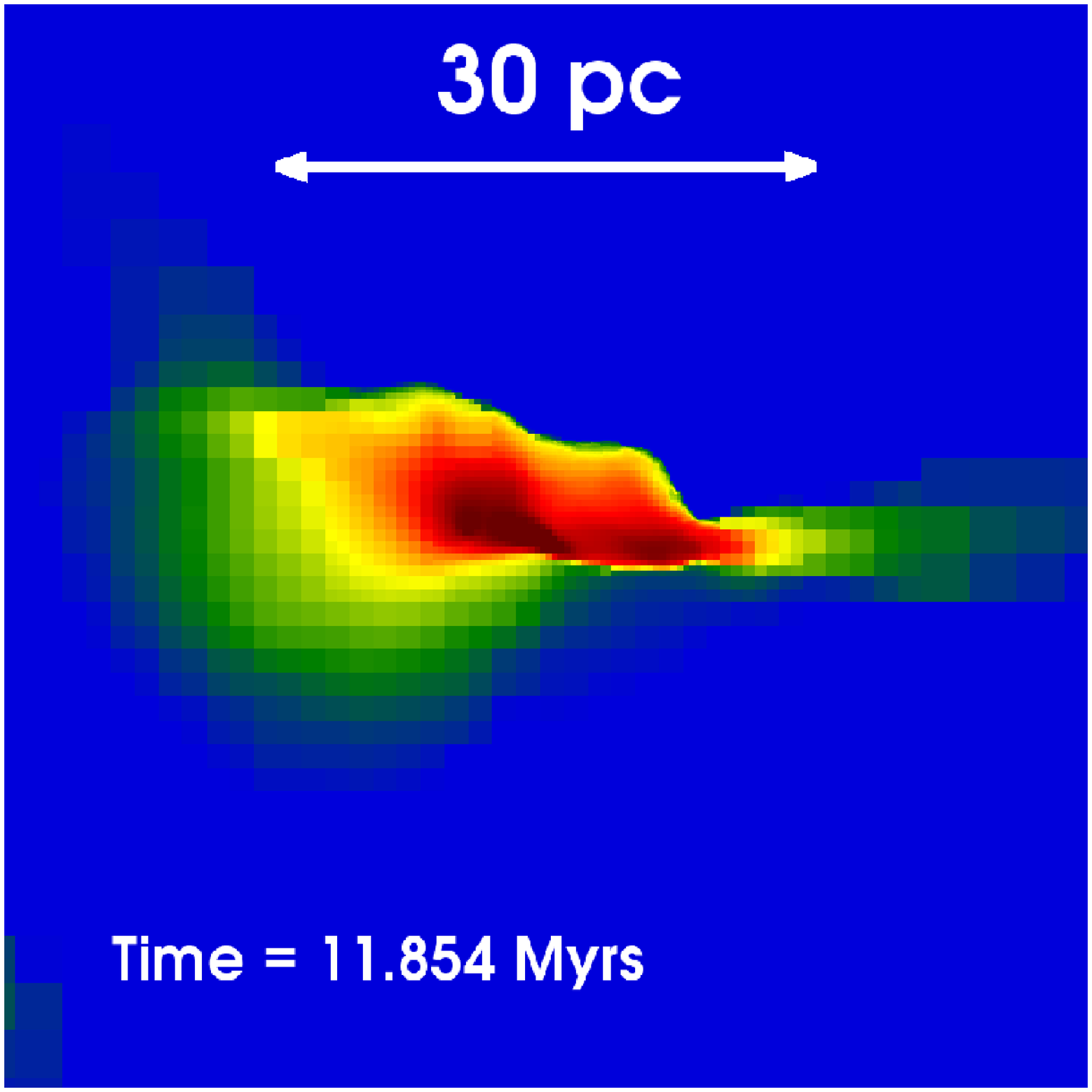}  
\includegraphics[height=4.0cm, width=5cm]{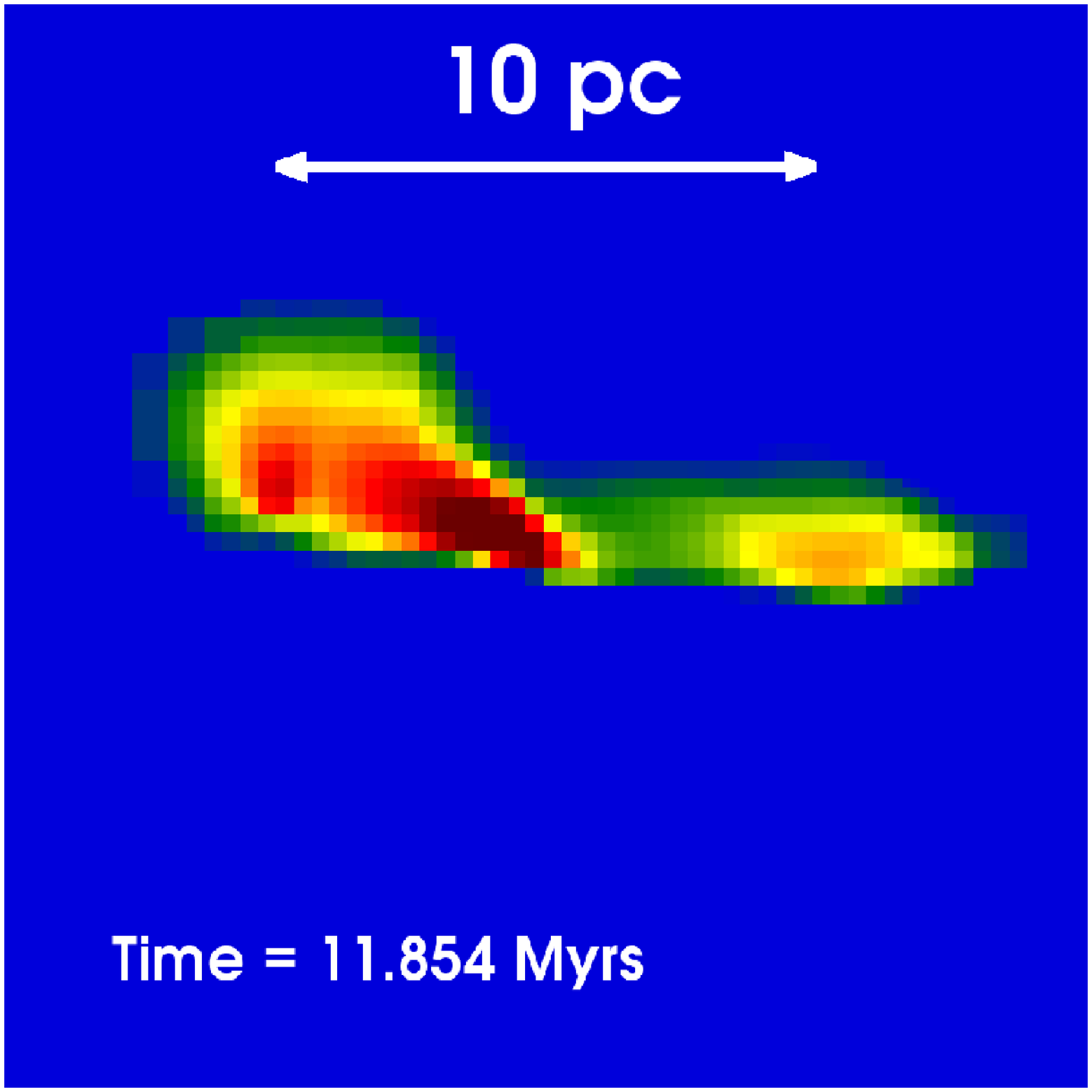}  
\includegraphics[height=4.0cm, width=5cm]{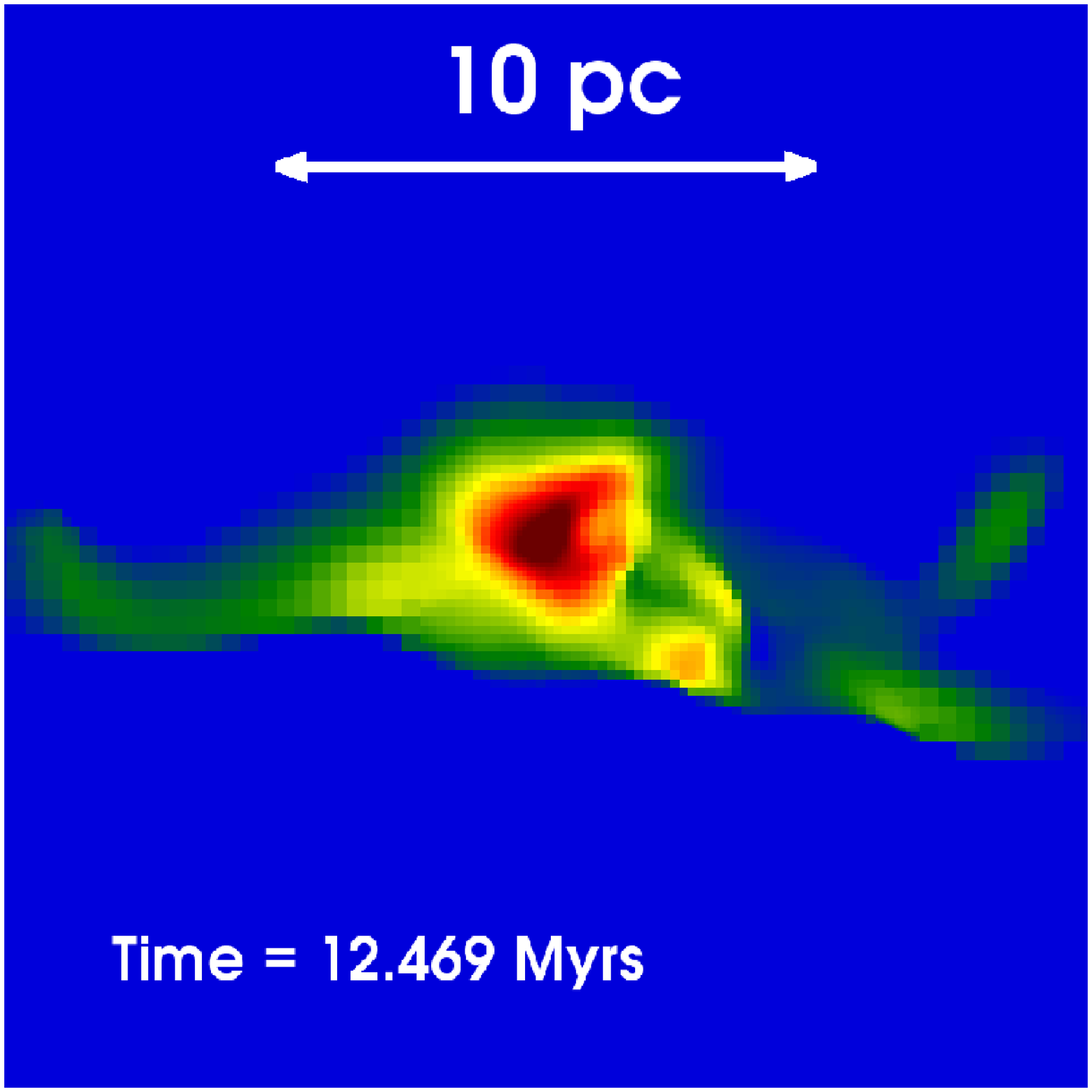}  
\includegraphics[height=4.0cm, width=5cm]{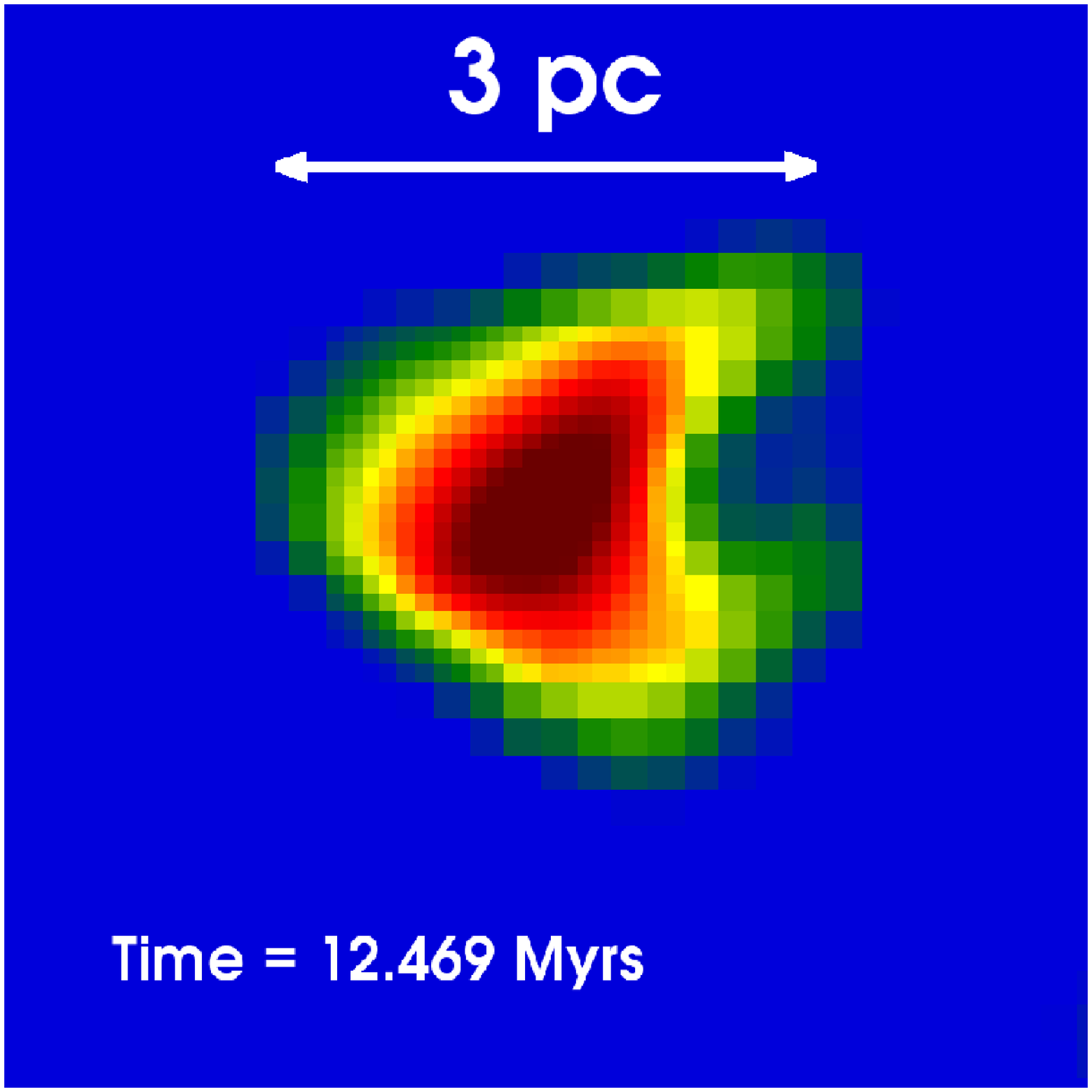}  
\includegraphics[height=4.0cm, width=5cm]{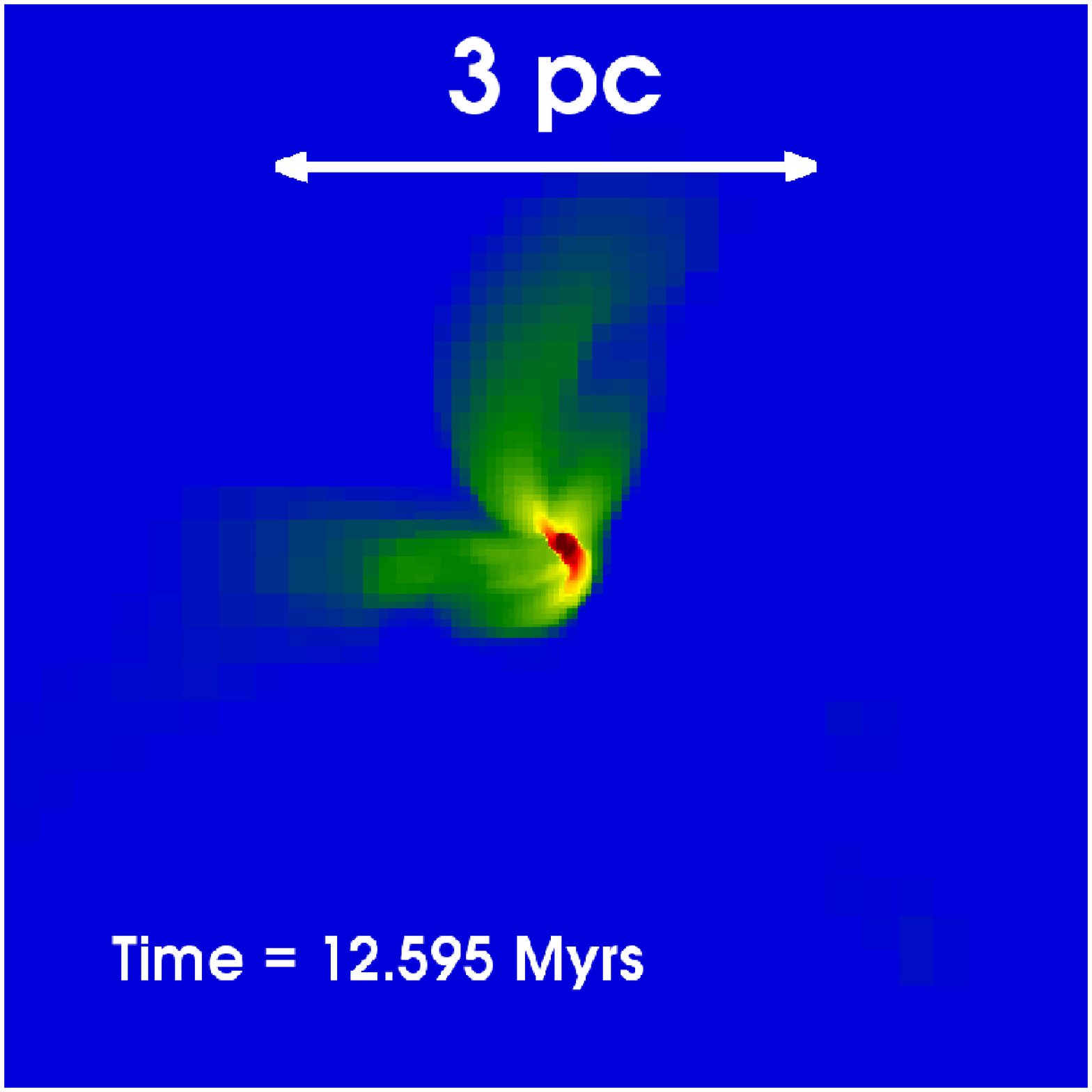} 
\includegraphics[height=4.0cm, width=5cm]{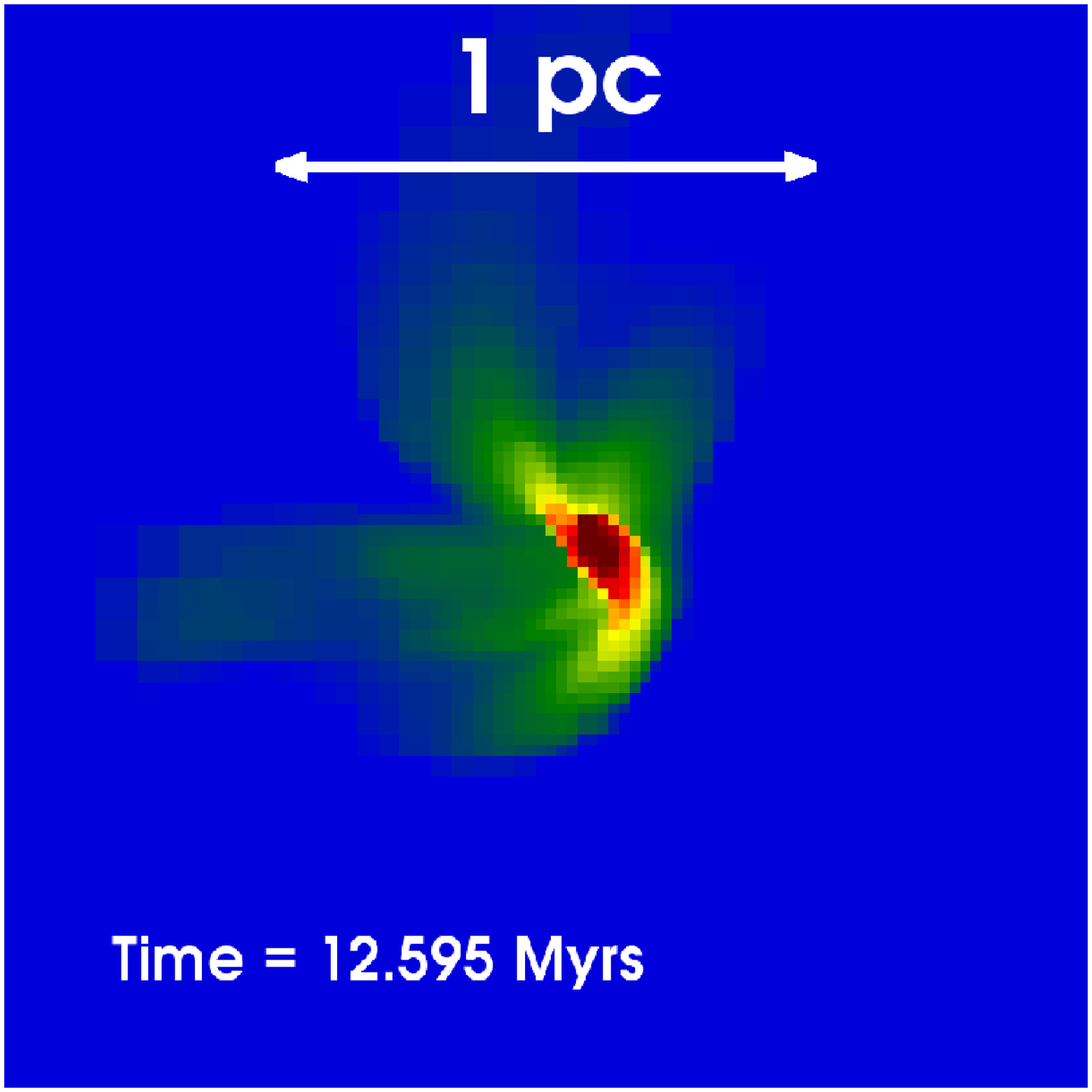} 
\includegraphics[height=4.0cm, width=5cm]{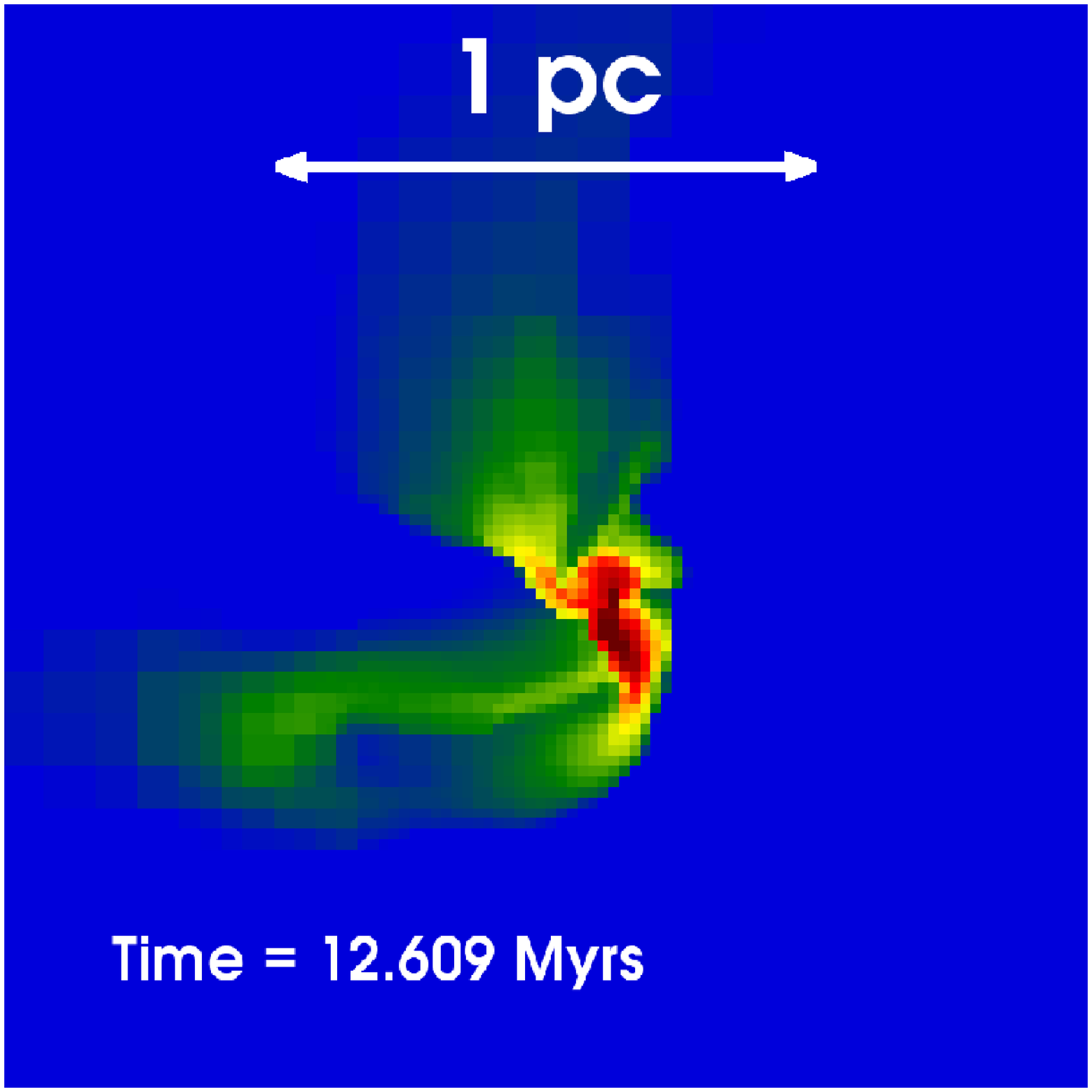} 
\includegraphics[height=4.0cm, width=5cm]{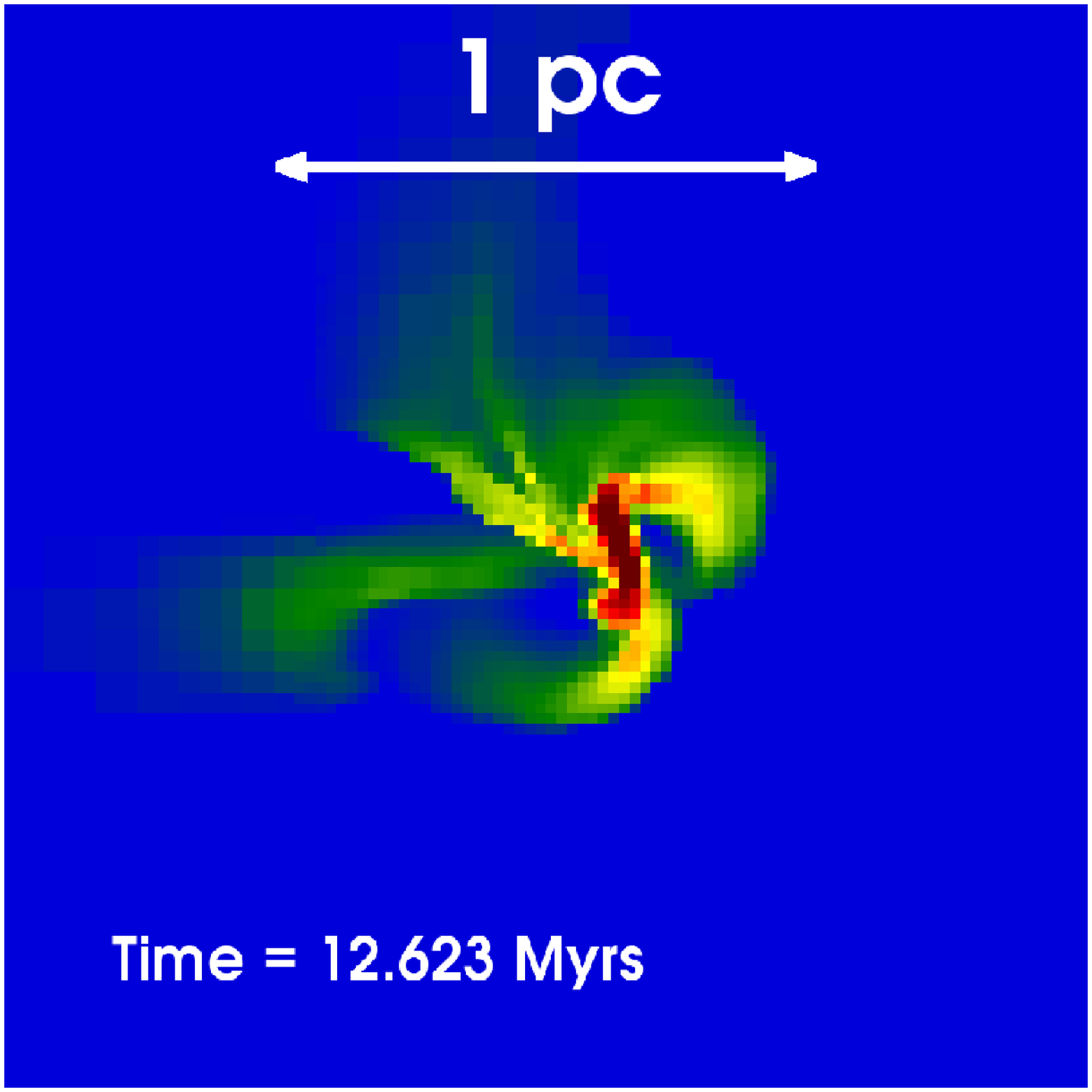} 
\caption[\label{SimAMap}]{\label{SimAMap}  
A sequence  of visualisations of the density distribution in
simulation A. The gas is collapsing in a halo of DM mass of $2.64\times
10^{8} M_{\odot}$, virial velocity of $\sim 30 \kms$ and virial
temperature of $\sim 32000 \, \rm{K}$. 
At the end of the simulation the gas at the centre of
the halo has settled into a rotationally supported disc. 
Each plot shows a thin slice of the density distribution
centered on the grid cell with the highest density in our halo of interest. 
Between panels either the density range  or the time of
the simulation output change. The colour range represents approximately an order of 
magnitude range in density in each plot.  The scales are proper distances.}

\end{figure*}



\begin{figure*}
\includegraphics[height=4.0cm, width=5cm]{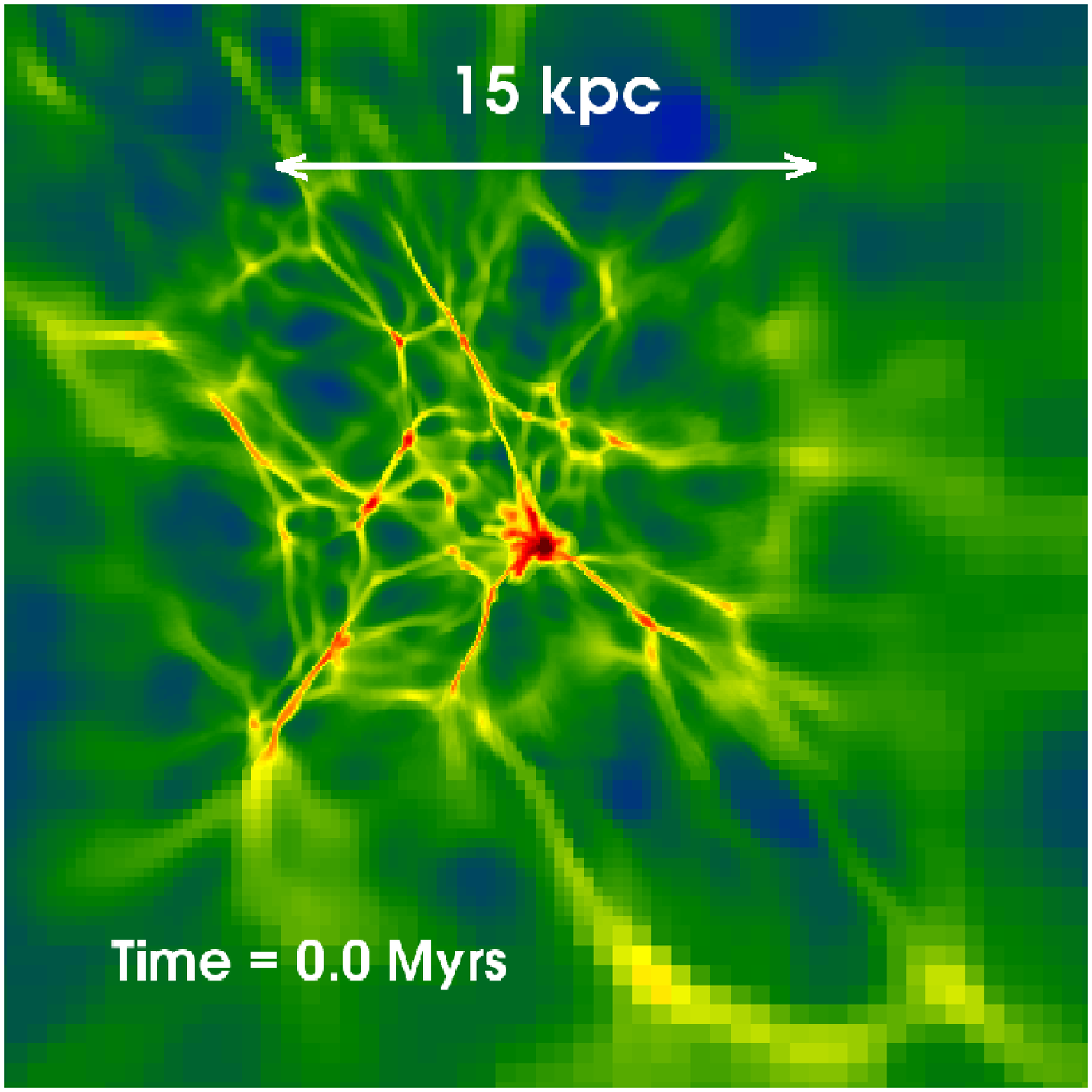} 
\includegraphics[height=4.0cm, width=5cm]{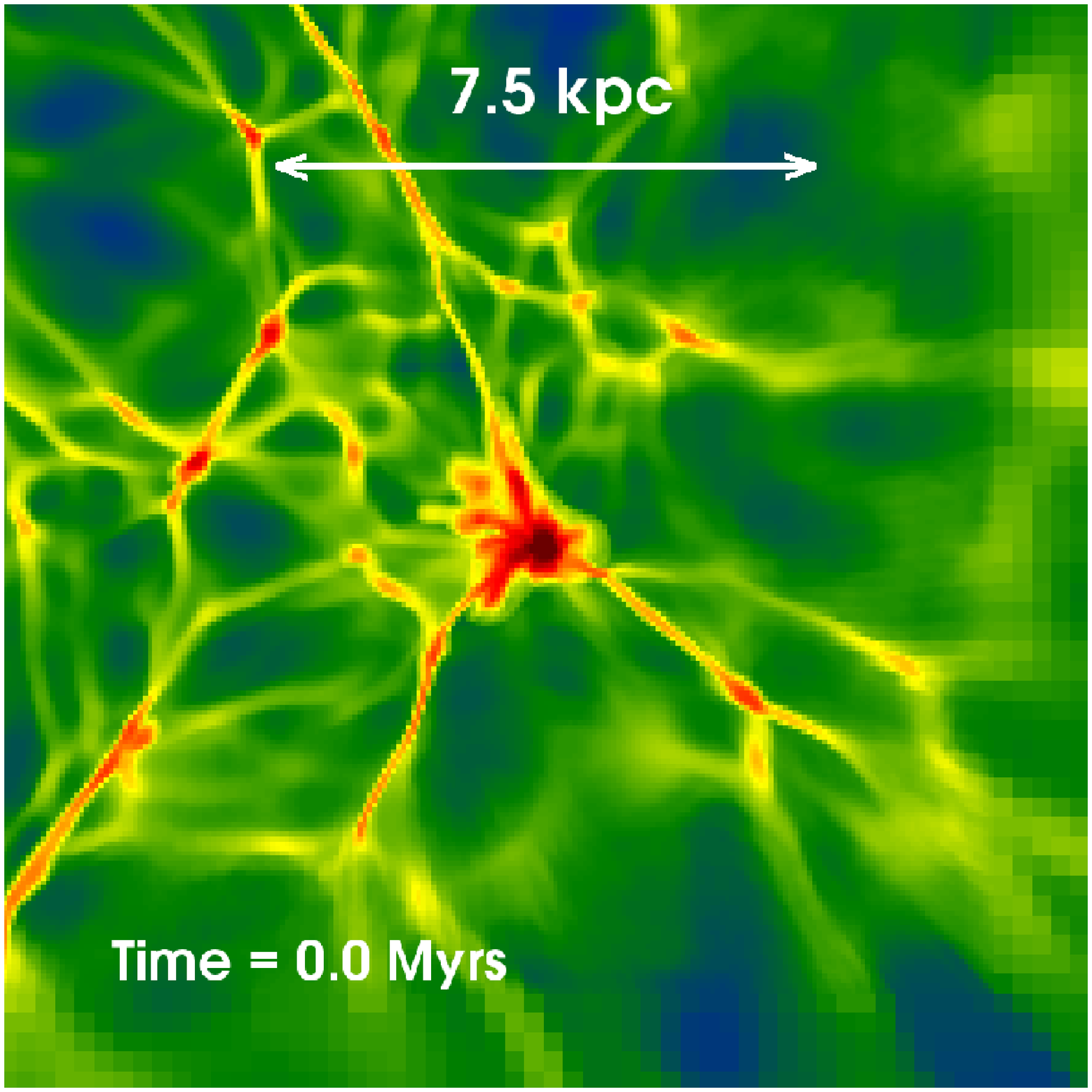} 
\includegraphics[height=4.0cm, width=5cm]{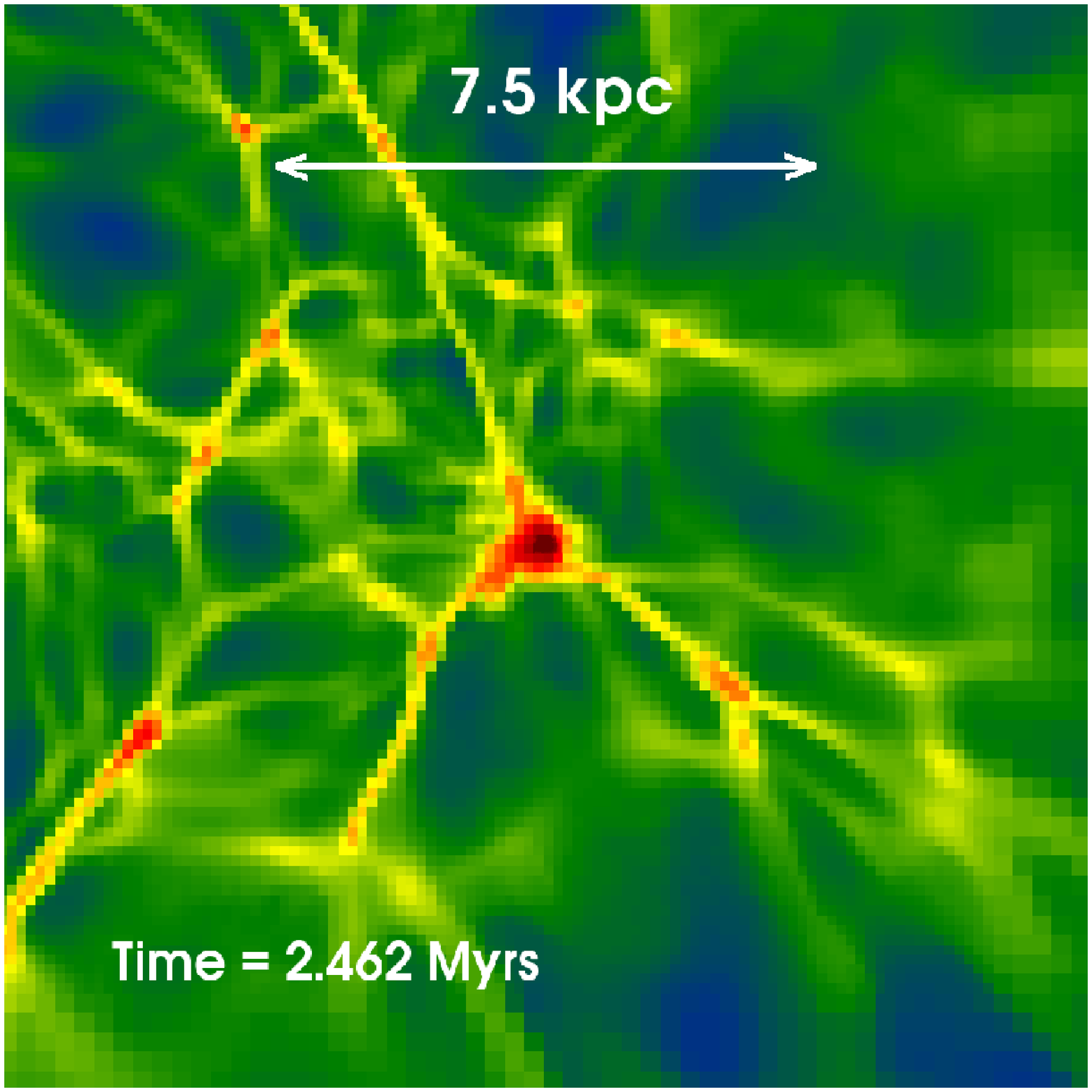} 
\includegraphics[height=4.0cm, width=5cm]{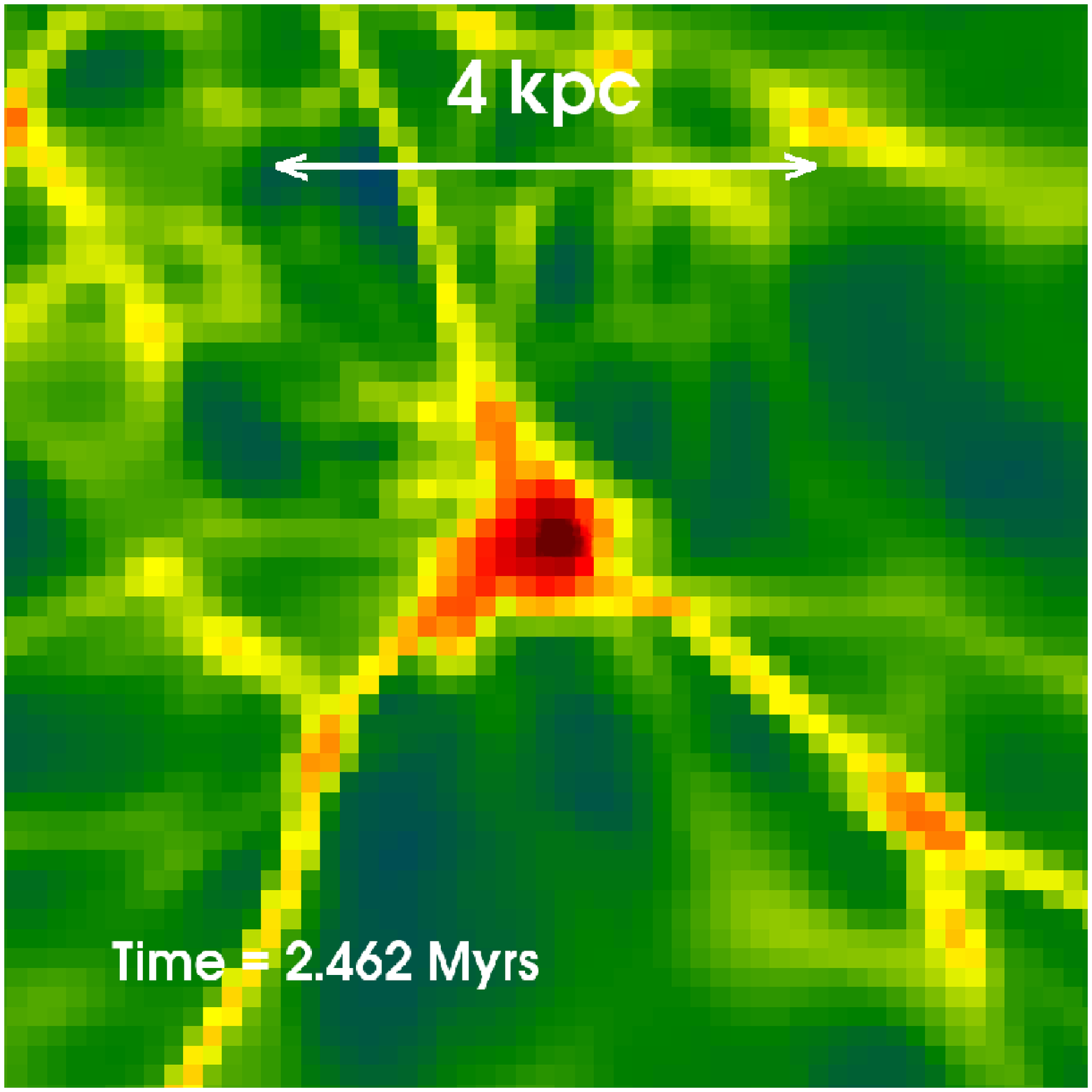} 
\includegraphics[height=4.0cm, width=5cm]{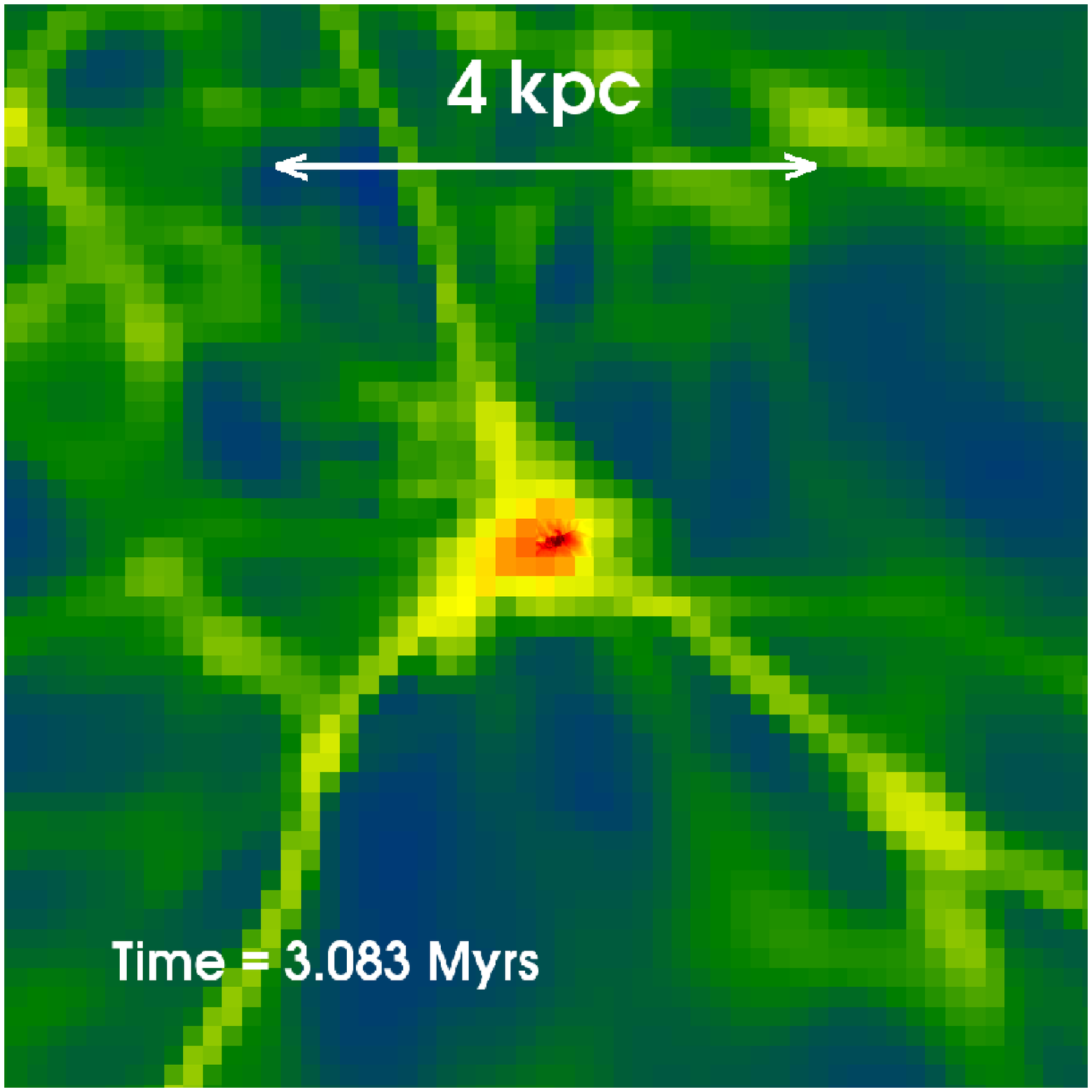} 
\includegraphics[height=4.0cm, width=5cm]{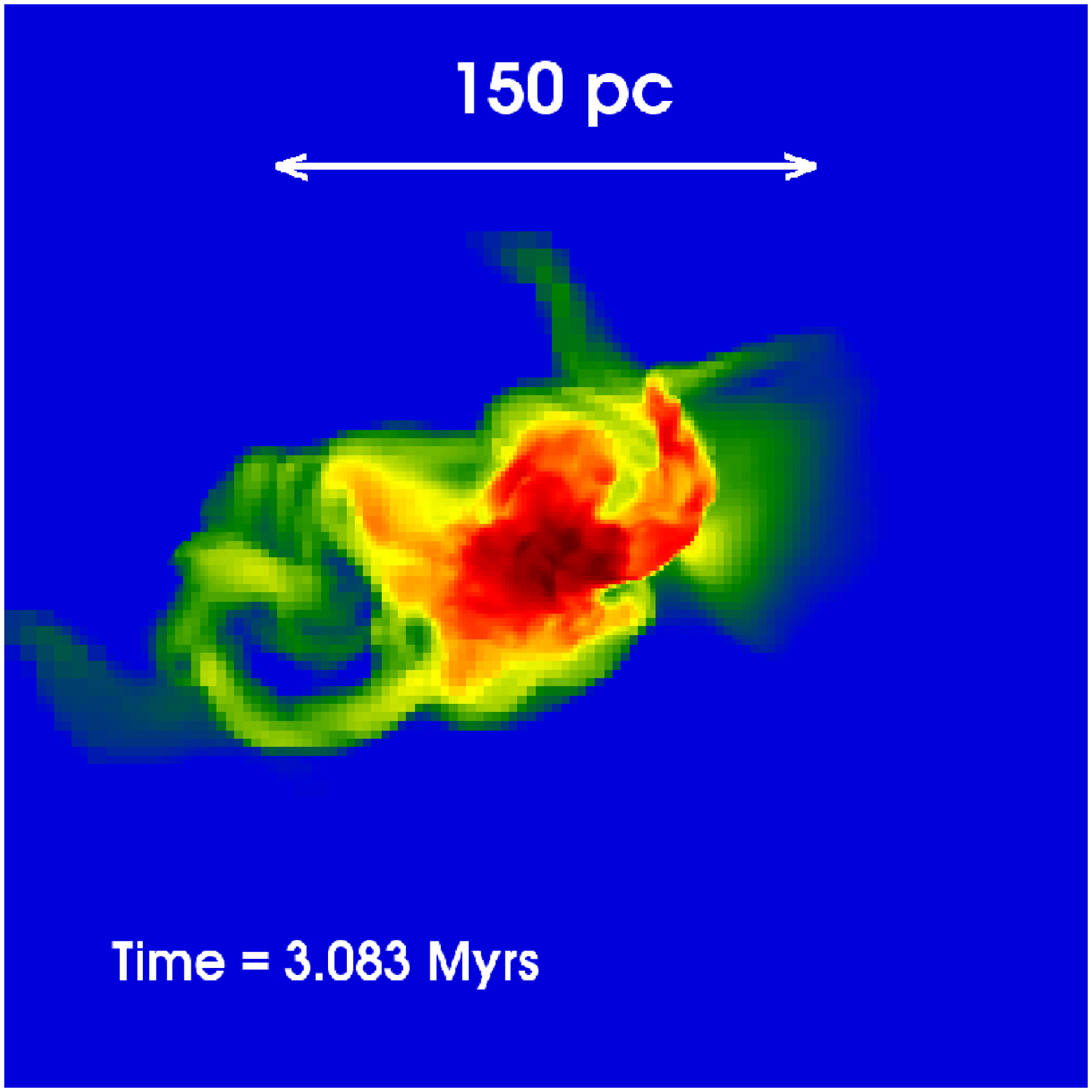} 
\includegraphics[height=4.0cm, width=5cm]{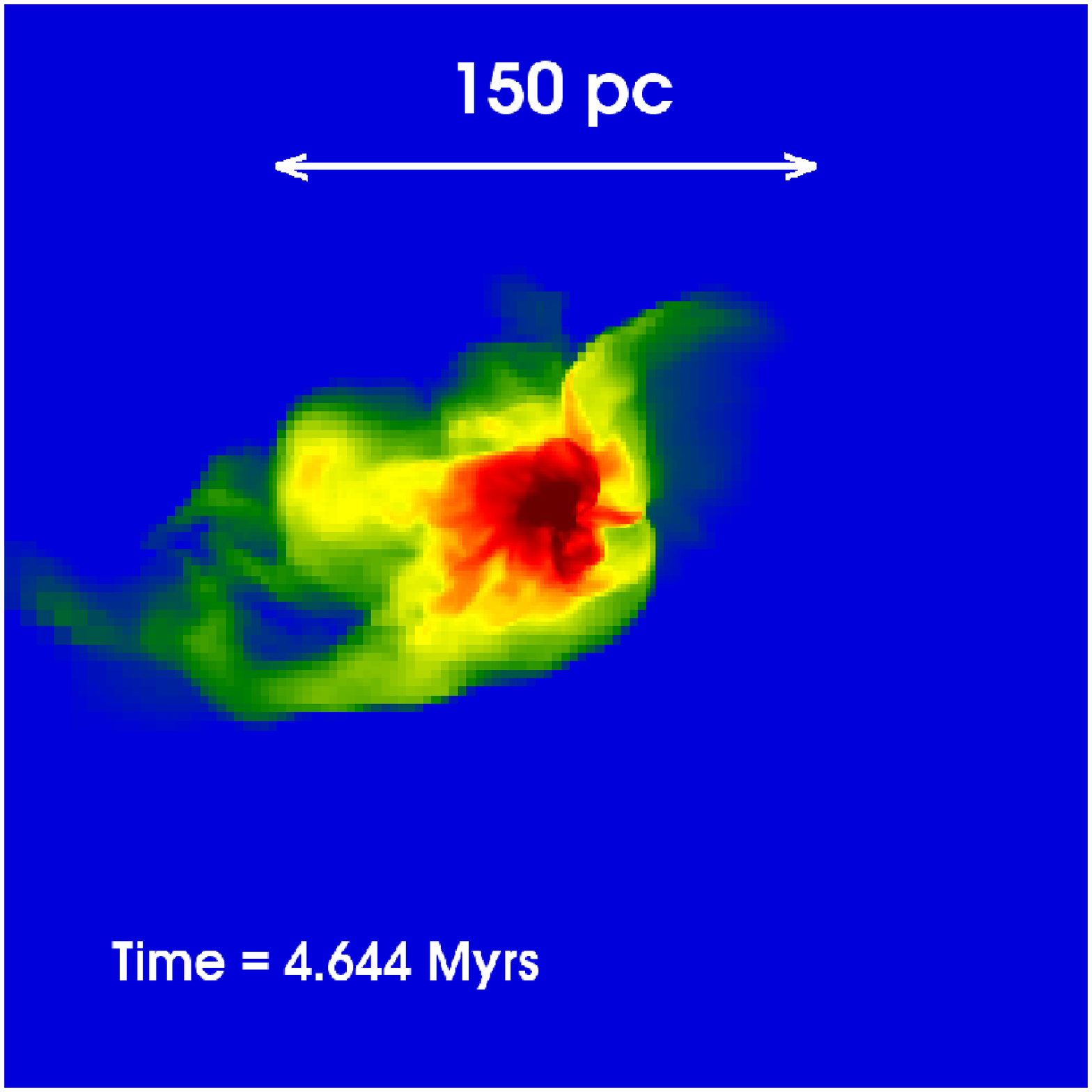} 
\includegraphics[height=4.0cm, width=5cm]{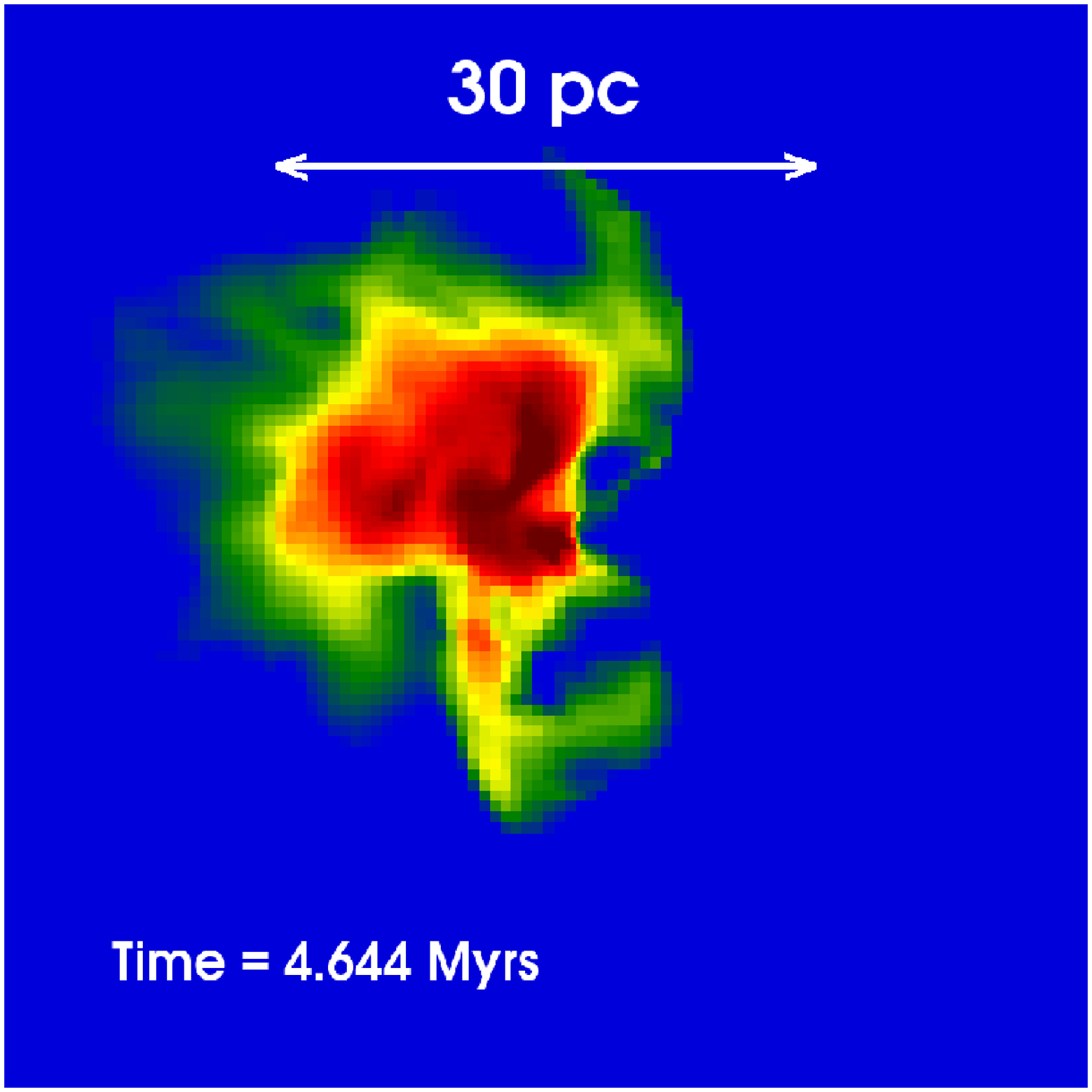}  
\includegraphics[height=4.0cm, width=5cm]{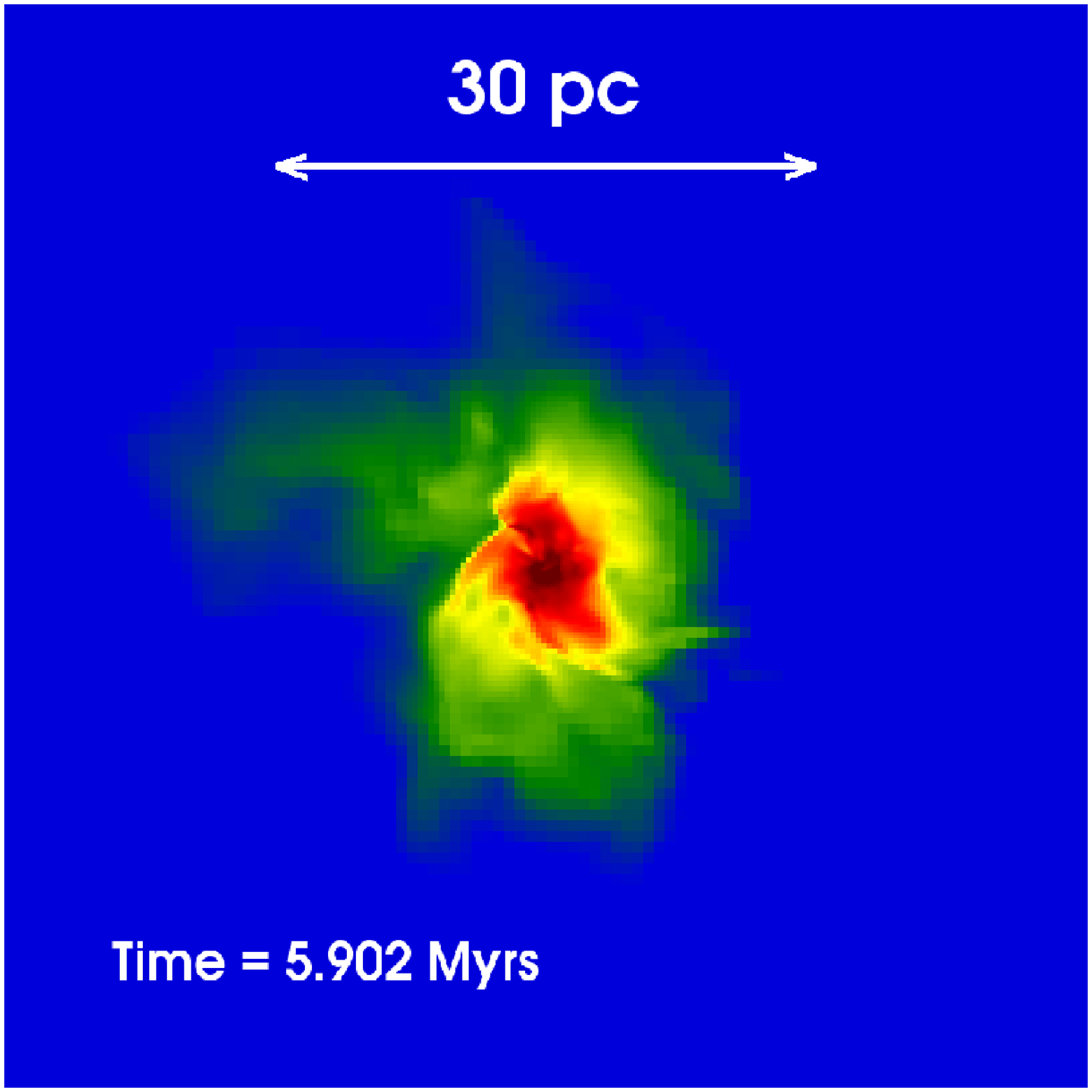}  
\includegraphics[height=4.0cm, width=5cm]{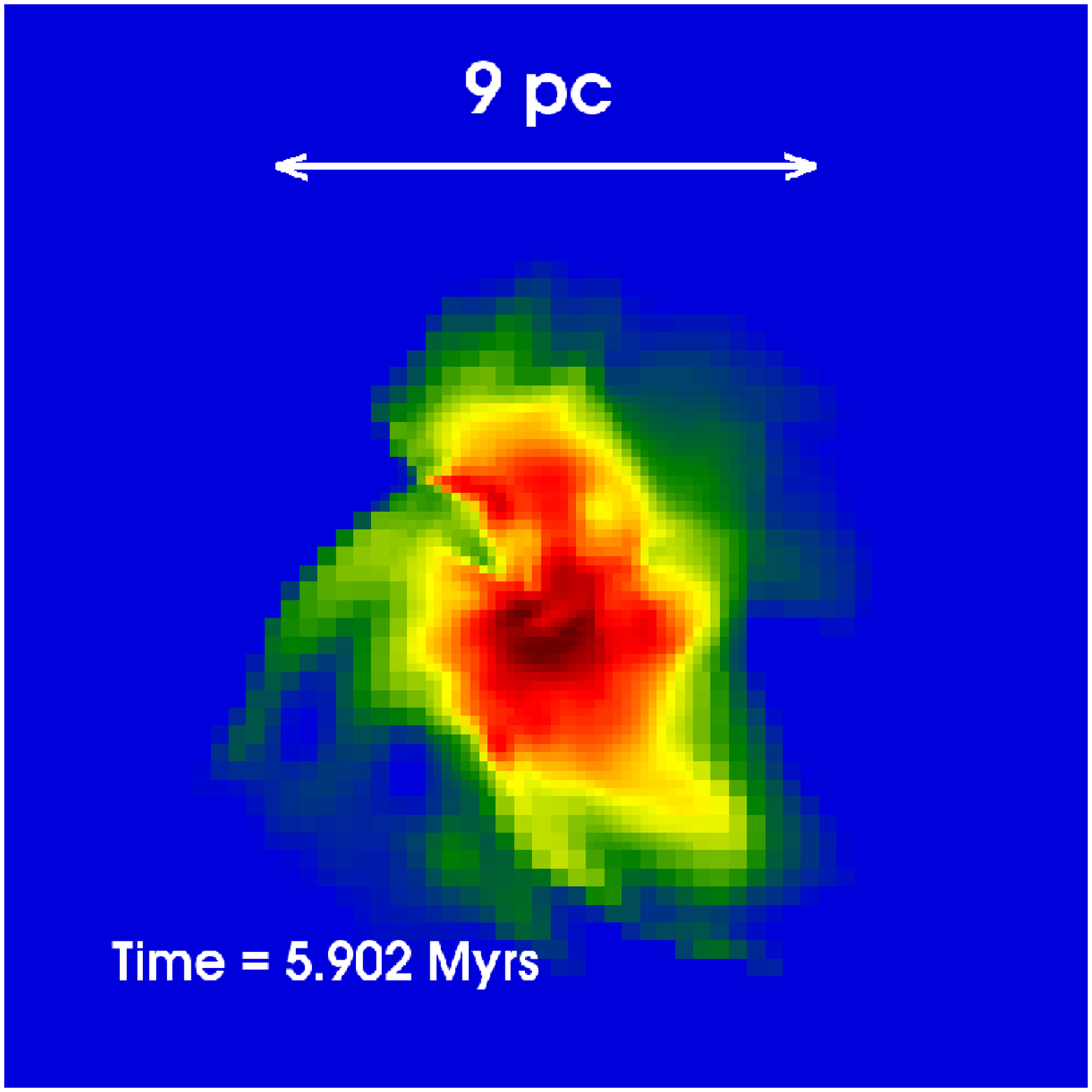}  
\includegraphics[height=4.0cm, width=5cm]{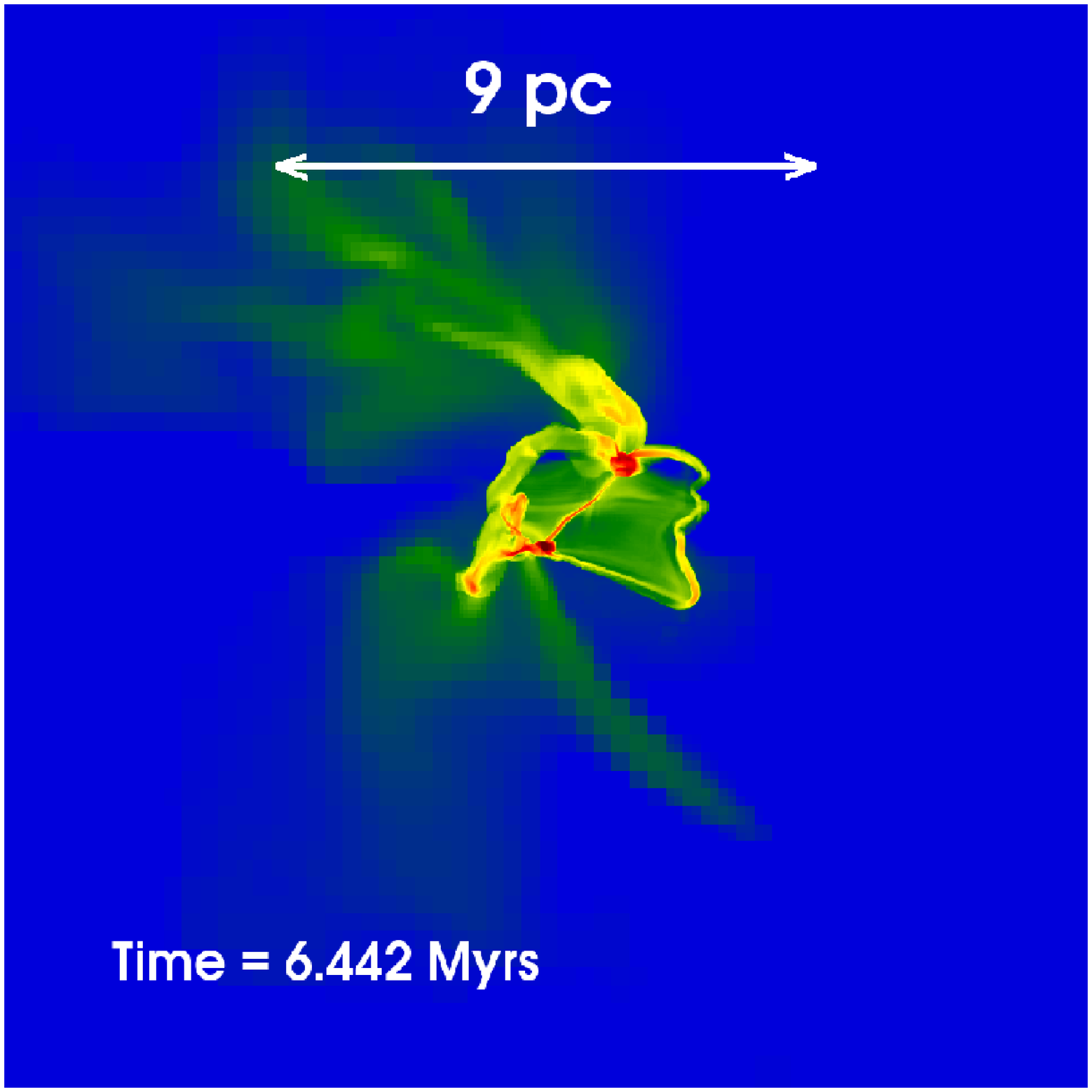}  
\includegraphics[height=4.0cm, width=5cm]{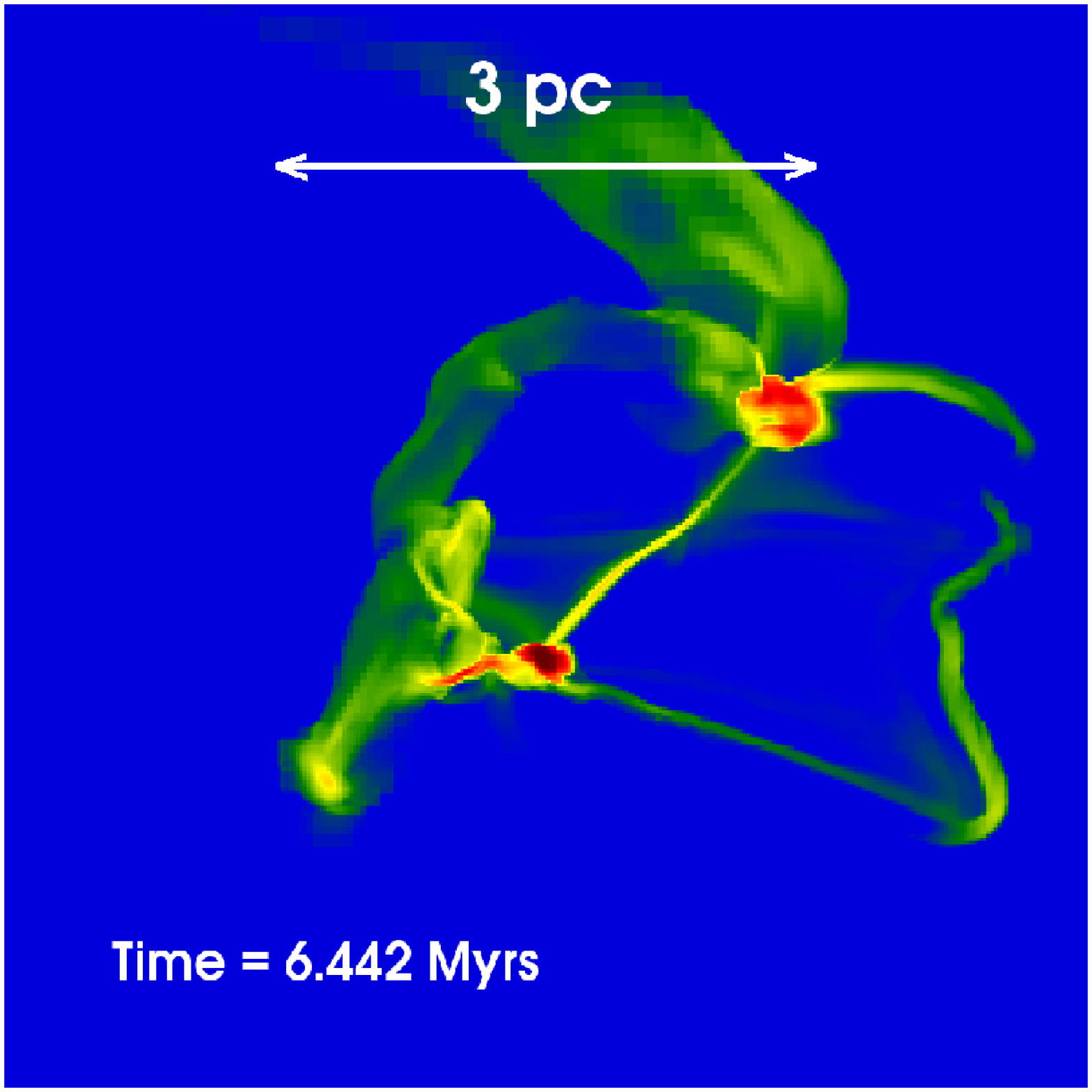} 
\includegraphics[height=4.0cm, width=5cm]{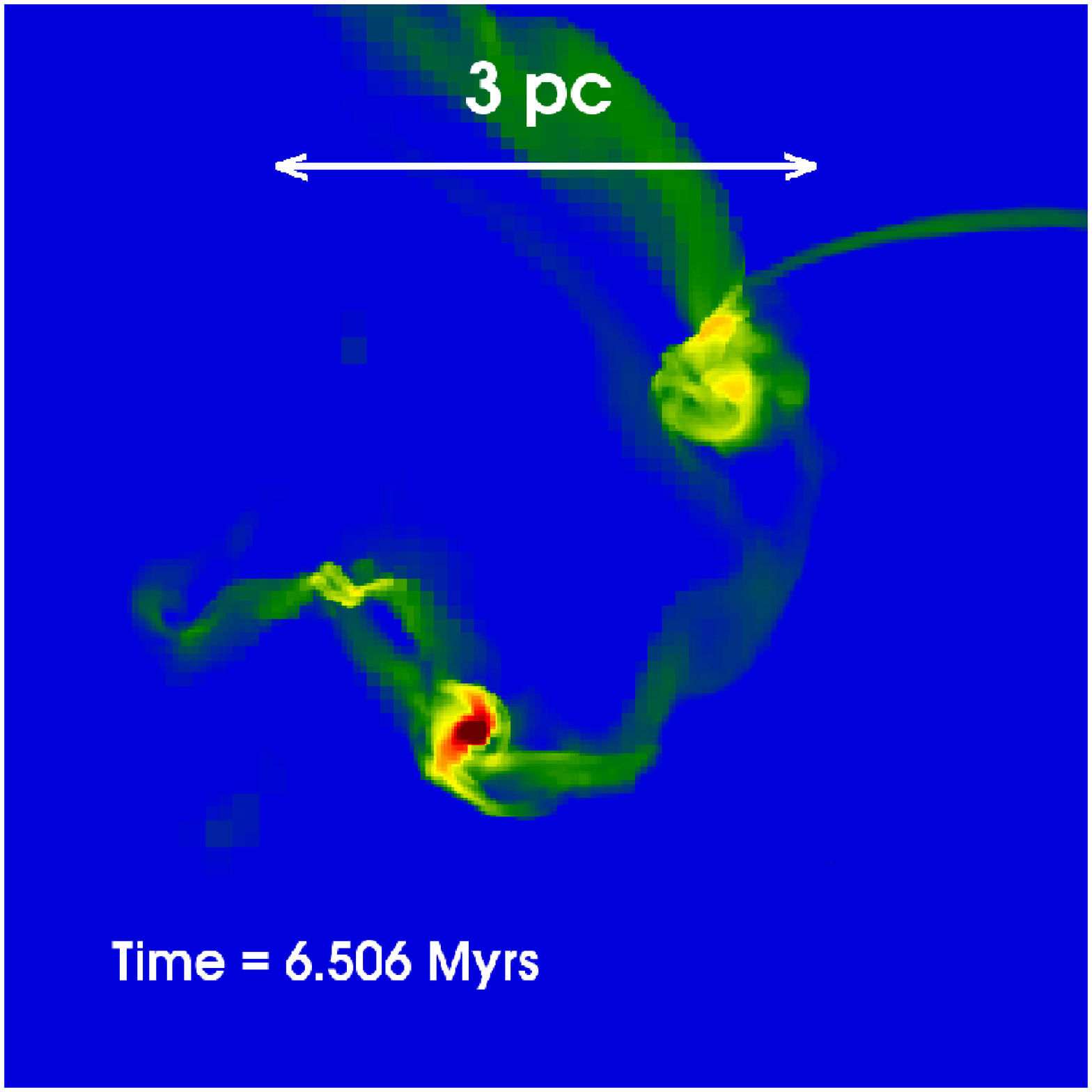} 
\includegraphics[height=4.0cm, width=5cm]{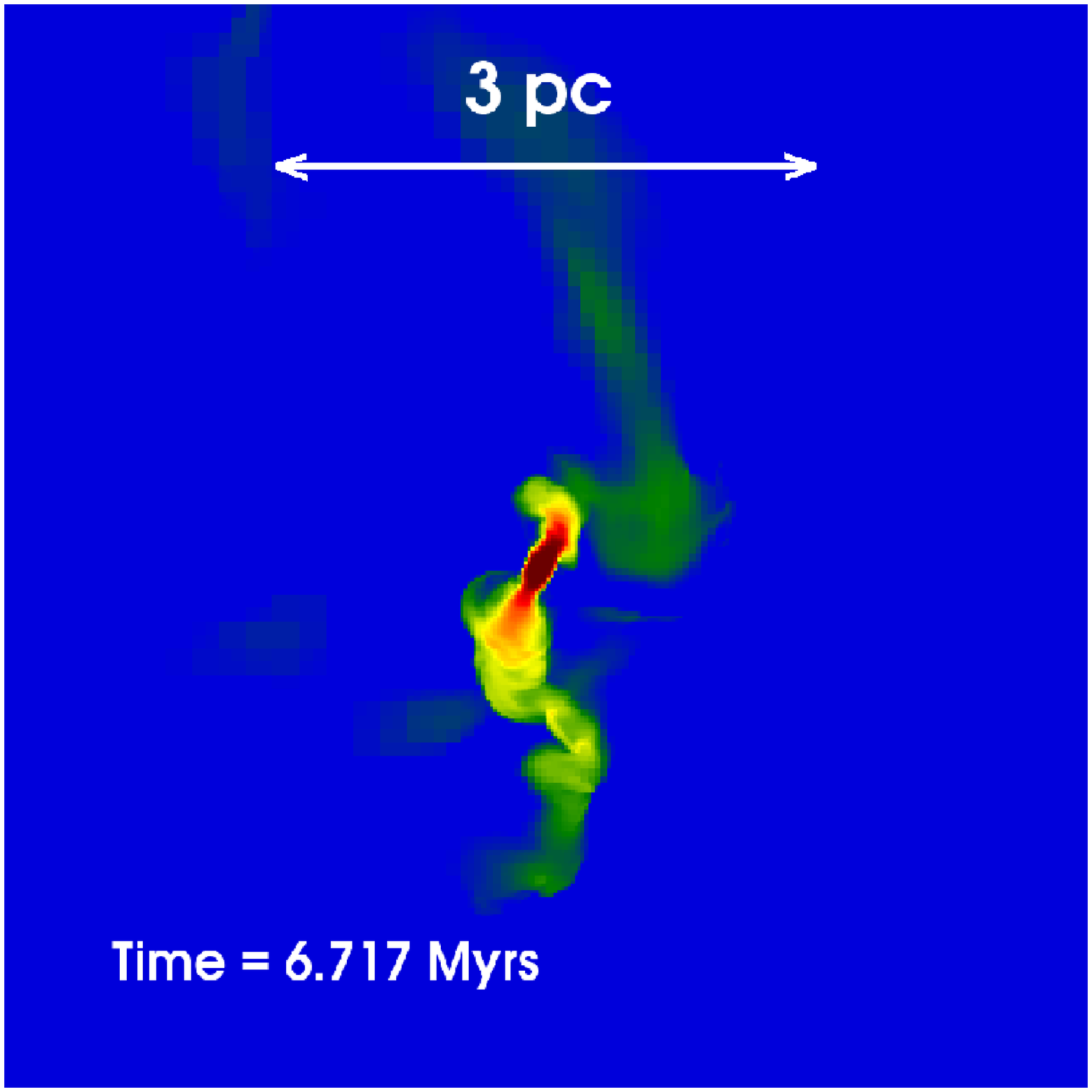} 
\includegraphics[height=4.0cm, width=5cm]{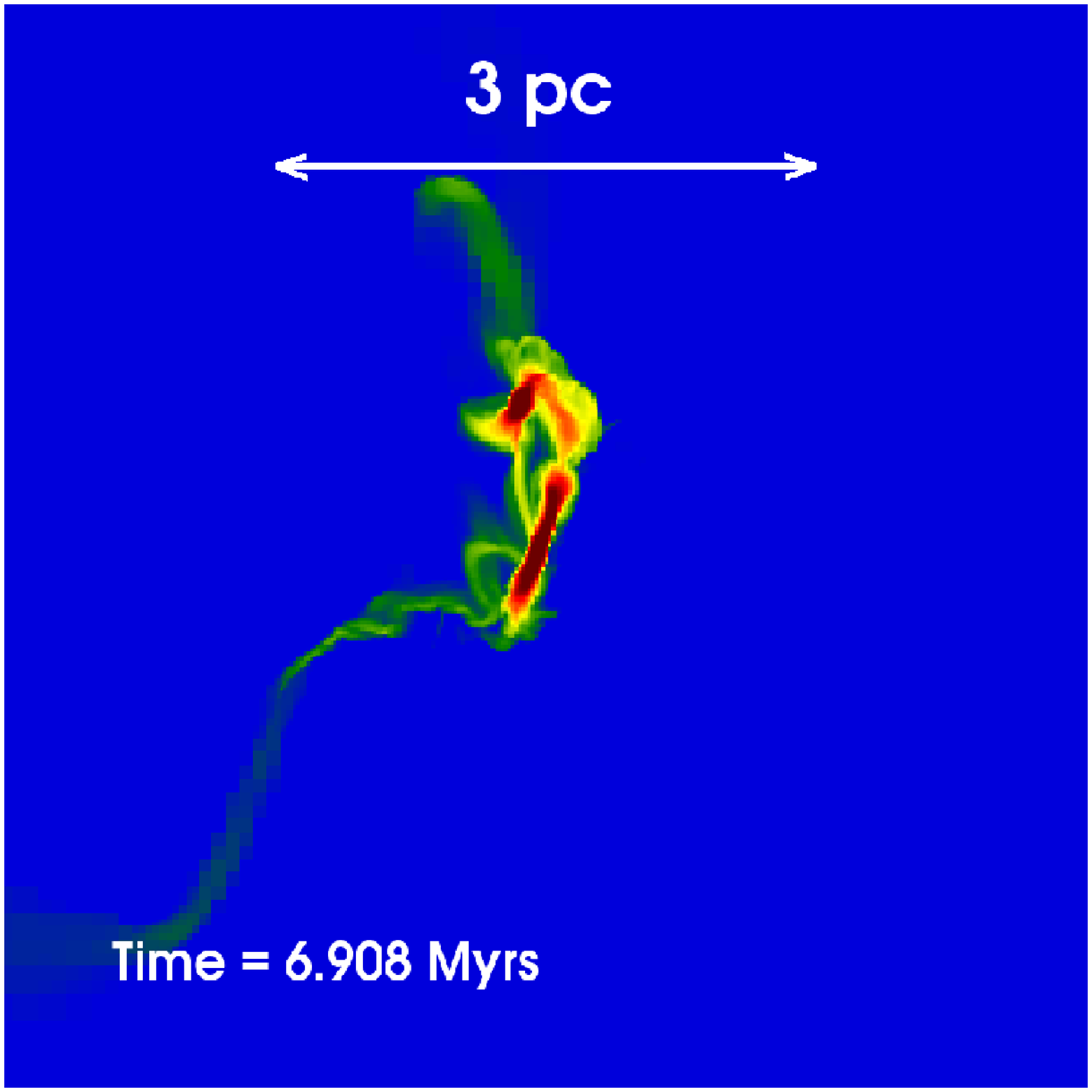} 
\caption[\label{SimBMap} ]{\label{SimBMap}
  Same as figure \ref{SimAMap} for simulation B. 
  The gas is collapsing in a halo of DM mass of $5.37\times
  10^{7} M_{\odot}$, virial velocity of $\sim 19 \kms$ and virial
  temperature of $\sim 13000 \, \rm{K}$. 
  Note how the gas fragments into  three clumps at $T = 6.442$ Myrs which tidally distort
  each other and undergo a violent dynamical interaction.}
\end{figure*}



\begin{figure*}
\includegraphics[height=4.0cm, width=5cm]{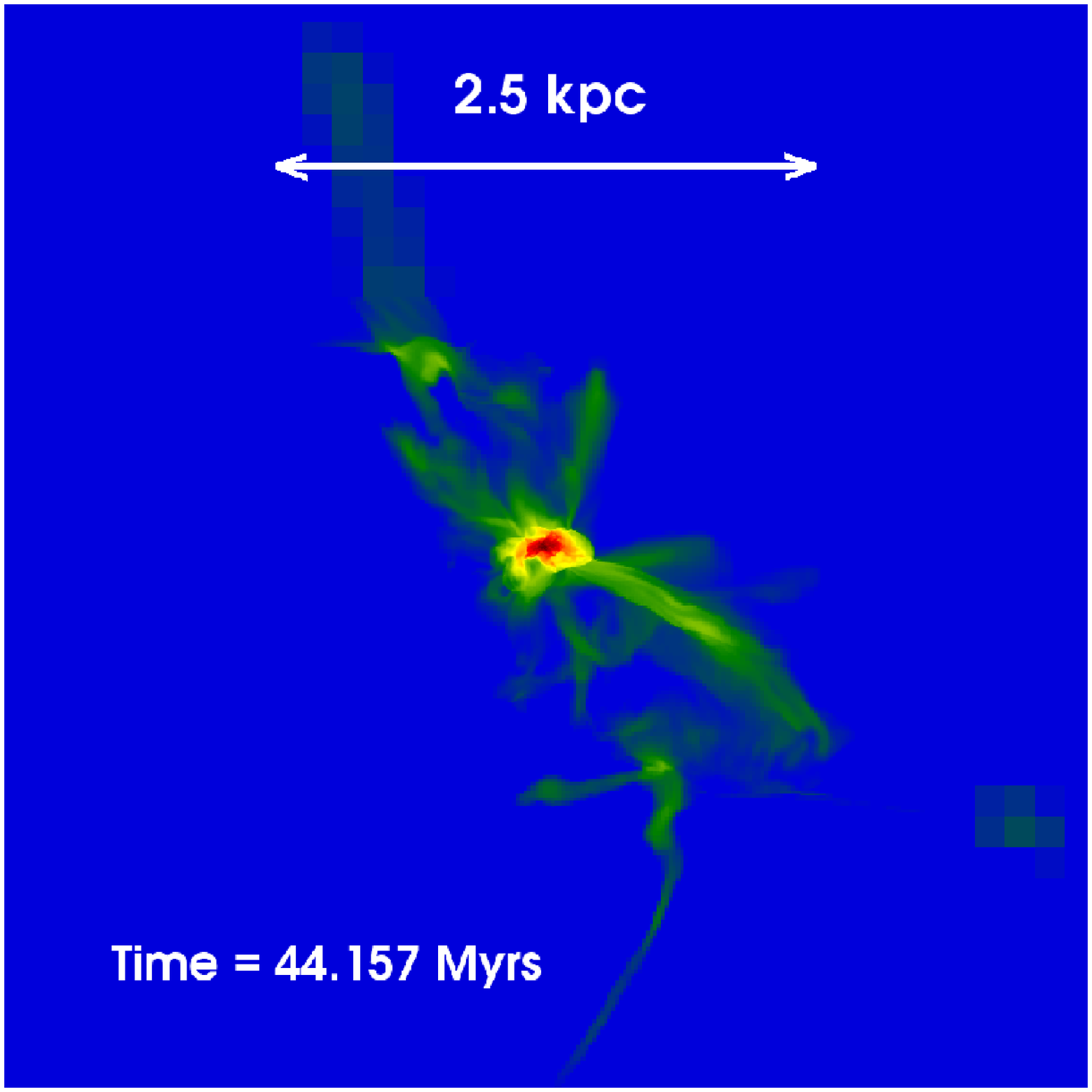} 
\includegraphics[height=4.0cm, width=5cm]{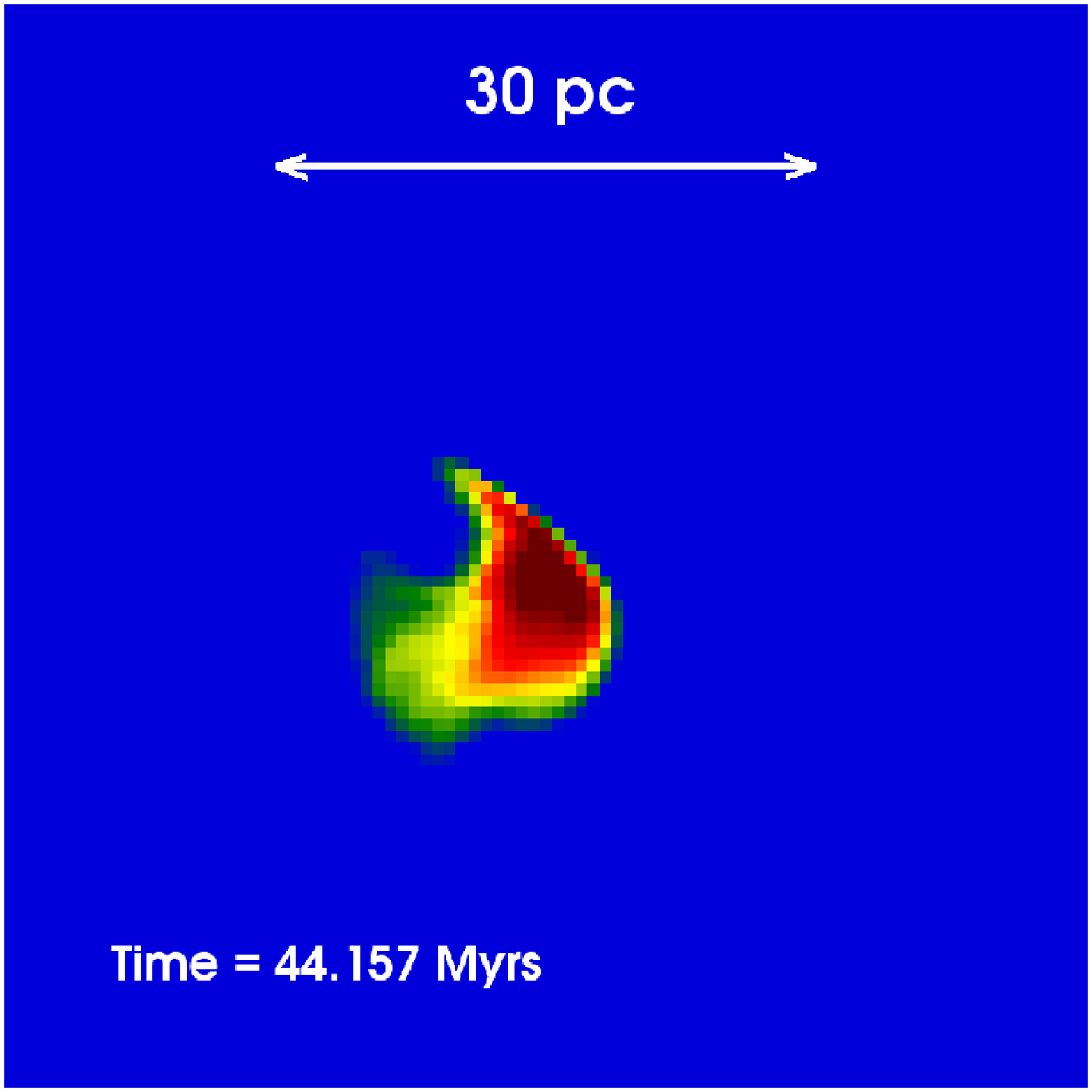} 
\includegraphics[height=4.0cm, width=5cm]{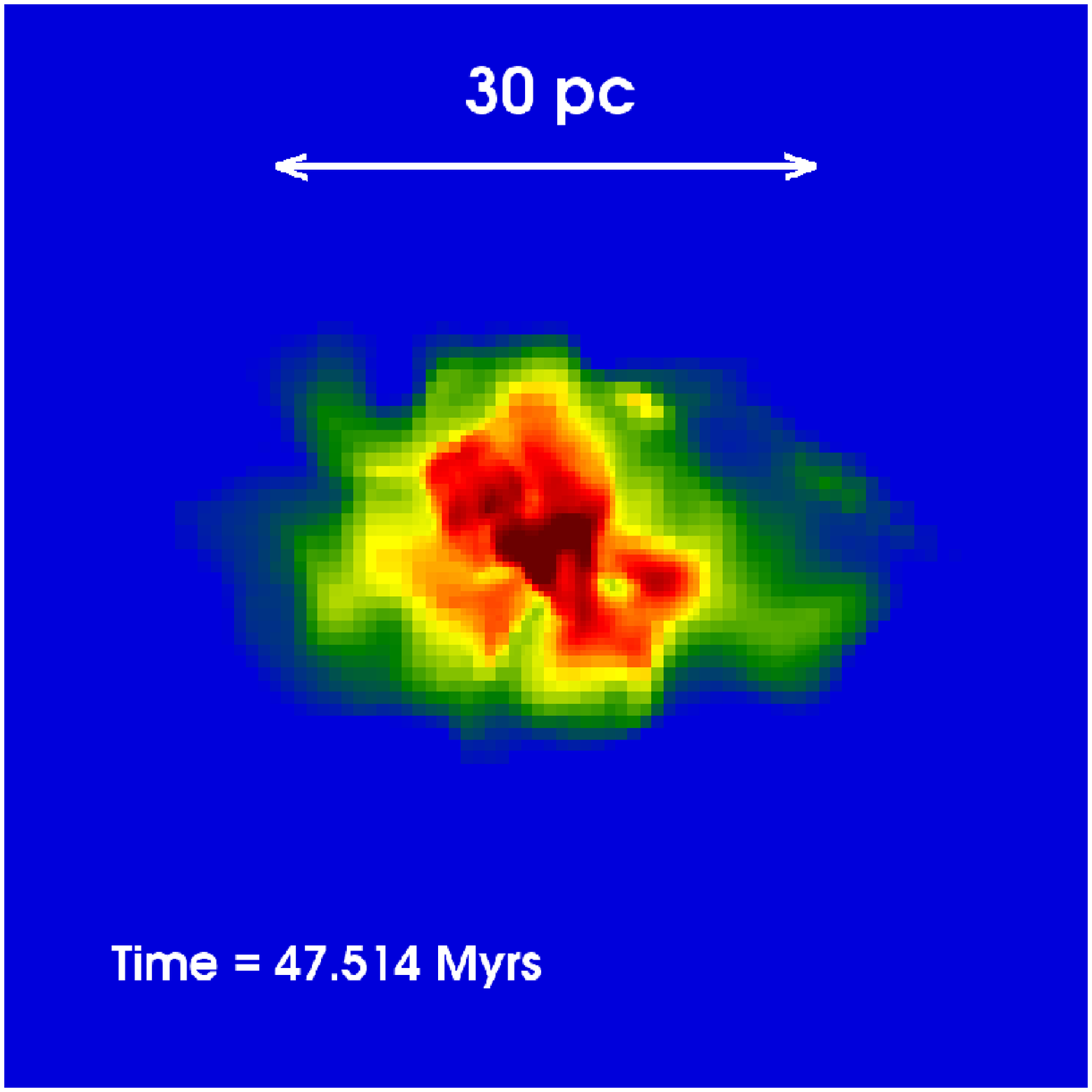} 
\includegraphics[height=4.0cm, width=5cm]{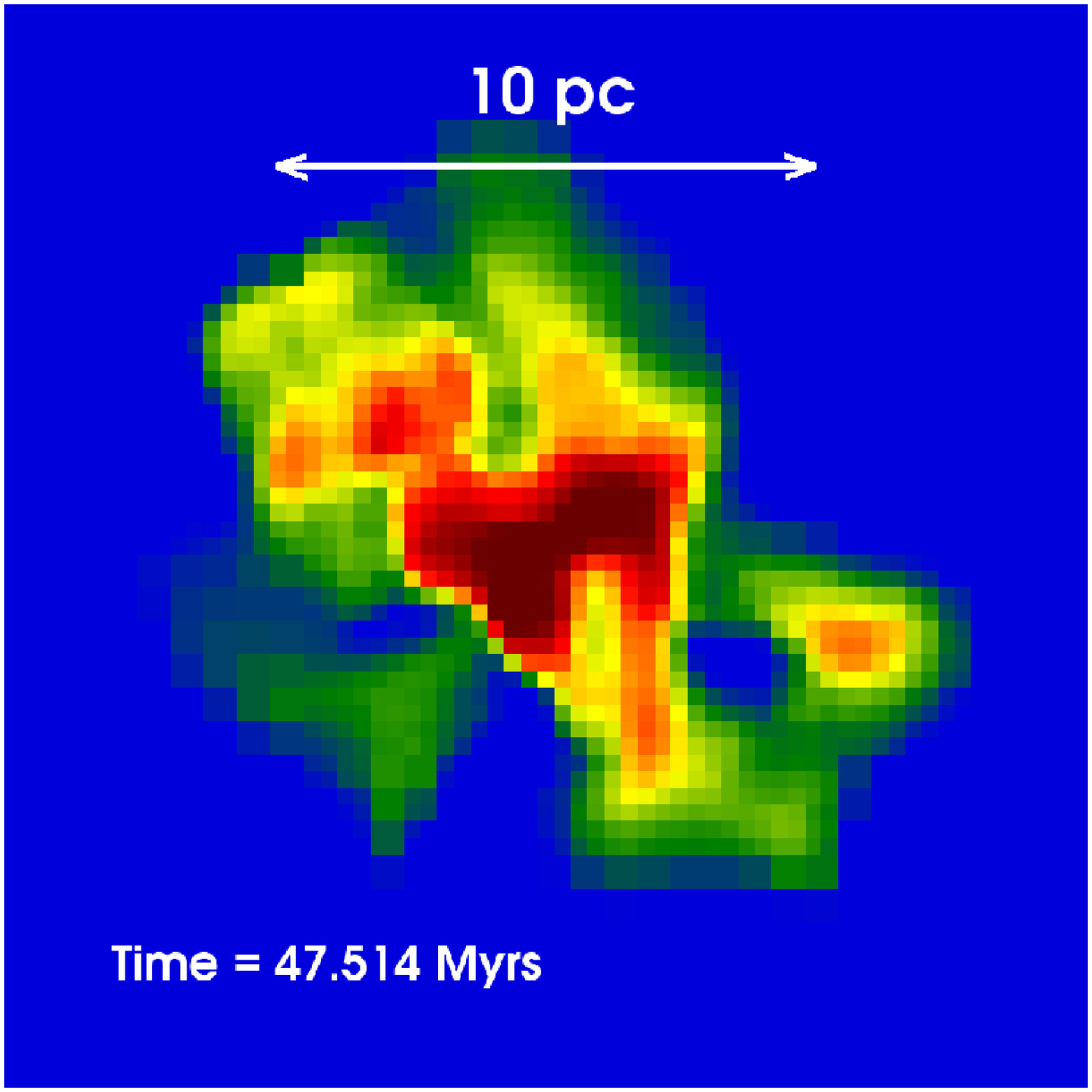} 
\includegraphics[height=4.0cm, width=5cm]{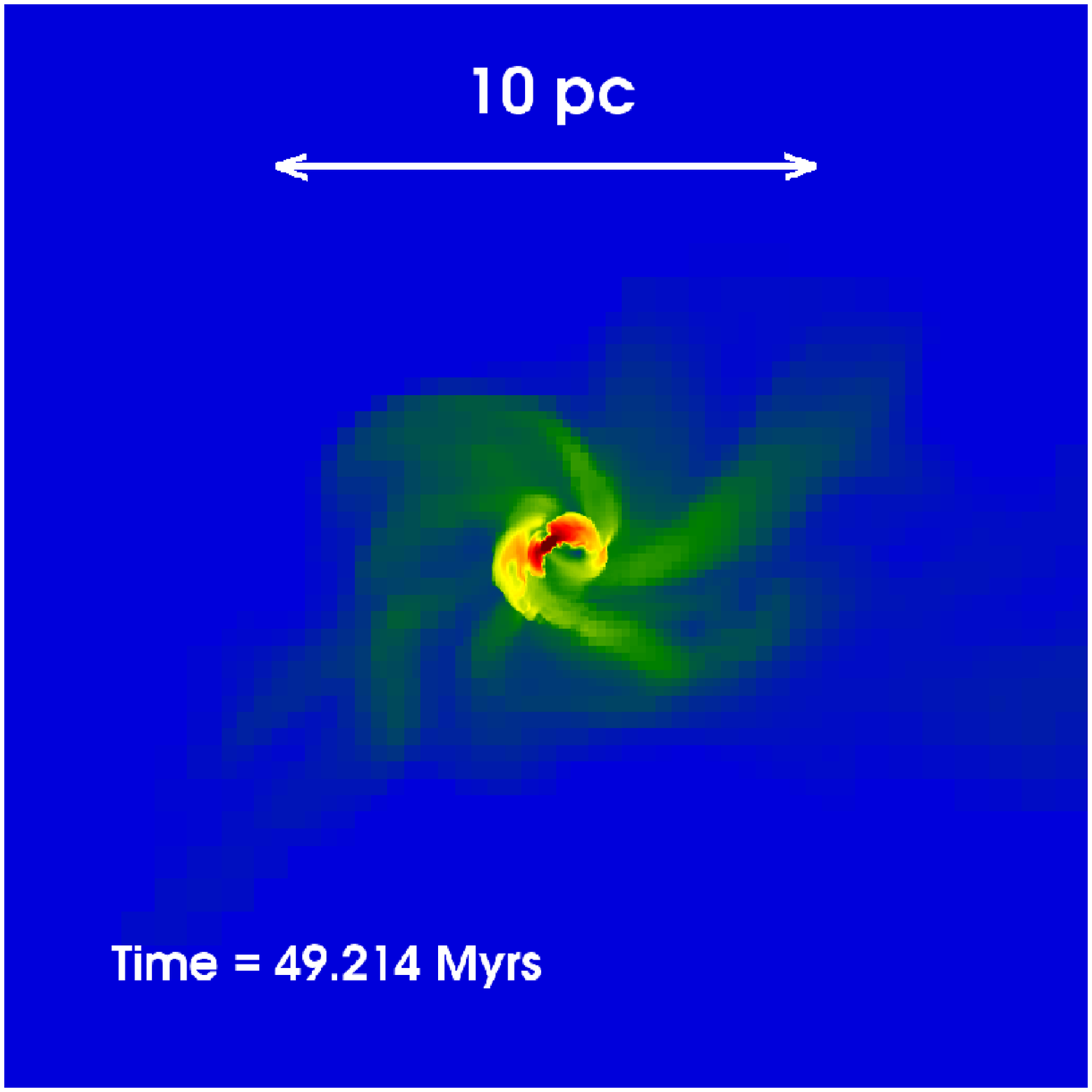} 
\includegraphics[height=4.0cm, width=5cm]{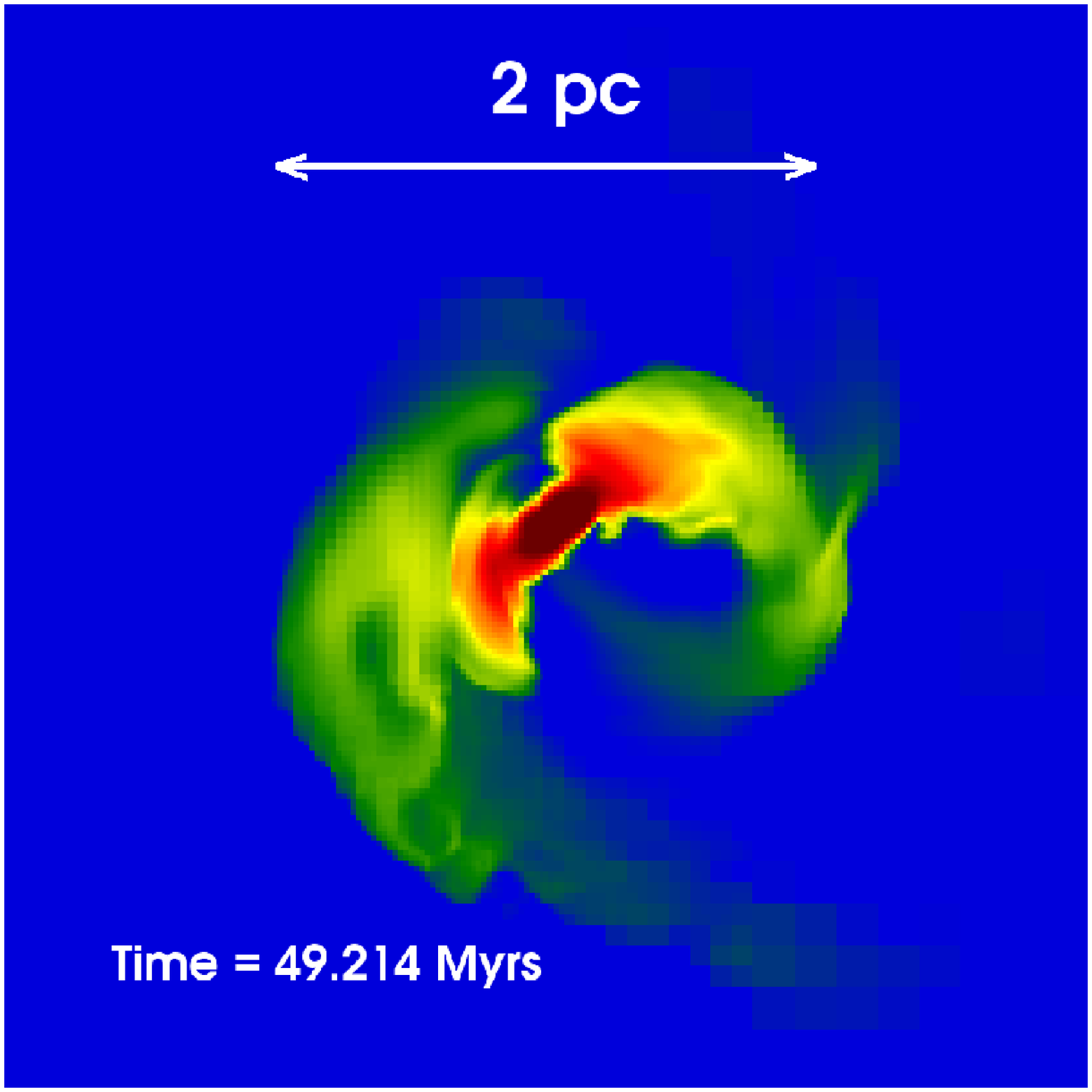} 
\includegraphics[height=4.0cm, width=5cm]{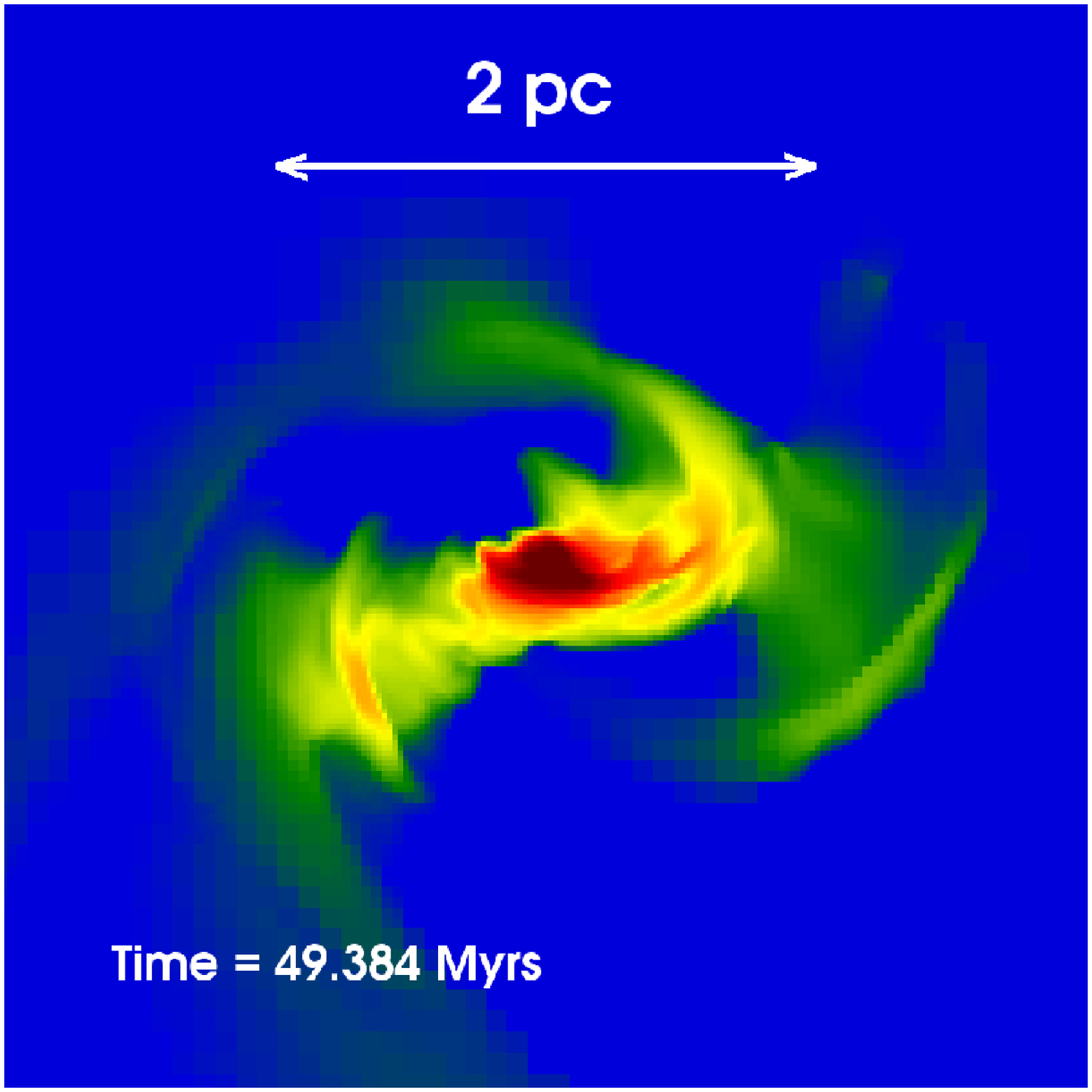} 
\includegraphics[height=4.0cm, width=5cm]{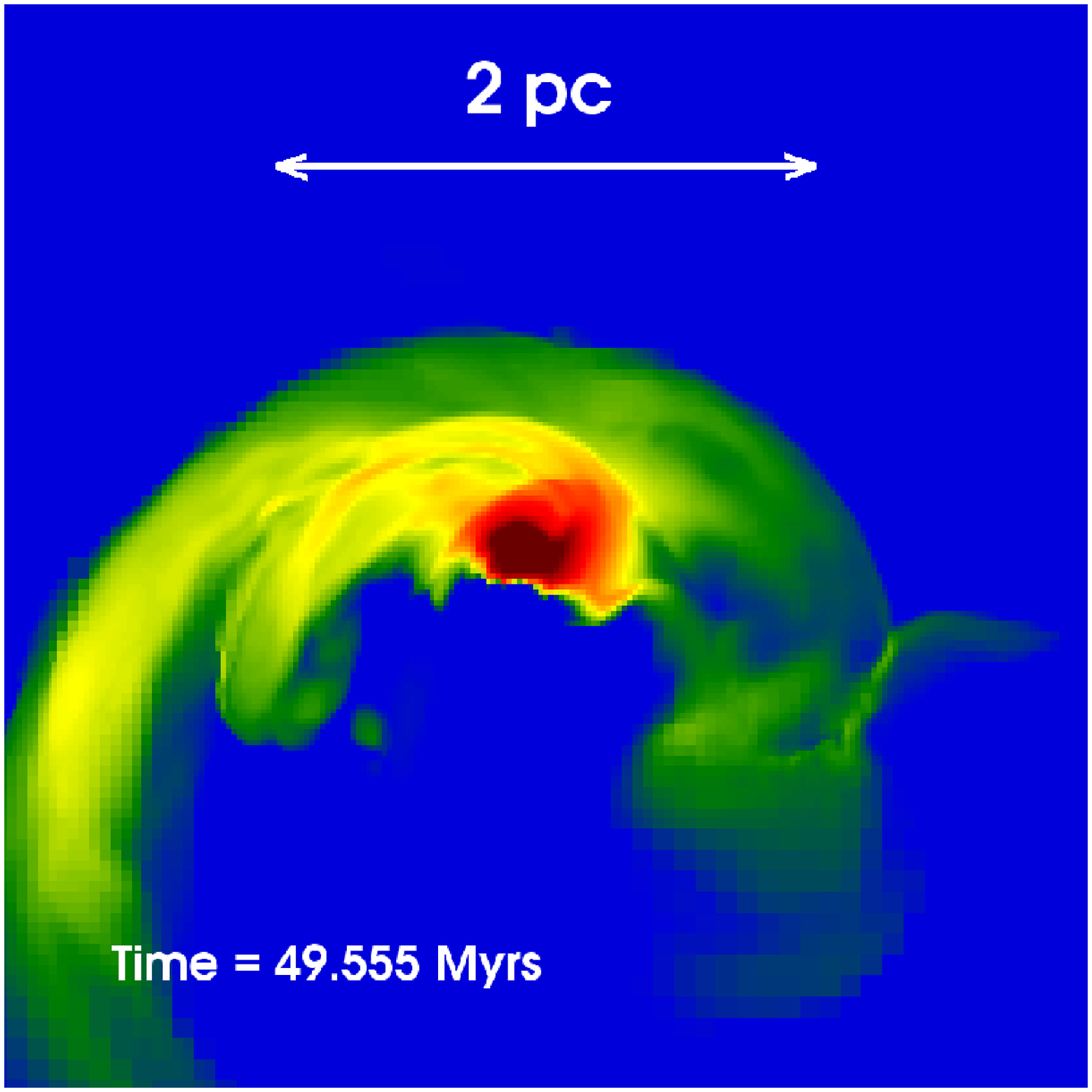}  
\includegraphics[height=4.0cm, width=5cm]{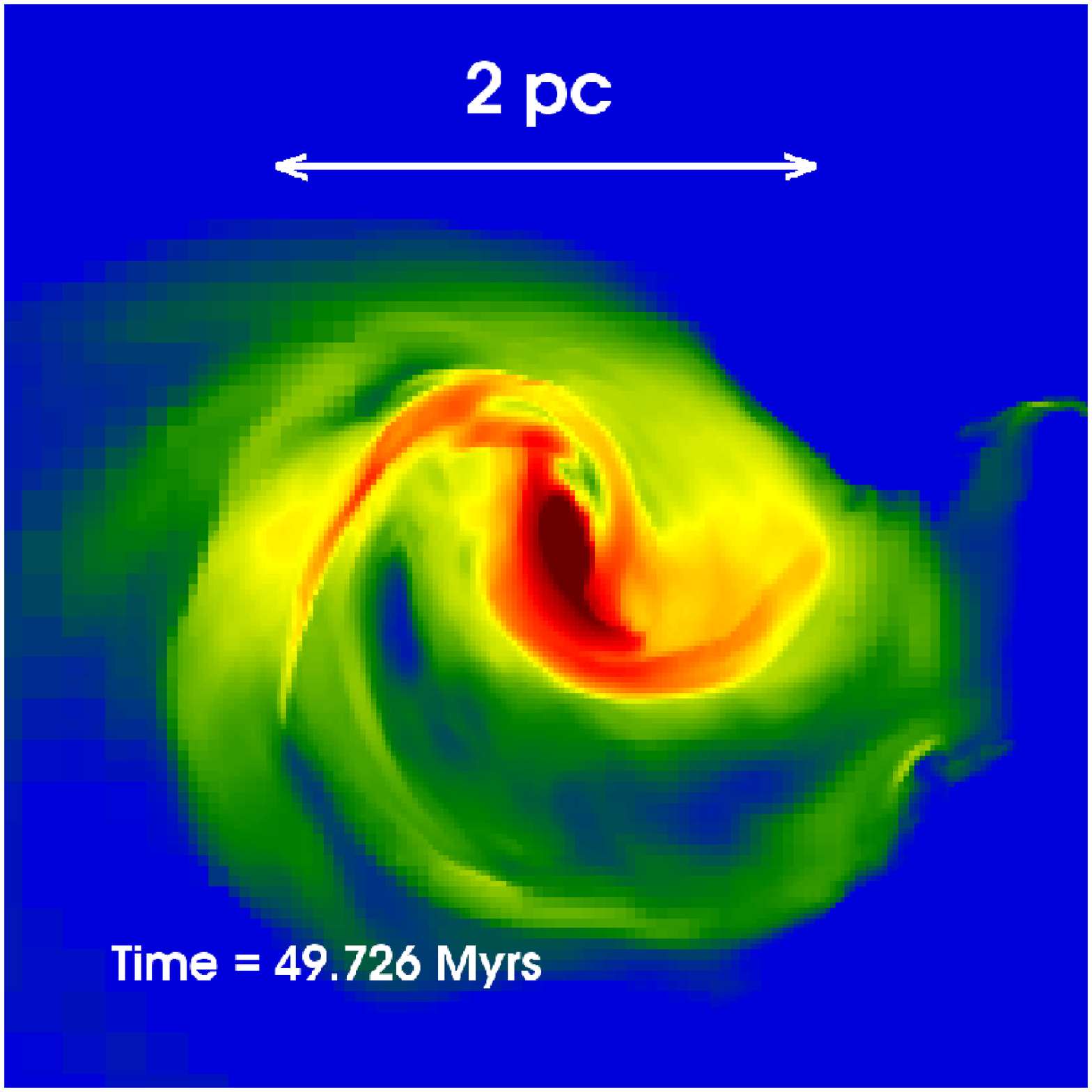}  
\includegraphics[height=4.0cm, width=5cm]{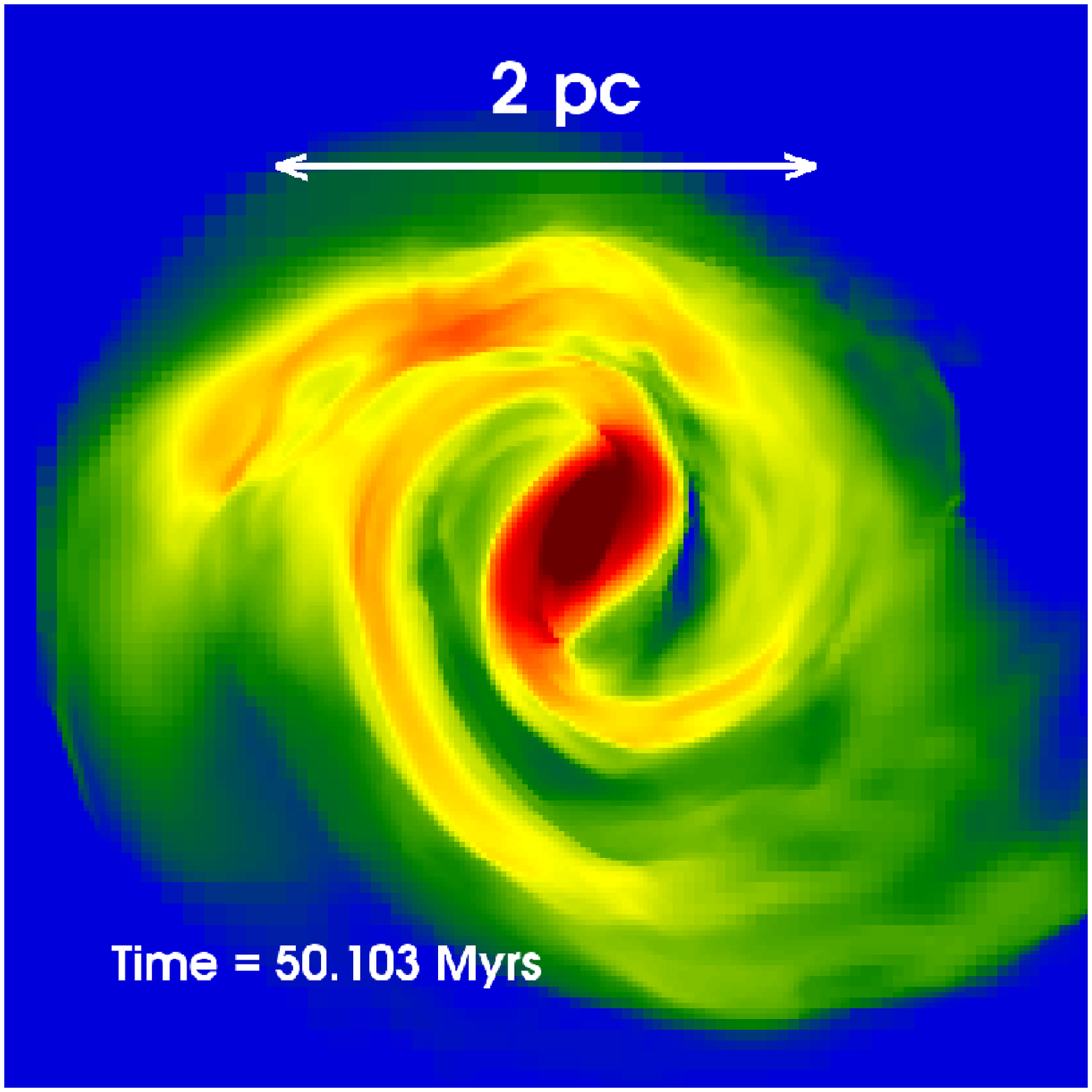}  
\includegraphics[height=4.0cm, width=5cm]{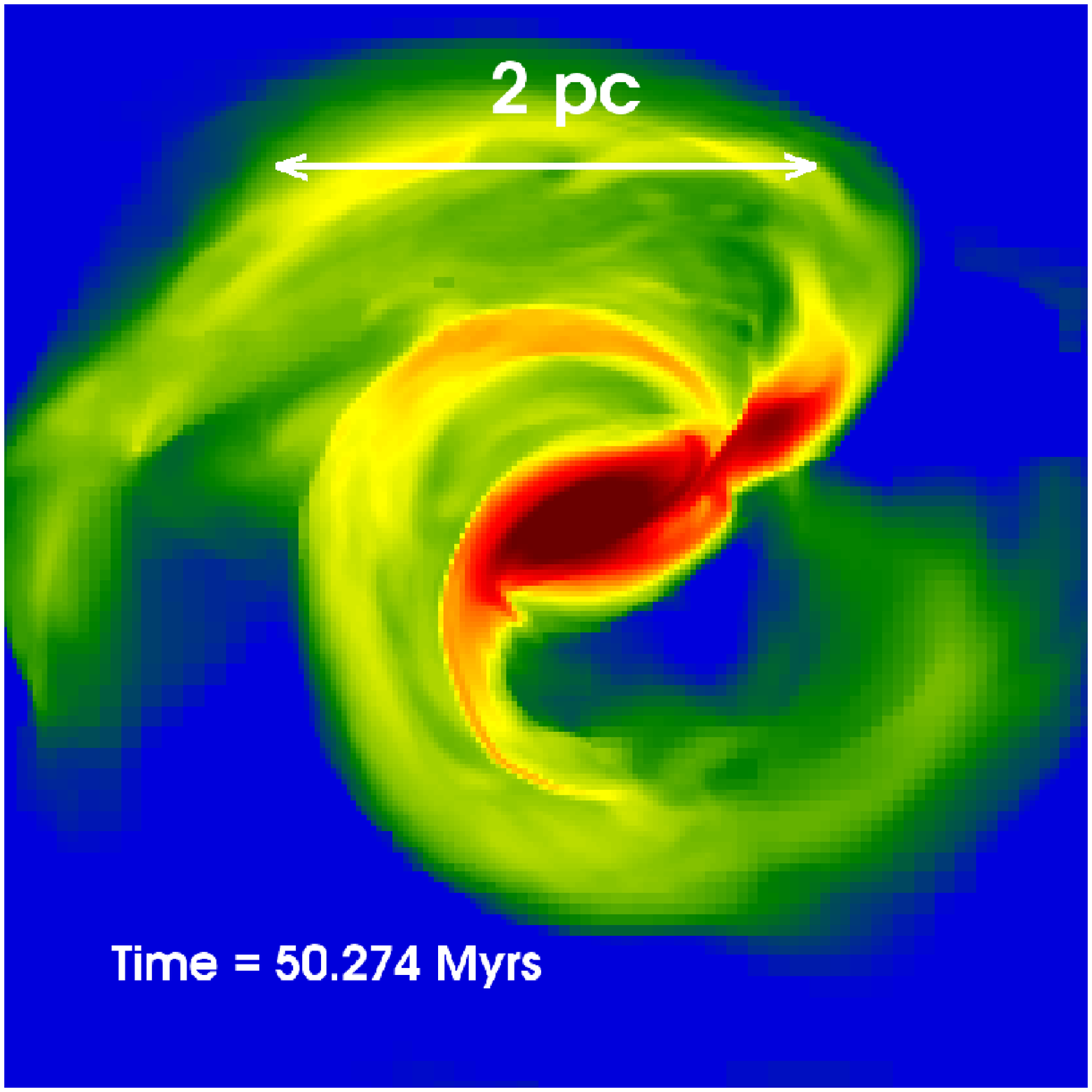}  
\includegraphics[height=4.0cm, width=5cm]{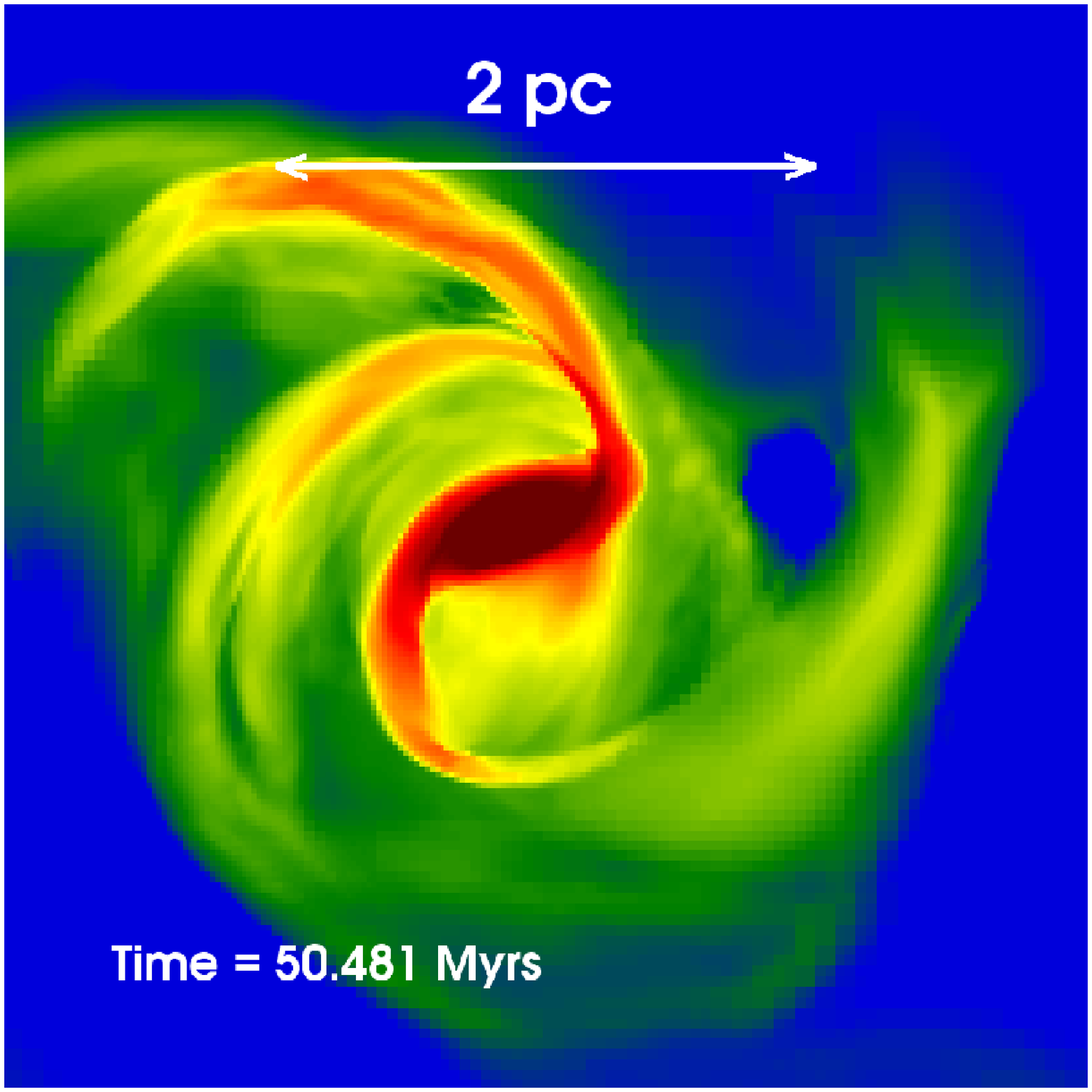} 
\includegraphics[height=4.0cm, width=5cm]{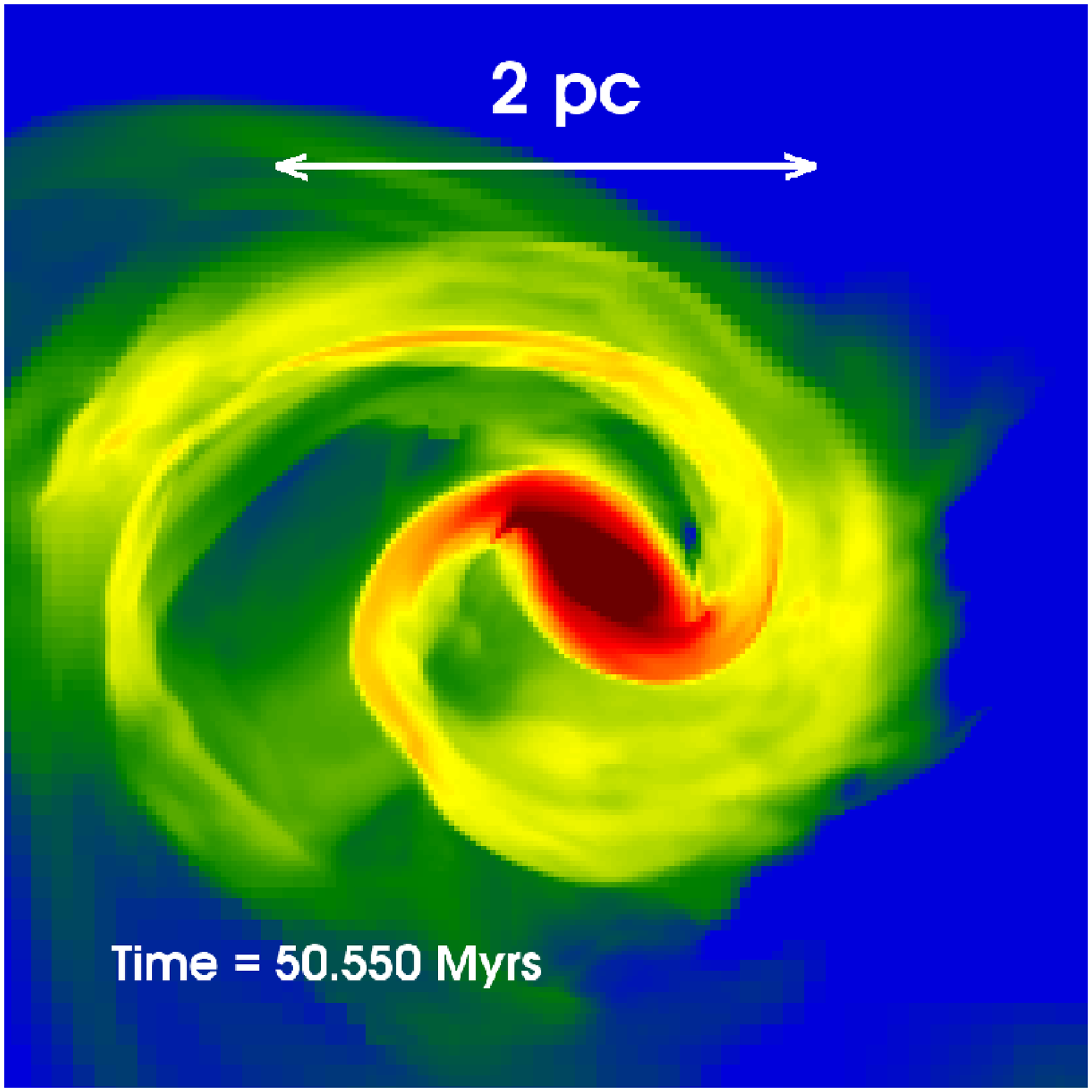} 
\includegraphics[height=4.0cm, width=5cm]{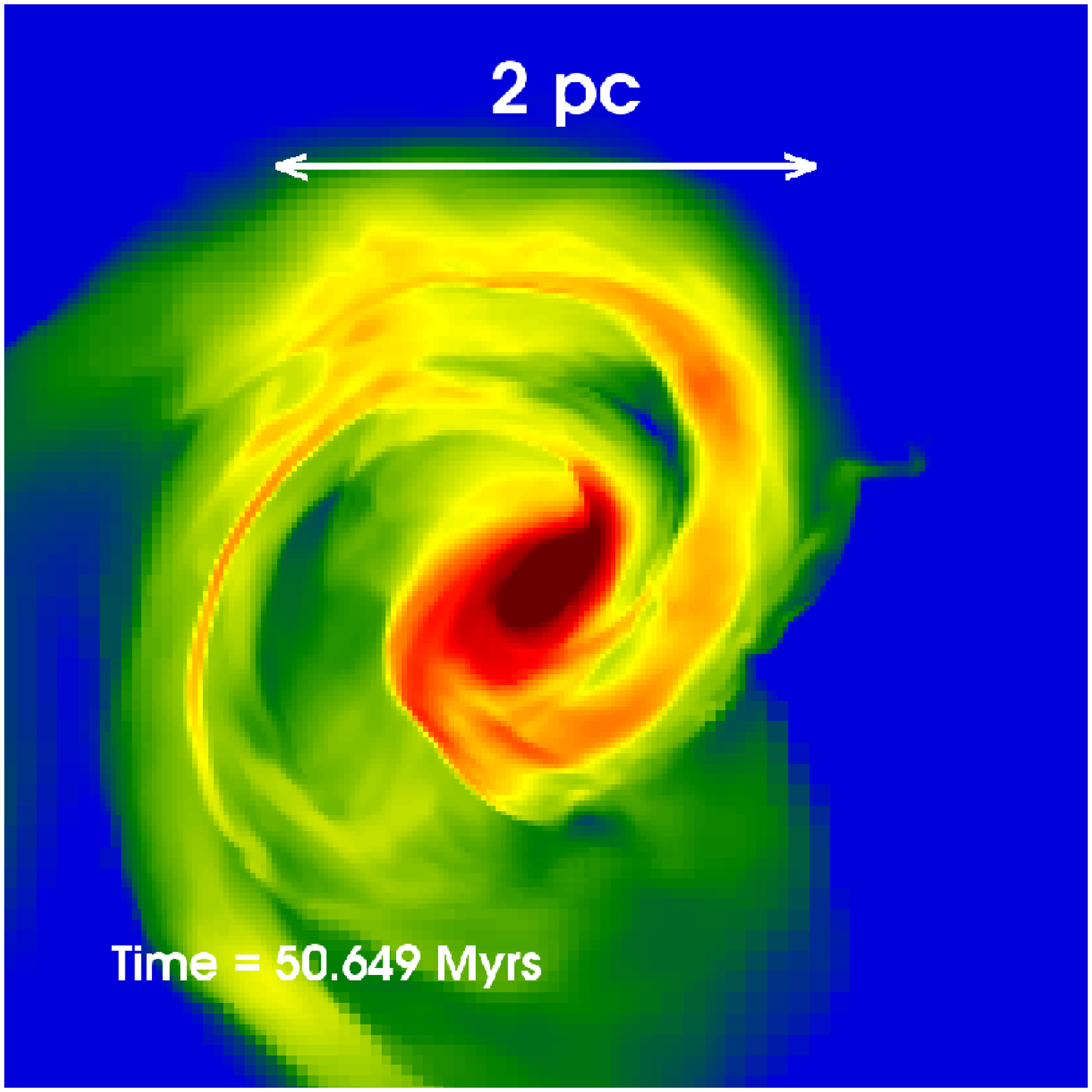} 
\includegraphics[height=4.0cm, width=5cm]{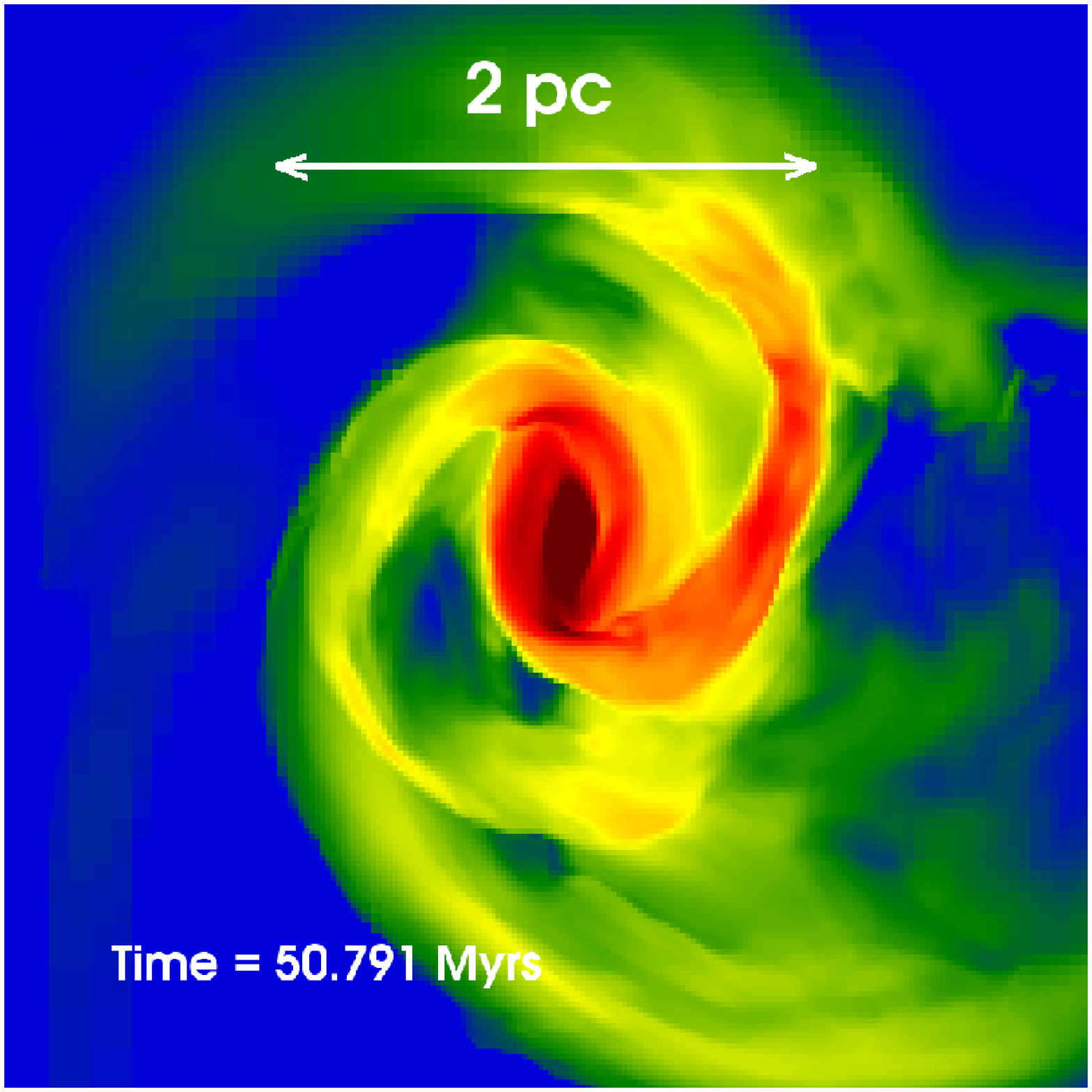} 
\caption[\label{SimCMap} ]{\label{SimCMap}
Same as figure \ref{SimAMap} for simulation C but only for the 
later stages of the evolution. 
The gas is collapsing in a halo of DM mass of $5.15\times
10^{7} M_{\odot}$, virial velocity of $\sim 19.3 \kms$ and virial
temperature of $\sim 13500 \, \rm{K}$. Panels 9 to 15 
illustrate  the dynamical evolution of the
rotationally supported gas at the centre of the halo for a few
dynamical times.}
\end{figure*}


\begin{figure*} \centering 
  \begin{minipage}{175mm}      \begin{center} {
\includegraphics[width=5.77cm]{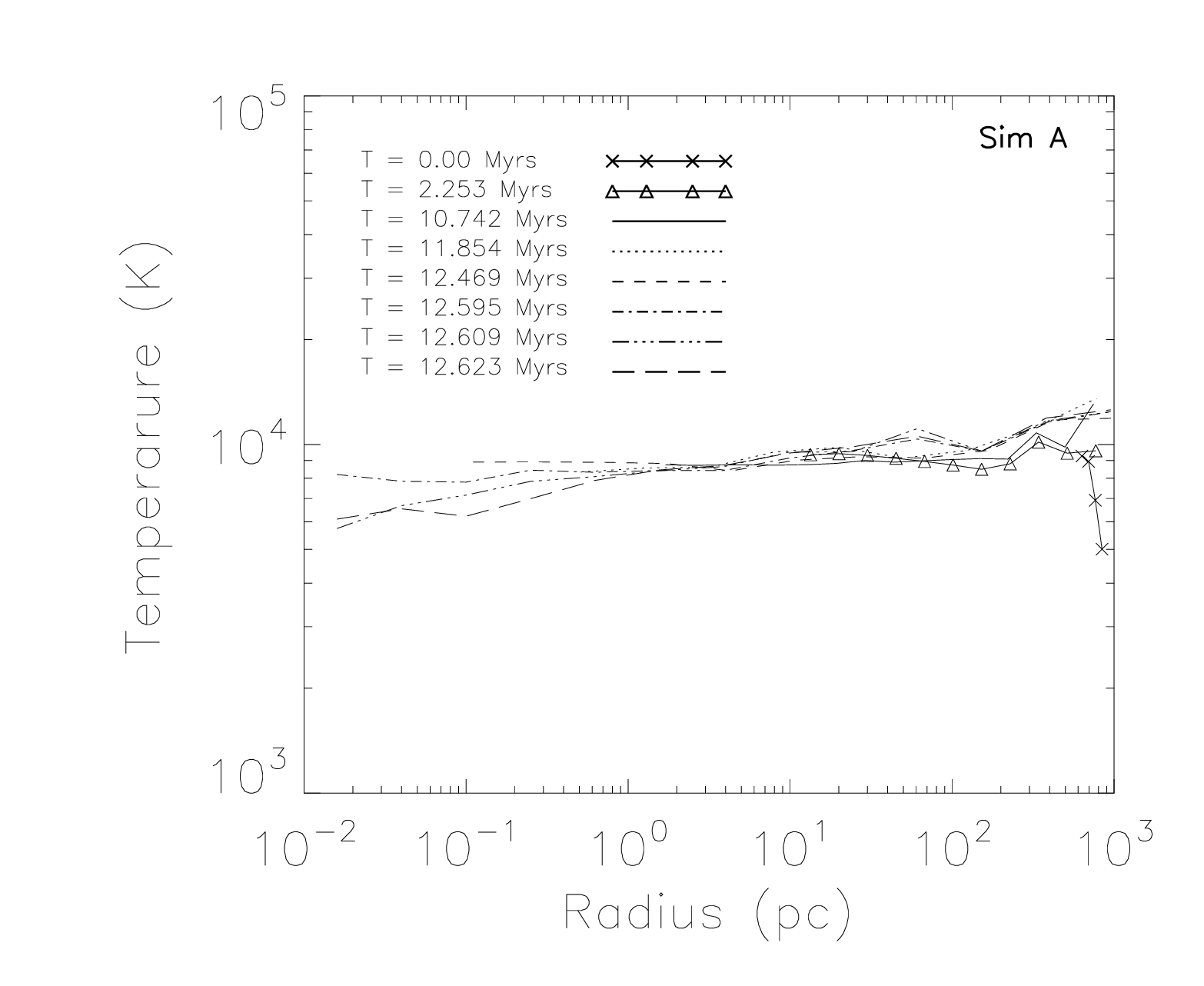}
\includegraphics[width=5.77cm]{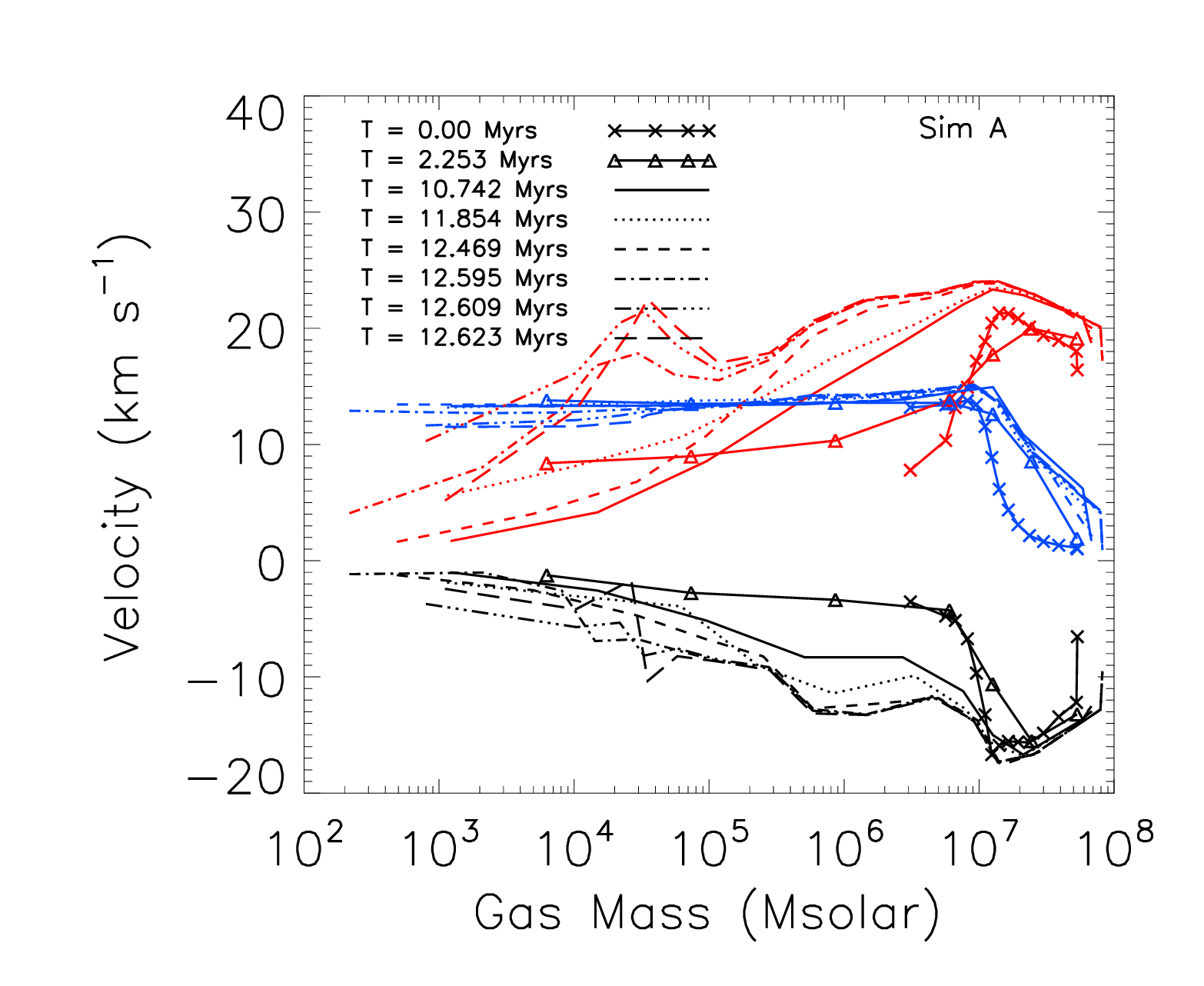} 
\includegraphics[width=5.77cm]{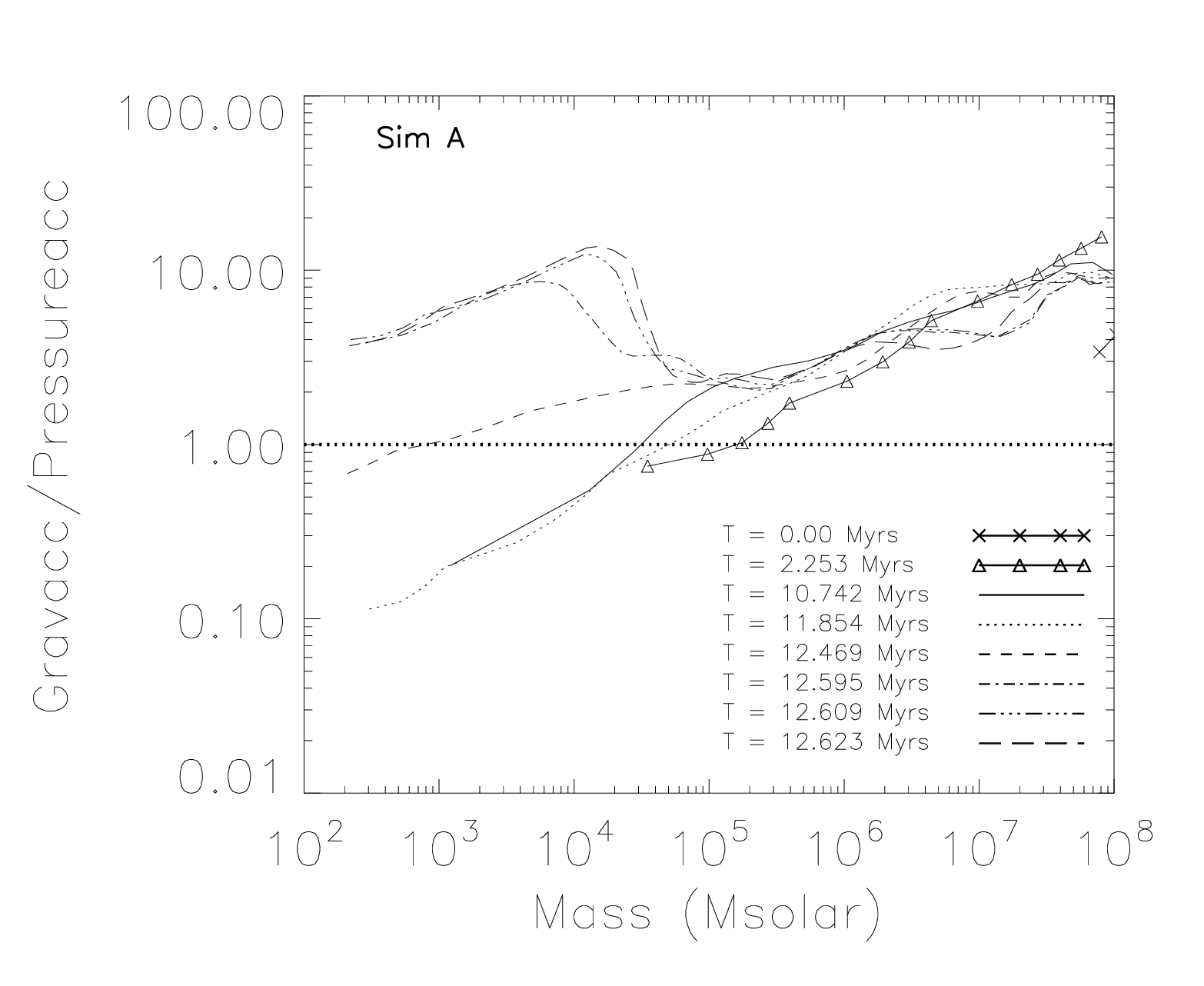}
\includegraphics[width=5.77cm]{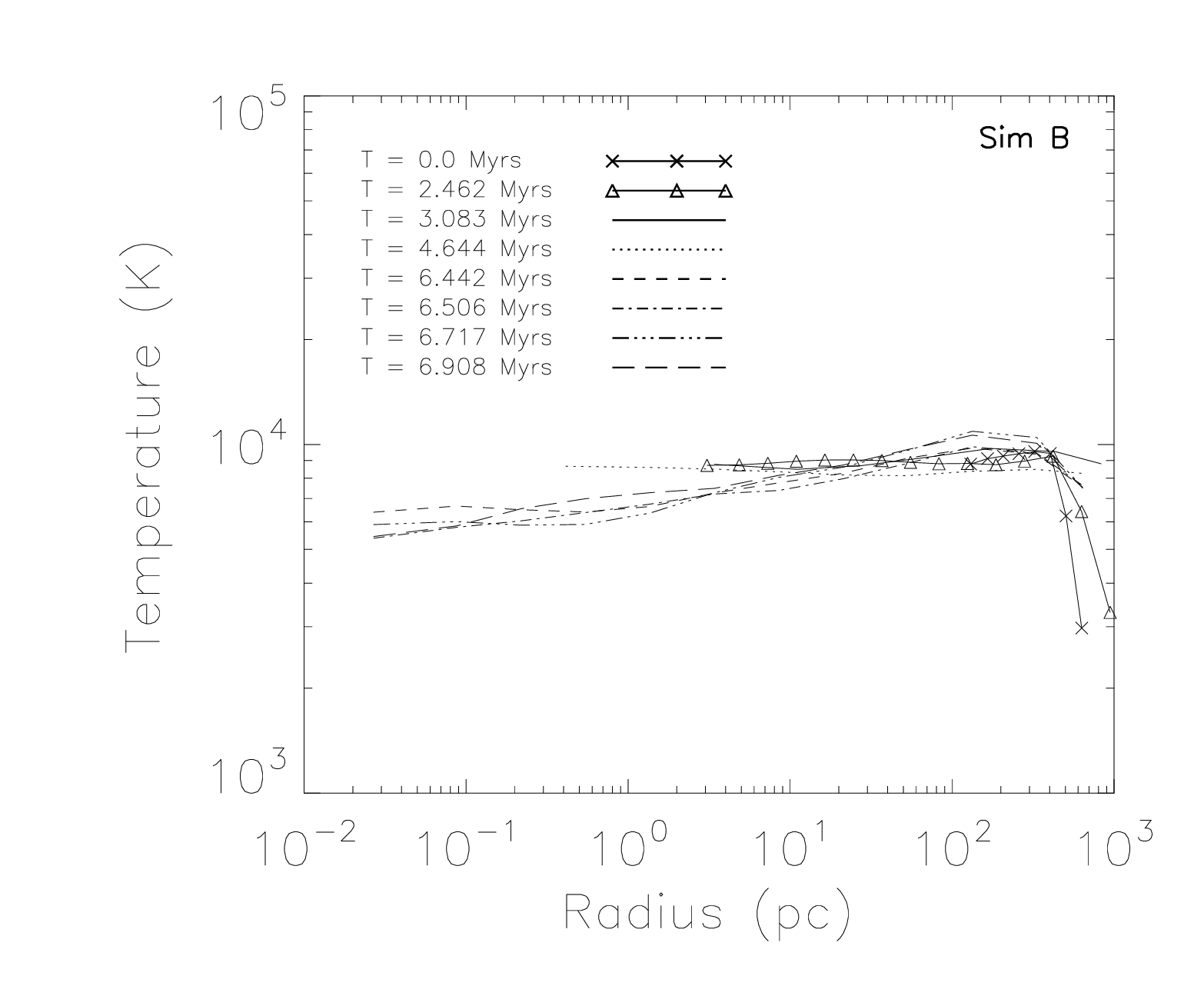}
\includegraphics[width=5.77cm]{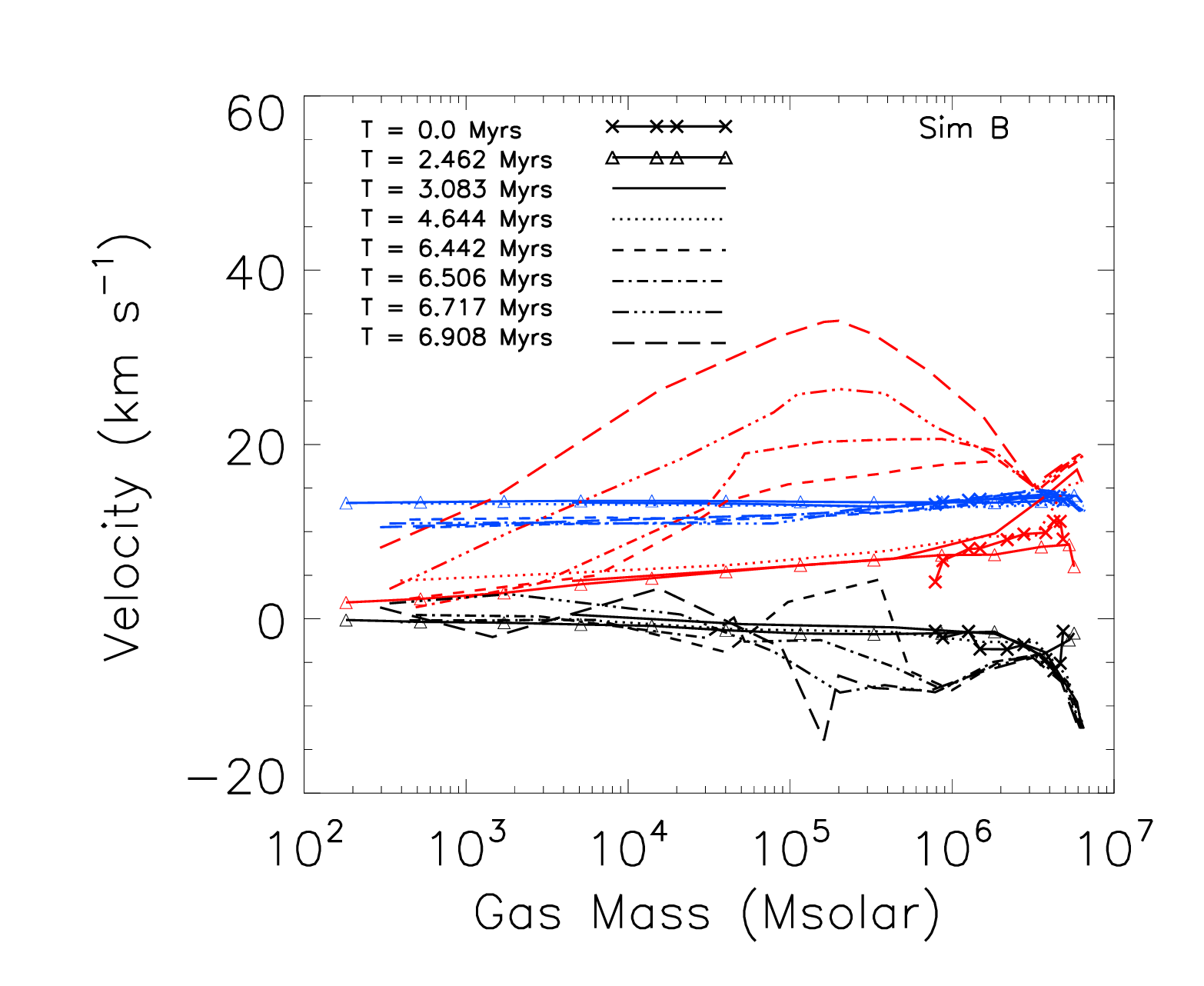} 
\includegraphics[width=5.77cm]{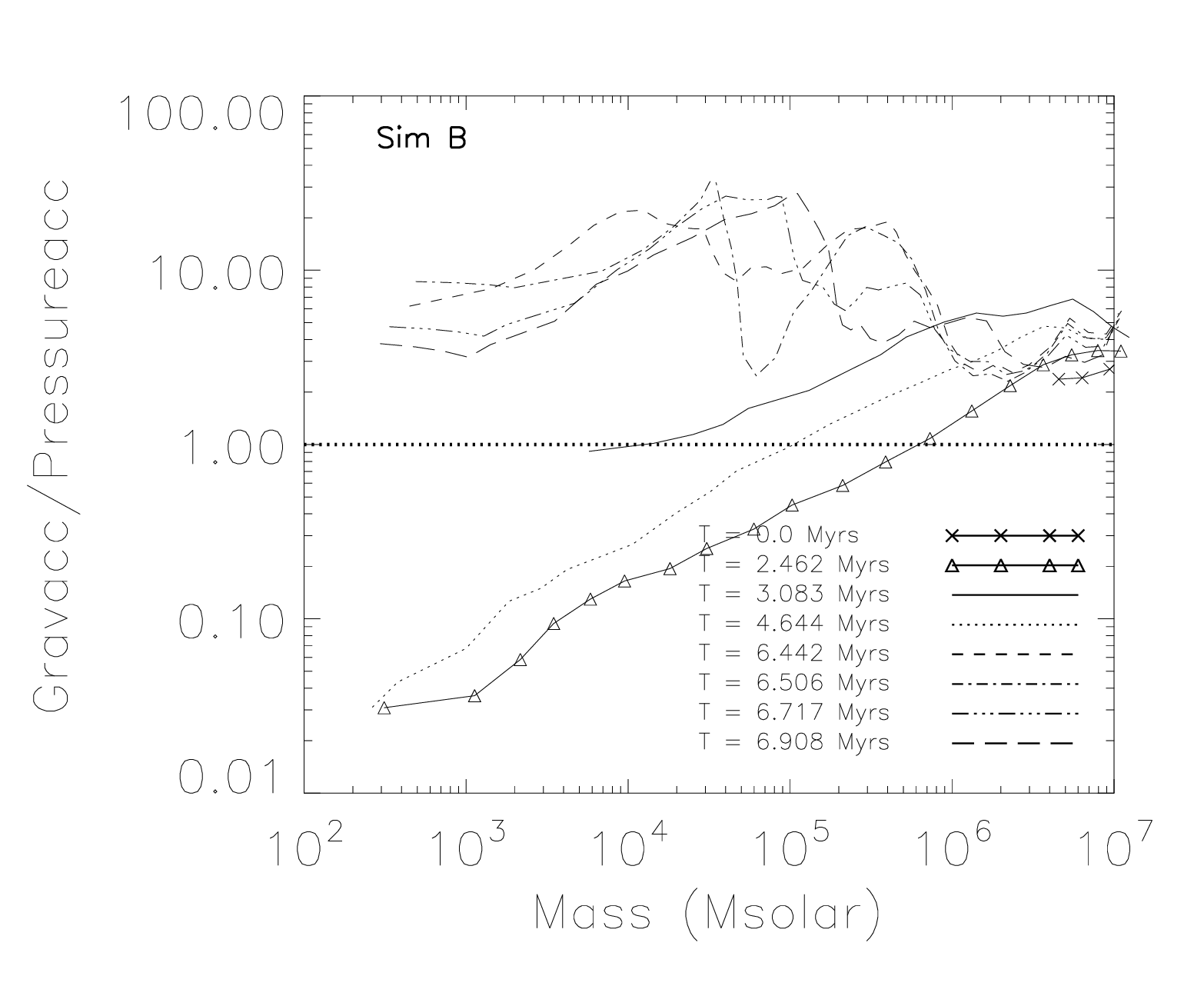}}
\caption[\label{Temp_evolution}]{\label{Temp_evolution} 
{\it Left-hand Panels}:  The temperature profile for the halo in simulations A and B 
as a function of radius.  {\it Middle Panels:} The time evolution of the radial velocity
(black), thermal velocity (blue)
and the turbulent velocity (red) as a function of total enclosed gas mass.
{\it Right-hand Panels:} Ratio of the gravitational
acceleration to the acceleration due to thermal pressure 
as a function of total enclosed mass. Note that 
initially the central regions are supported by thermal pressure.} 
\end{center} \end{minipage}
\end{figure*}


\begin{figure*} \centering 
  \begin{minipage}{175mm}      \begin{center}
{
\includegraphics[width=8.7cm]{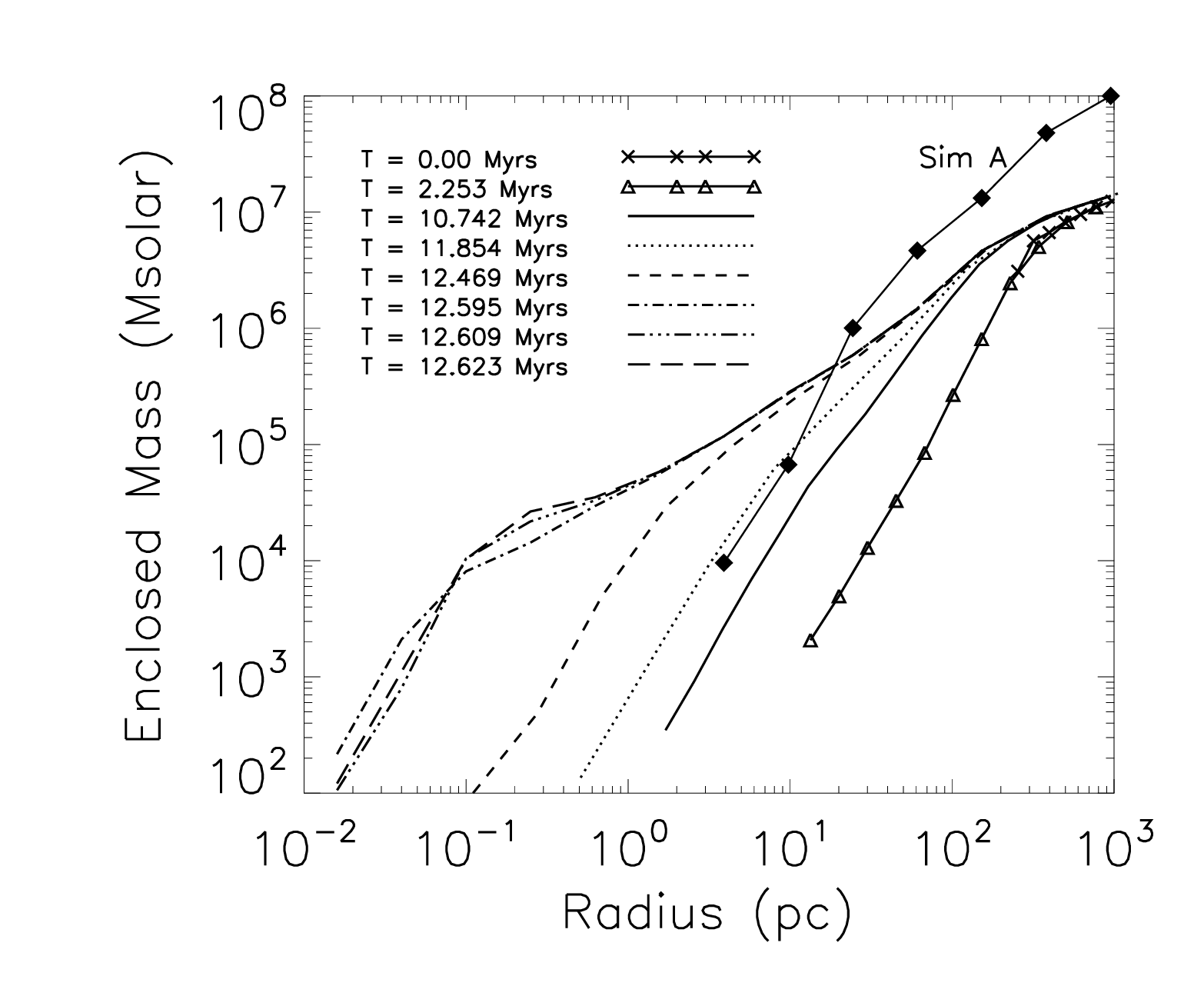}
\includegraphics[width=8.7cm]{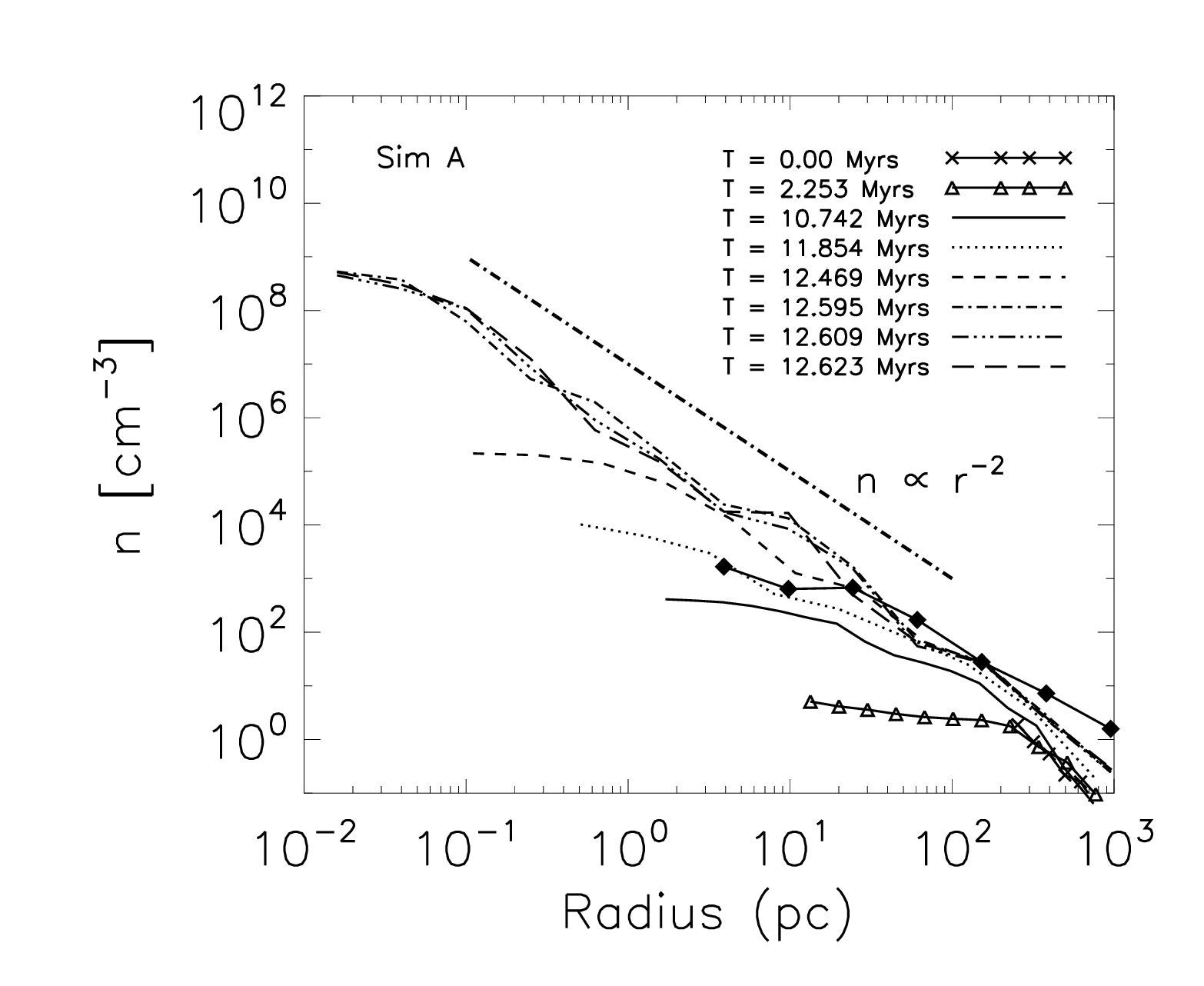}
\includegraphics[width=8.7cm]{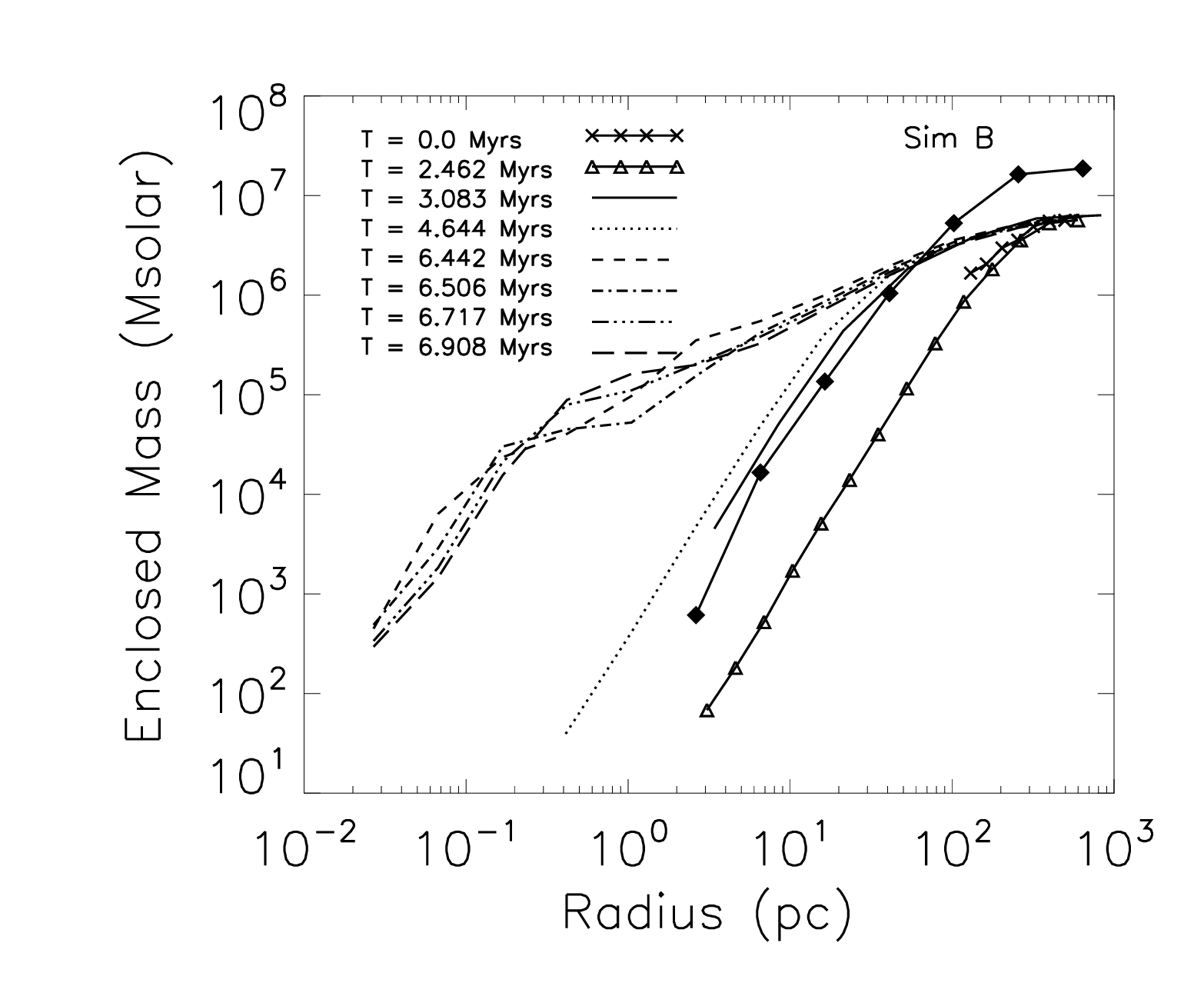}
\includegraphics[width=8.7cm]{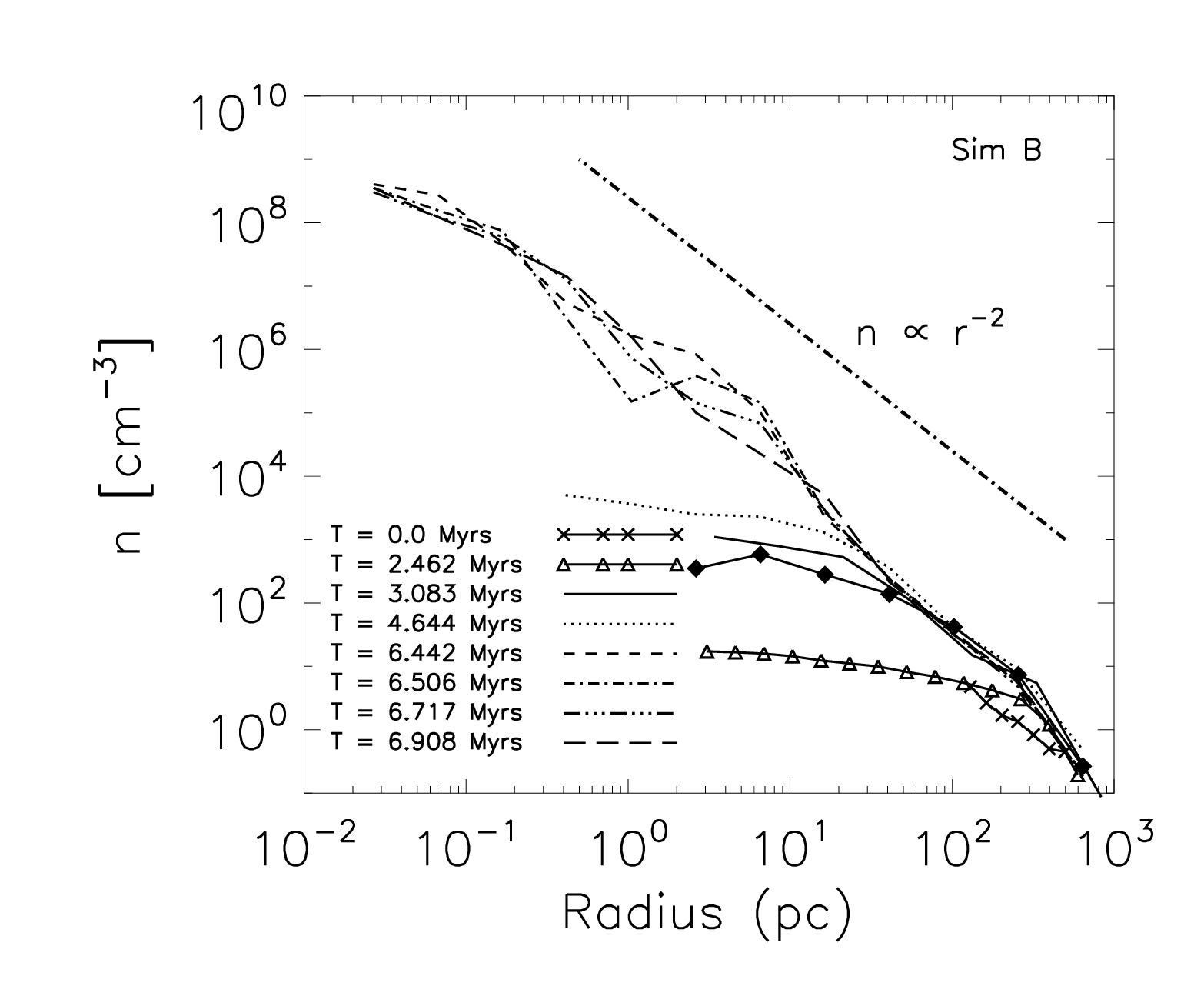}
}
\caption[\label{density_evolution}]{\label{density_evolution} 
{\it Left-hand Panels}: The evolution of the enclosed (gas) mass within spherical shells around
the centre of the halo  as a function of radius. 
The enclosed DM mass profile at the end of the simulation is  shown
by the filled diamonds. 
{\it Right-hand Panels}: The evolution of the (gas) density profile during the collapse. 
The gas settles into an close to  isothermal ($r^{-2}$) density profile
(thick dot-dashed line) over many decades in radius. The DM density at the end of the simulation 
is shown by the filled diamonds.   
 }
\end{center} \end{minipage}
\end{figure*}


\begin{figure*} \centering 
  \begin{minipage}{175mm}      \begin{center}
\centerline{\includegraphics[width=9cm]{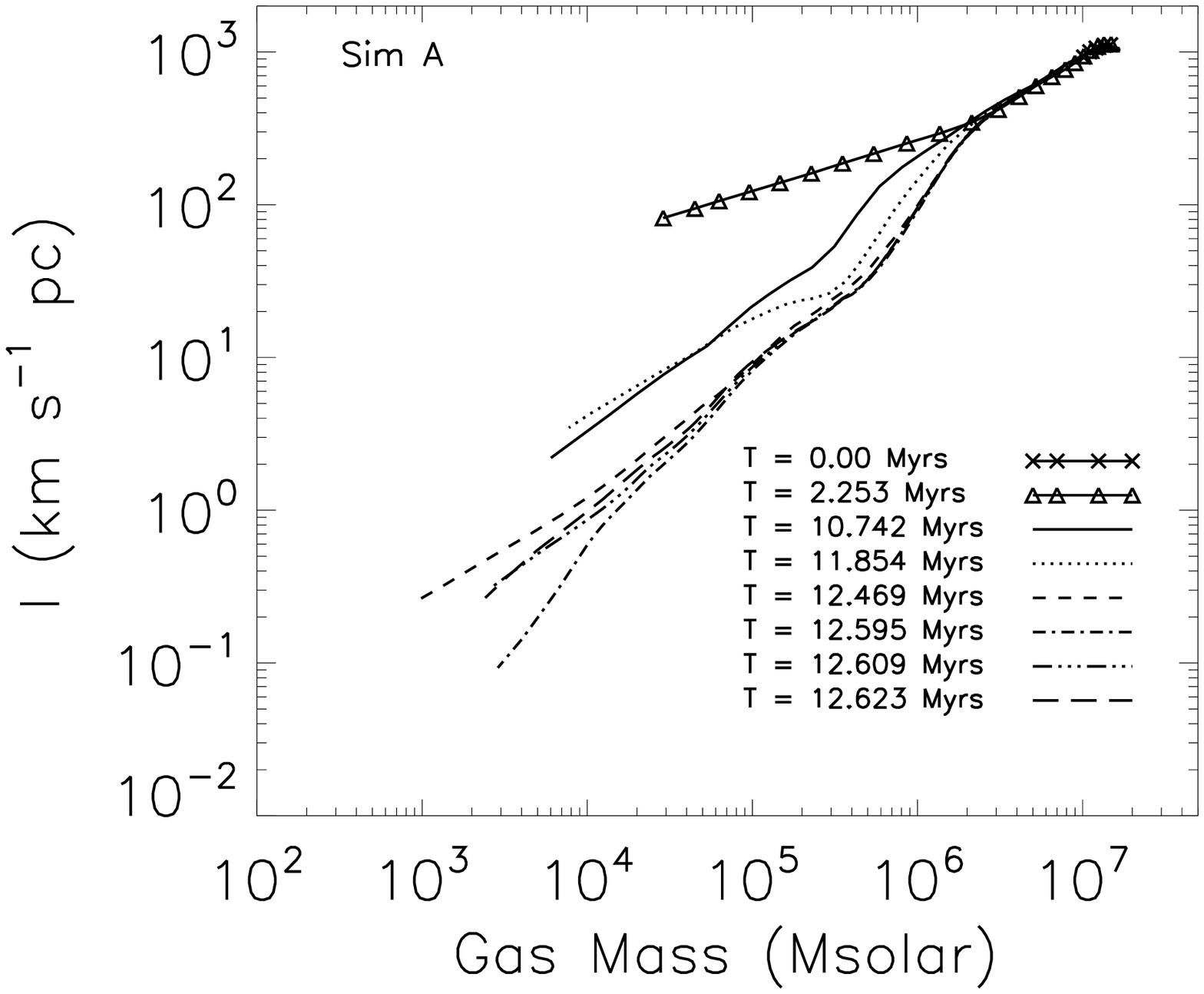} 
  \includegraphics[width=9cm]{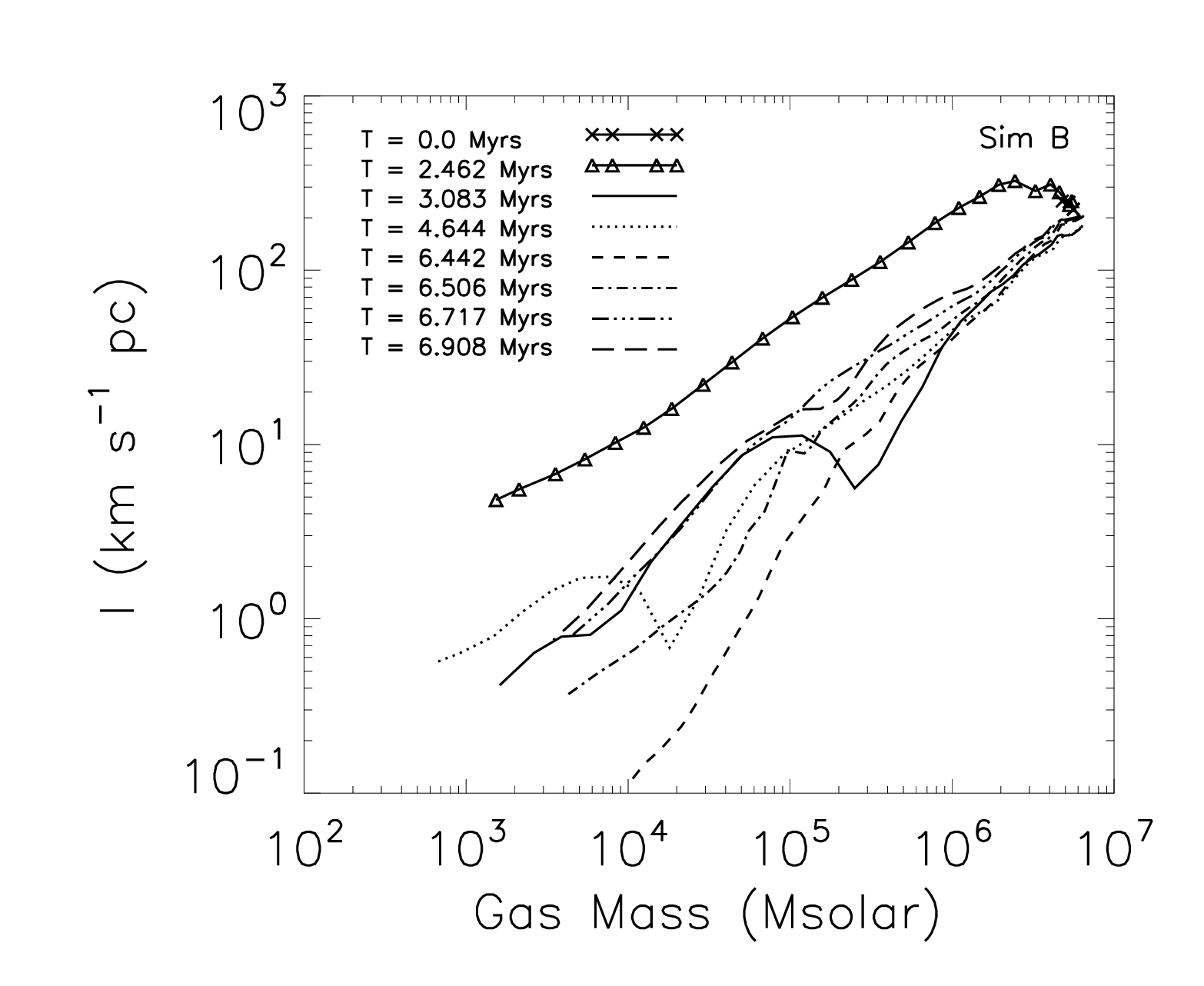}}
    \caption[\label{angularmomentumevolution}
]{\label{angularmomentumevolution}: {\it Left-hand Panel:}
The  evolution of the  total specific angular momentum
($l$) of the gas  as a function of enclosed mass for eight of the
simulation outputs in  figure \ref{SimAMap}.
{\it Right-hand Panel:} The same for simulation B for eight simulation
outputs shown in  figure \ref{SimBMap}. }

   \end{center} \end{minipage}
\end{figure*}


\begin{figure*} \centering 
  \begin{minipage}{175mm}      \begin{center}
{
\includegraphics[width=8.7cm]{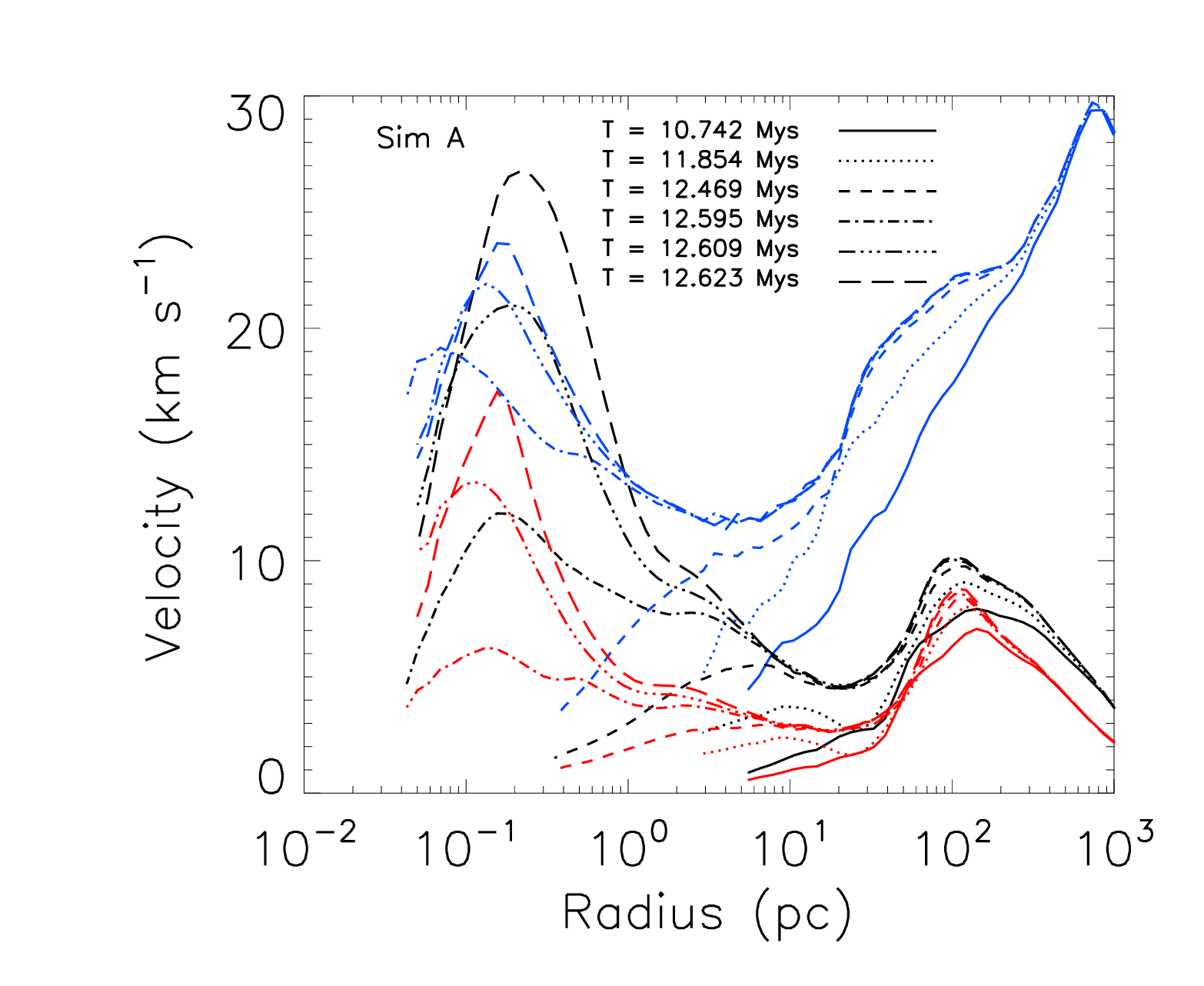}
\includegraphics[width=8.7cm]{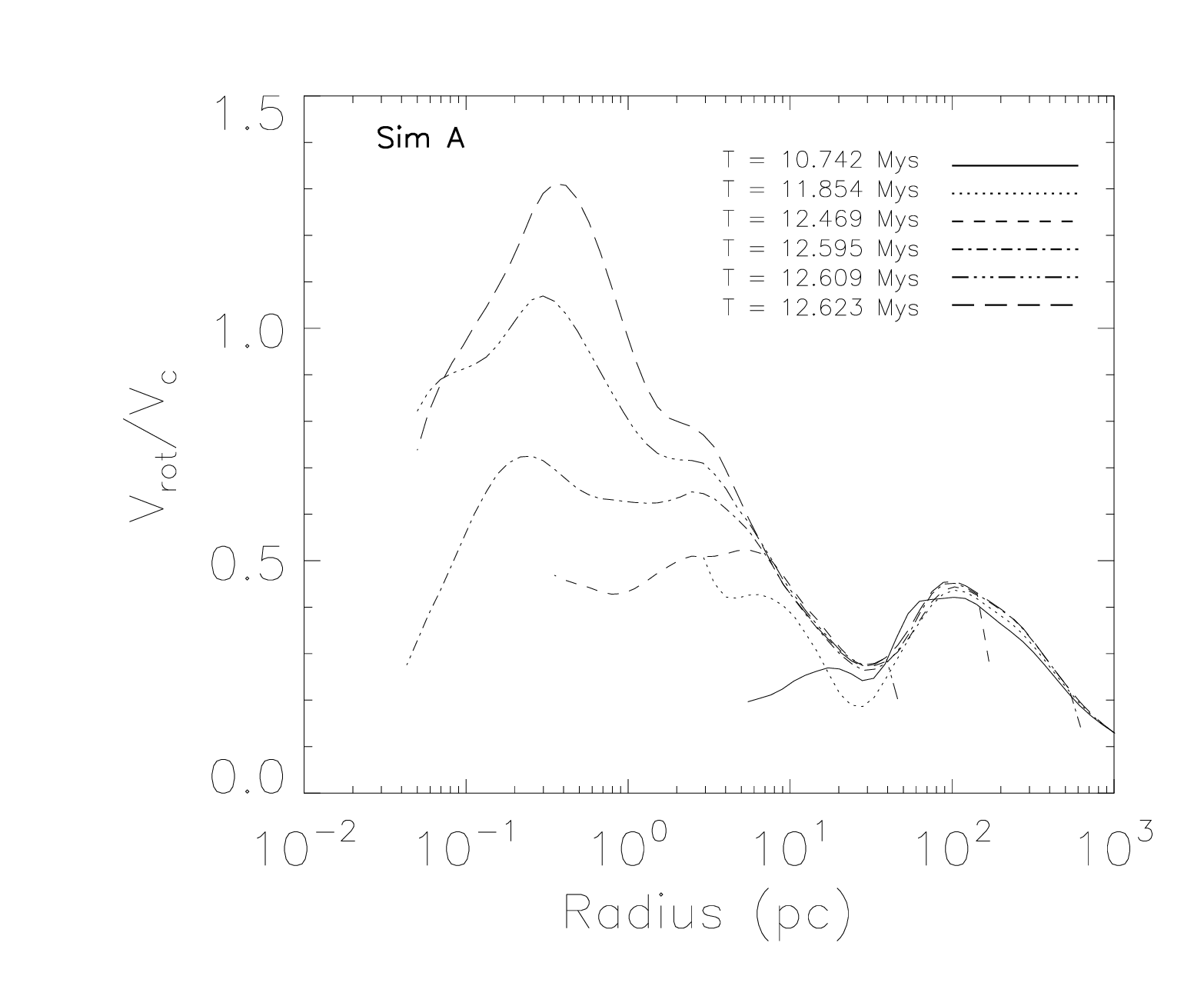} 
\includegraphics[width=8.7cm]{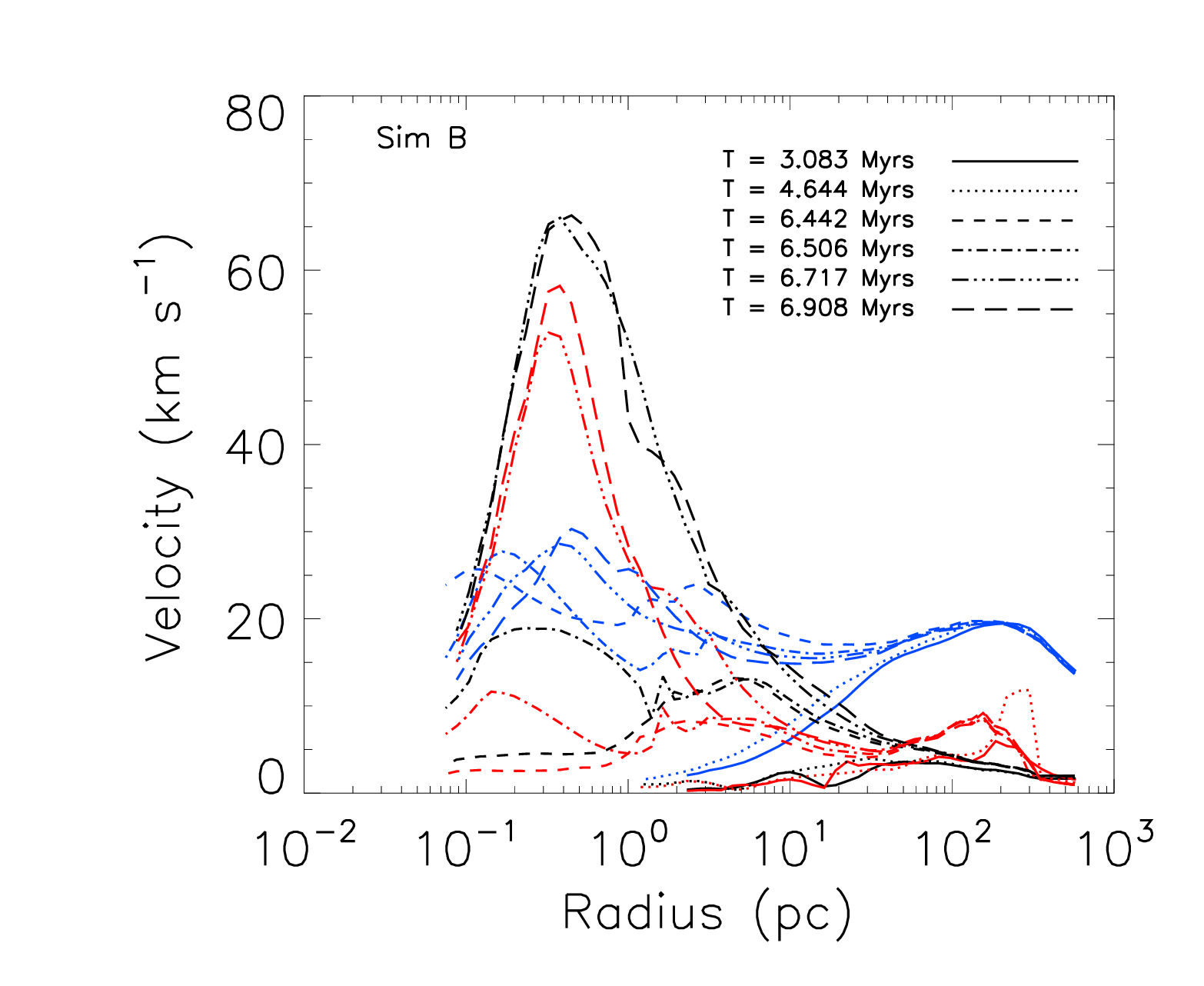}
\includegraphics[width=8.7cm]{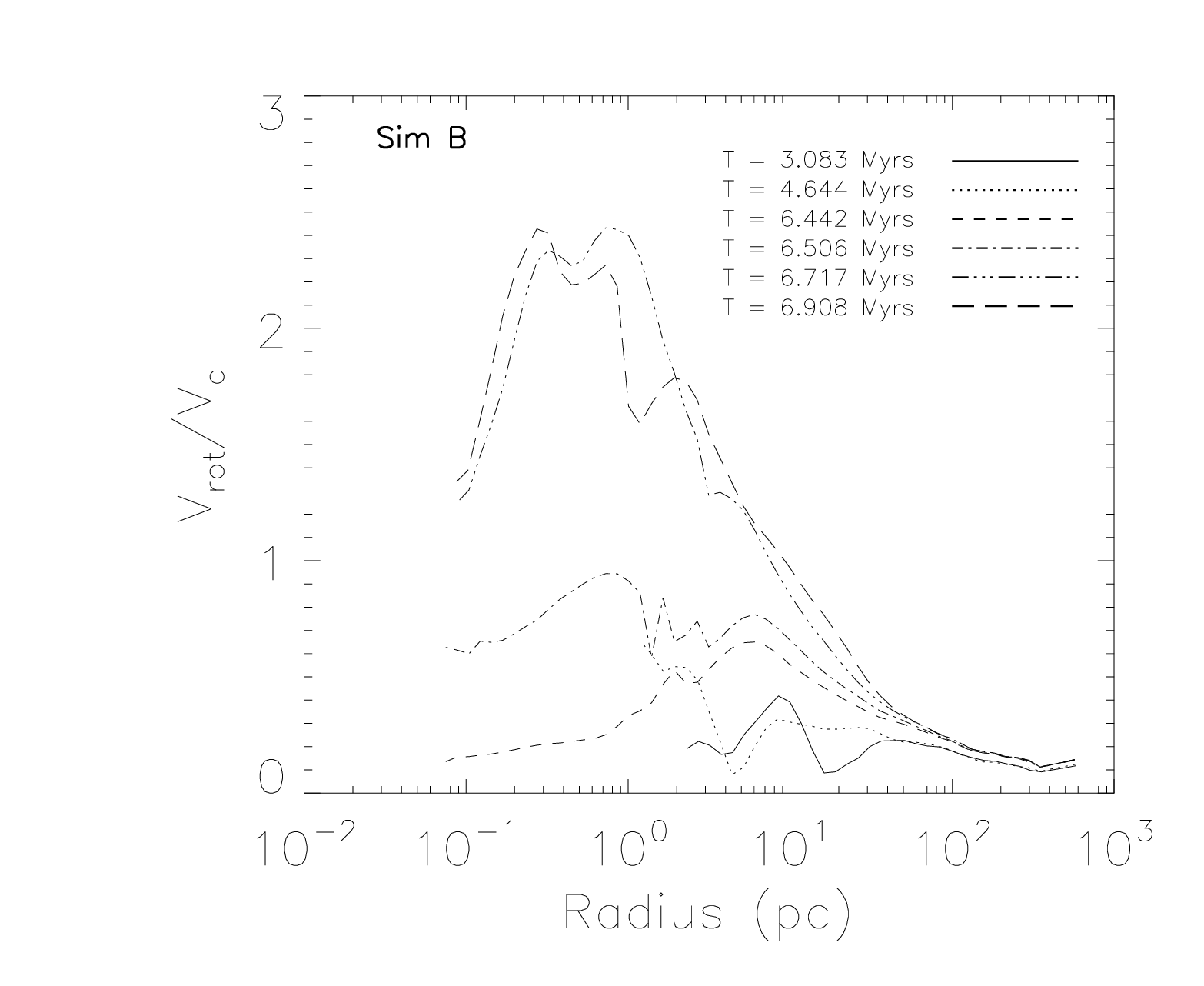} 
}
\caption[\label{rotsupport_evolution}]{\label{rotsupport_evolution} 
 {\it Left-hand Panels}: The circular velocity $V_{\rm c}$(blue),  and our two estimates of
 the rotation velocity $\overrightarrow{l}/a_{1}$ (black) 
 and $\overrightarrow{l}/r$ (red) are plotted as a function of
radius. As the collapse proceeds the inner part of the halo settles into
centrifugal support. {\it Right-hand Panels}: The
ratio of $V_{\rm rot}=  \overrightarrow{l}/a_{1} $ to $V_{\rm c}$,
for the same six output times plotted against radius. 
The central region attains rotational support, after $\approx 12.6$
Myrs and  $\approx 6.7$ Myrs, in simulations A and B, respectively.
} 

   \end{center} \end{minipage}
\end{figure*}


\subsection{Following the  collapse of the gas and dark matter  with
\enzo}
\label{Collapse} As mentioned previously \enzo uses adaptive grids to
provide increased resolution where it is required. 
For the simulations discussed in this paper we have used
four refinement criteria implemented in \textsc{enzo}: DM over-density, baryon over-density, 
Jeans length and cooling time. The first two criteria introduce 
additional meshes when the over-density of a grid cell with respect to
the mean density exceeds 3.0  for baryons and/or DM. The
third criterion is the  Truelove criterion \cite[]{Truelove_1997}
which in its original form states that at least 4 grid cells should be used
to resolve the Jeans length  to ensure that no artificial
fragmentation takes place. The cooling time is often shorter than the
dynamical time and the  cooling time refinement criterion helps to ensure that
the Jeans mass is properly resolved.
As other authors \cite[e.g.][W07]{OShea_2008} we are conservative here
and set this criterion to $16$ grid cells in our simulations.  The fourth
criterion ensures that the cooling time in a given grid cell is always
longer than the sound crossing time of that cell.  We also set the
\emph{MinimumMassForRefinementExponent} parameter to $-0.2$ making the
simulation super-Lagrangian and therefore reducing the criteria for
refinement as higher densities are reached \cite[]{OShea_2008}. We 
furthermore set the \emph{MinimumPressureSupportParameter}   equal to 
ten as we have restricted the maximum refinement level in our 
simulations \cite[e.g.][]{Kuhlen_2005}. With the
\emph{MinimumPressureSupport} option the code assumes in each computational cell 
the minimum temperature  necessary to make the cell  Jeans stable at the highest refinement level. \\
\indent We first advance our simulation forward in time using a maximum
refinement level of 4.  If we would  have run the  simulation
with a high maximum refinement level from the outset the simulation
would have stalled once the virial temperature of the halo has 
reached 10000K in the attempt to follow the dynamical evolution of
a small dense region before a sufficiently massive DM
halo had formed. We are effectively delaying  the collapse 
by limiting the resolution. 
We use a friends of friends (FoF) algorithm to keep
track of how our halo is  progressing through time. Once a
sufficiently massive halo (corresponding to a  DM particle number of
about 30000 particles in simulation A and 120000 particles in 
simulations B and C)  has formed  we halt the simulation. We
then  restart the simulation with the maximum level of refinement 
set to 18 (16 in simulation B and C). We will refer to this point as the collapse redshift. \\
\indent The simulation then continues at a much higher resolution
albeit at a significantly slower speed. We  allow the simulation to
evolve until the code no longer tracks the hydrodynamic evolution of
the gas  accurately  due to a lack of resolution at the highest densities 
caused by restricting the highest allowed refinement level. 
The \enzo code issues warnings alerting  the user when the code begins to 
produce spurious results due to a lack of resolution. 
At this  point we terminate the simulation. \\
\indent We have also followed the recommendation in \cite{Lukic_2007}
for the starting  redshift of  our  simulations. As a result the
initial redshift is quite a bit higher than that of other simulations
described in the literature. Experimenting with the  value of the
initial redshift suggests, however, that this increased initial
redshift  has little effect on our results. 


\section{Characteristic Formulae}
\label{formulae}
 
As mentioned before we use a FoF algorithm to identify DM haloes in our simulation outputs. 
We adopt a linking length of 0.2 for all of our analysis \cite[see][]{Jenkins_2001, 
Lukic_2007}. We compute physical  properties of the virialised haloes using 
standard formulae \cite[see e.g.][]{Mo_2002}. For convenience  we
summarise some of the  formulae below.  We calculate characteristic properties
for a sphere for which the mean enclosed density  is 200 times the
mean cosmic value $\overline{\rho}$. The characteristic ``virial'' radius is then
related to the  mass as,  
\begin{equation}
R_{200} = \Big[ {GM \over {100 \Omega_{\rm m} (z) H^2 (z) } }
\Big]^{\frac{1}{3}}, 
\end{equation}
while the circular velocity, $V_{\rm c}$, can be written as 
\begin{equation}
V_{\rm c}  = \Big( {GM(r)  \over r} \Big)^{\frac{1}{2}}, 
\end{equation}
where $H(z)$ is Hubble's constant at redshift $z$ and $\Omega_{\rm m}(z)$ is
the density parameter of non-relativistic matter, $M$
is the mass of our halo as determined by FoF group finder. Rewriting the equation for the
virial velocity in terms of the mass, $M$, only gives 
\begin{equation}
V_{\rm c} = M^{1/3} \Big(\sqrt{100 \Omega_{\rm m} (z)} G H(z) \Big)^{1/3}.
\end{equation}

The Hubble constant and density parameter  are related to their present-day values by

\begin{equation}
H(z) = H_0 E(z)
\end{equation}
and
\begin{equation}
\Omega_{\rm m}(z) = {\Omega_{\rm m,0}(1 + z)^3 \over {E^2(z)}},
\end{equation}
where E(z) is given by 
\begin{equation}
E(z) = \big[ \Omega_{\Lambda, 0} + (1 - \Omega_{\rm m,0}- \Omega_{\Lambda, 0} )(1+z)^2 + \Omega_{\rm m,0}(1+z)^3 \big]^{\frac{1}{2}}.
\end{equation}
We can now define the virial  temperature of the halo as 
\begin{equation}
T_{\rm vir} = {\mu V_{\rm c}^2 \over 2k} = 3.24 \times 10^4 \Big[ {V_{\rm c} \over 30 \kms} \Big]^2 K,
\end{equation}
where $\mu = 0.6m_p$, with $m_p$ being the proton mass, is the mean molecular weight. \\
\indent The level of rotational support is often  characterized by the
dimensionless angular momentum parameter $\lambda$ which can be  
written as \cite[e.g.][]{Bullock_2001} 
\begin{equation} \lambda = {| \overrightarrow{L} | \over \sqrt{2} M
    V_{\rm c} r}, 
\end{equation} 
where $| \overrightarrow{L} |$ is the angular momentum inside a sphere
of radius $r$ containing  mass $M$ and  $V_{\rm c}$ is the  virial 
velocity of the halo. The mean value of $\lambda$ is $\sim 0.04$ for
typical haloes in cosmological simulations \cite[]{Barnes_1987, Bett_2007}.
We have included the above properties for the three virialised haloes in
table \ref{TableSims} for reference. \\
\indent Also of interest is the dynamical
time which we express in terms of the enclosed mass $M(r)$ at radius $r$,  
\begin{eqnarray}
t_{\rm dyn} &=& \Big( { 3 \pi \over 16 G \rho} \Big)^{1/2} \nonumber \\
       &\sim& 2.09 \times 10^{5} \Big( {r \over 1 \, \mathrm{pc}}
       \Big)^{3/2} \Big( {10^4 \, M_{\odot} \over M(r) } \Big)^{-1/2}
       \, \mathrm{yr}. 
\end{eqnarray}


\section{Results of the numerical simulation}
\label{results}
\subsection{The global dynamics of the collapse of the gas and dark matter}

We first discuss the global dynamical evolution of the gas and 
dark matter  in the three haloes we have simulated. Note that the dynamical 
evolution in simulation C is similar to that in simulation A 
and we will thus show only a subset of the plots for simulation C
in the following.  
We have simulated  boxes with $5 \, h^{-1}$ Mpc and 
$2 \, h^{-1}$ Mpc on a side, respectively.  Figure 1 and 2   
show the location of the haloes in simulation A and B within 
the typical characteristic web of filaments and sheets which is already well 
developed  at $z \sim 15$ (simulation C looks very  similar in this respect). 
The haloes are located at the intersection of 
several filaments. 
In the top row of figure 1 and 2 we zoom in several steps 
onto the haloes which are approximately spherical and have a virial 
radius of about 1.5 and 0.5 kpc.
In Figure 3 we focus on  the later  stages of the evolution of the gas 
at the centre of simulation C. From one panel to the next we either 
change the output time {\it or} the density/length scale. 
We have selected time 0 as the time at which
we re-start the simulation with increased resolution, with a maximum
refinement level of 18 for simulation A and a maximum refinement level of 16 for simulations B \& C.  
Recall that up to now we had run the simulation 
limiting the refinement to 4 levels in order to allow the halo
to build up.  We will 
discuss possible  limitations of this approach later. Once the resolution is
increased the inner part of the haloes starts to collapse. \\
\indent As discussed by W07 the collapse is highly turbulent and  
dynamically complex.  In each halo we are able to follow the collapse of the 
gravitationally unstable gas at the centre of the halo into a centrifugally
supported disc (or disc-like object) and follow the dynamical evolution of the disc 
for several dynamical times. In simulation A this disc has a mass of  
$2 \times 10^4 M_{\odot}$.   In the second of our haloes 
the gas at the centre of the halo fragments  into three 
gravitationally bound clumps  with masses of a few times $ 10^4 M_{\odot}$ 
and a  mass ratio of approximately 3:1:1.
The clumps tidally distort
each other and subsequently undergo a violent dynamical interaction. One of the
smaller clumps eventually merges with the most massive clump 
and forms a disc of  $\sim 1 \times 10^5 M_{\odot}$. The second small
clump is tidally stripped of most of its mass and has a mass 
of a few times  $10^3 M_{\odot}$ at the end of the simulation. 
The (temporary) formation of multiple systems appears,
however, not  to be generic. Neither simulation A with its five times
more massive halo nor simulation C with its halo
of similar mass show the formation of a similar multiple system. 
The disc forming in simulation C has a mass of $\sim 1 \times 10^5 M_{\odot}$.

\subsection{The onset of gravitational instability, mass inflow and
  the evolution of the density profile}

The gas at the centre of our haloes attains an almost isothermal temperature profile
with a  characteristic temperature of 6000-7000K the temperature below
which atomic cooling due to hydrogen cooling becomes inefficient. The
temperature  rises gently to  9000-10000 K at larger radii. 
The left-hand panels in figure \ref{Temp_evolution} show the temperature
profile of the gas in the haloes and its evolution for simulation A
(top panel) \& B (bottom panel), respectively.\\
In order to understand the  dynamical evolution of the gas 
it is illustrative to consider the  thermal and turbulent velocities
of the gas. In the middle panel of figure \ref{Temp_evolution}
we show the evolution of the radial velocity (black), the thermal velocity (blue) and the
turbulent velocity (red) of the gas. As the collapse gets under way
the gas develops turbulent velocities comparable  to the virial velocities
of the halo starting in the outer parts of the halo  and progressively moving inwards.  
The turbulent velocities are calculated by computing the root mean
square velocity of the gas after subtracting the centre of mass velocity of the
halo  and the velocity due to the radial inflow of the gas. We also subtract
an estimate of the  rotation velocity (based on the
specific angular momentum and the inertia tensor as
described in section 4.3) in quadrature.
The turbulent velocities in the later stages of the collapse
become significantly larger than  the  thermal velocities of the gas which we calculate as
$V_{\rm th} = \sqrt{3kT/ \mu m_{H}}$, where  $\mu$ is the mean molecular
weight, chosen to be  $1.22$  and $m_{H}$ is the mass of the hydrogen 
atom. As in the simulations of  W07  the turbulent velocities are supersonic. 
There is a net inflow of gas with velocities 
 about a factor three smaller than the turbulent velocity component. \\
\indent In the right hand panel of figure \ref{Temp_evolution} we compare 
the ratio of the inward acceleration due to gravity  to the
outward acceleration due to the thermal pressure (gradient) for both simulation A (upper panel)
and simulation B (bottom panel). 
Once the DM halo has built up, the mass of the halo is above the Jeans
mass and the inward gravitational acceleration comfortably exceeds the 
outward acceleration due to the thermal pressure. The peak which
builds up at an enclosed mass of a few times $10^4 \, M_{\odot}$ for simulation A
and $\sim 10^5 \, M_{\odot}$ for simulation B
is due to the formation of a centrifugally supported disc at the
centre. The double peak at certain output times for simulation B is due to the presence 
of multiple gas clumps within the system. \\
\indent In figure \ref{density_evolution}
we show the evolution of the enclosed gas and DM mass as a function
of radius and the density profiles of  the gas and DM distribution. 
The DM density (in units of $m_H \, \rm{cm^{-3}}$)  
and the  DM fraction are shown for the final output times only 
as filled diamonds. 
The density profiles are calculated by averaging over
spherical shells centred on the densest point in the halo. Note that in  simulation B the 
densest point is always found in the most massive of the three clumps.
The haloes collapse  within 12 and 7 Myrs, respectively,  after we have 
restarted 
the simulation with the increased refinement  level.  This corresponds
to about a few free-fall  times for the gas in the inner few hundred
parsecs which initially has pretty much constant density. 
As discussed already, we initially held the refinement level at a maximum of four 
until the halo had a mass corresponding to $\sim 30000$ ( $\sim 120000$) DM particles. 
The gas completely decouples from the
dark matter during the collapse and becomes self-gravitating. 
The density profile of the gas in the 
inner part steepens dramatically and settles into a close to $r^{-2}$
density profile over many decades in radius as expected for an
isothermal collapse \cite[e.g.][W07]{Larson_1969}.
The inner few times $10^{4} \, M_{\odot}$ $(10^5 \, M_{\odot})$ collapse further and settle
eventually into rotational support. 
In the bottom right hand panel of figure \ref{density_evolution} (simulation B) 
the secondary peak in the density profile is due to the secondary clump which eventually
merges with the primary clump. The peak is absent in the final two
output times as one of the smaller clumps has merged with  the primary clump
and the other smaller clump  has had most of its 
mass stripped away. Note that the continuous accretion onto the virialised haloes from the filaments
occurs on much longer time scales and thus has no visible effect. 
\begin{figure*}
  \centering 
  \begin{minipage}{175mm}      \begin{center}
    \centerline{\includegraphics[width=9cm]{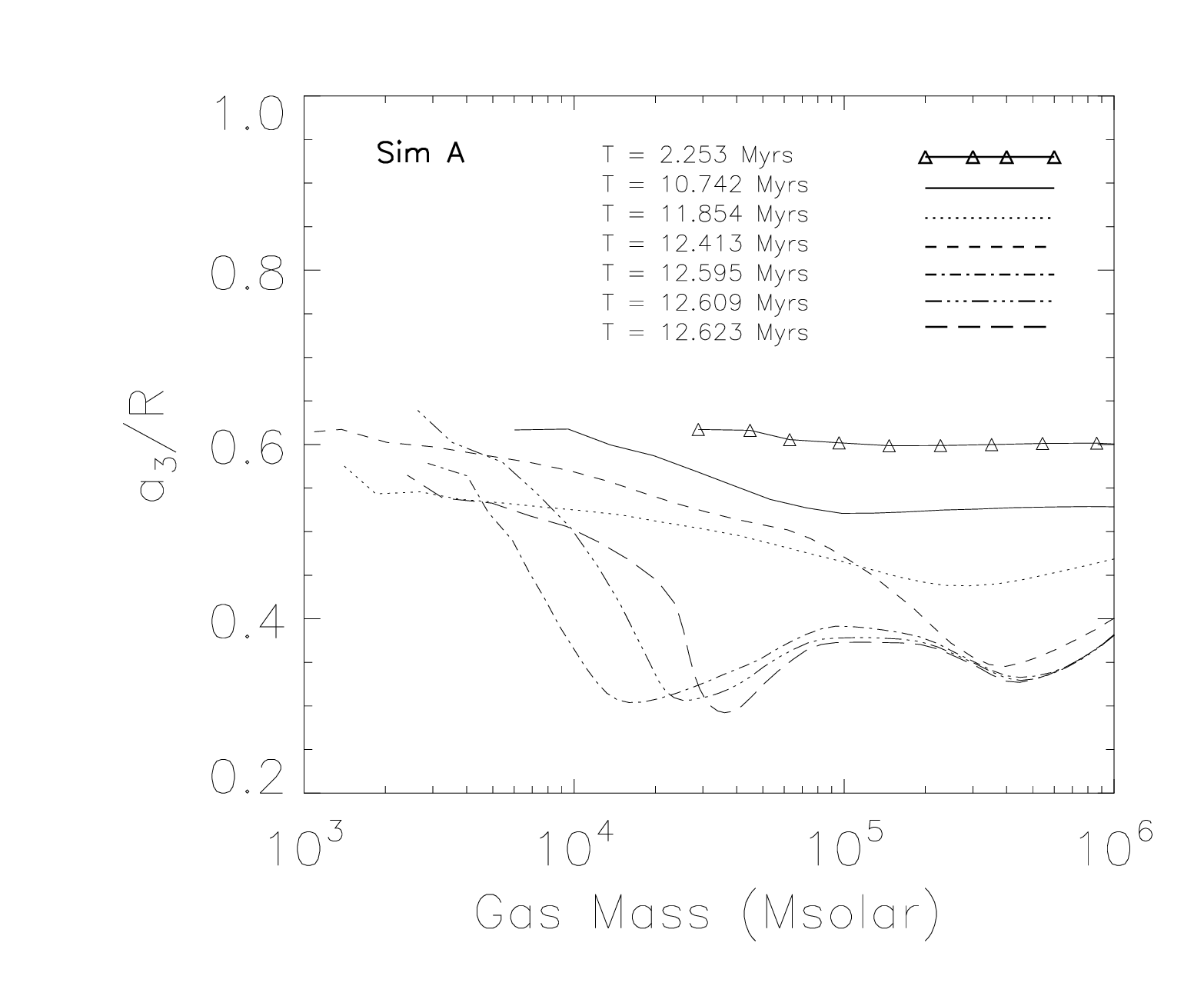}
      \includegraphics[width=9cm]{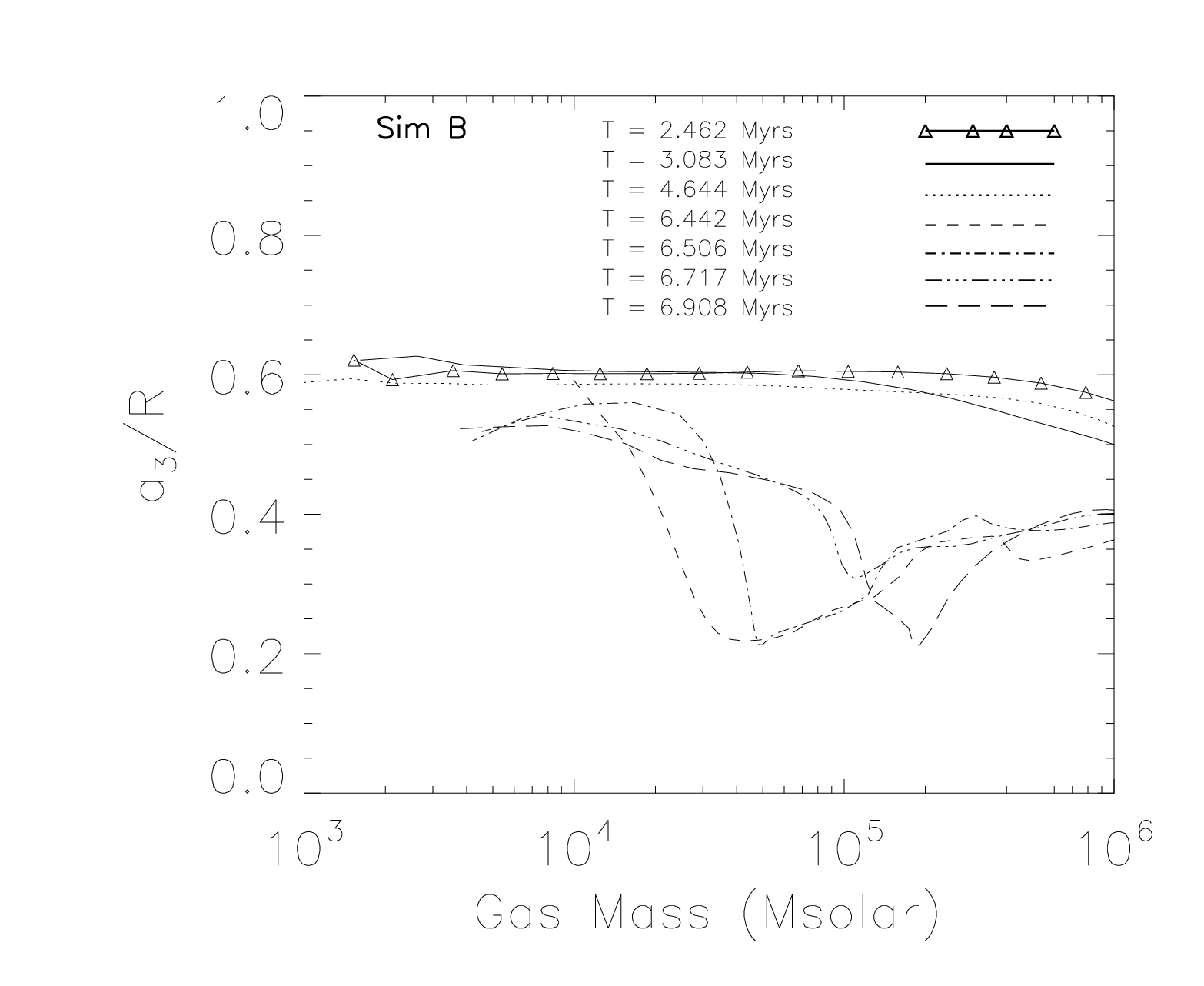}}
    \caption[\label{h/R} ]{\label{h/R}
      The ratio of the square root of the smallest eigenvalue, $a_3$, of
      the inertia tensor to  the distance to the centre of the halo. 
      Note the dip at a few times $10^4 \, M_{\odot}$ where the gas
      settles into a rotationally supported fat disc in simulation A.
      For simulation B the dip at the end of the simulation occurs at
      a much higher mass of 
      $\sim 1 \times 10^5 \, M_{\odot}$. }

   \end{center} \end{minipage}
\end{figure*}
 
\subsection{Angular momentum loss and settling into rotational support} 

In Figure \ref{angularmomentumevolution} we show the evolution of the specific angular
momentum, $|\vec{l}|$,  as a function of
enclosed mass for both simulation A and simulation B.
The angular momentum vector is calculated in the
usual way via the cross product 
$\overrightarrow{L}  = \vec{r} \times \vec{p} $ where  $\vec{p}$ is
the momentum vector. We then obtain the specific angular momentum
vector as a function of enclosed mass centred on the centre of mass by  dividing by the
enclosed mass. 
In both haloes there is very significant angular momentum loss. In simulation A the angular momentum
drops by a factor of 20 or more between the initial output and the  final output 
in the inner few times $10^4 \, M_{\odot}$. The evolution in
simulation B is more complicated due to the complex dynamical
interaction of the triple system. Initially the angular momentum of the gas drops by a
factor 20-100 in the primary clump and then increases again by a
factor three to five when one of the smaller clumps merges with the primary
clump. \\
\indent We now want to discuss in more detail to what extent the gas 
settles into rotational support. During the turbulent collapse of the
gas, it is not obvious how
best to define rotation velocities. We follow W07 who 
use the ratio $|\vec{l}|/|\vec{r}|$, where $\vec{l}$ is the specific angular momentum 
vector and  $\vec{r}$ is the position vector from the centre of mass as a simple but rough measure of the
rotation velocity of the gas.  We also calculate a second estimate of
the rotation velocity based on the angular momentum  $\vec{l}$
and the inertia tensor, $\tilde{I}$.

The nine components of the inertia tensor are given by 

\begin{equation} I_{xx} = \sum_j m_j ( y_j^2 + z_j^2),
\end{equation}

\begin{equation} I_{xy} = -\sum_j m_j x_j y_j,
\end{equation}

and the corresponding cyclic permutations, where the sum is over
computational cells.  
The off-diagonal components are called the products of inertia while the diagonal
components are referred to as the moments of inertia. The matrix is
symmetric which guarantees that the eigenvalues are real. 

The  angular momentum and the inertia tensor are related as,

\begin{equation} \vec{l} = \tilde{I} \vec{\omega},
\end{equation}

where $\vec{\omega}$ is the angular  velocity. Using the square root of
the largest eigenvalue of the inertia tensor, $a_{1}$, we  
then estimate the  rotation velocity as 
\begin{equation}
V_{\rm rot} \approx  { |\overrightarrow{l}| \over a_{1}}.
\end{equation}
\indent In figure \ref{rotsupport_evolution}  we compare our two
estimates of the rotation velocity  
$|\vec{l}|/|\vec{r}|$ (red) and  $|\vec{l}|/a_{1}$  (black) to the
circular velocity (blue), $V_{\rm c}(r) \equiv \sqrt{GM(r)/r}$, for simulation A (top panel)
and simulation B (bottom panel). 
Early on in the collapse the gravitational potential is dominated by
the dark matter halo and the gas is only slowly rotating with a ratio 
of rotation to circular  velocity of about $1:20$. As the 
gas in the inner part of the haloes collapses and becomes self-gravitating 
both the circular velocities and the rotation velocities rise.  
As shown in the right hand panel, the inner few times 
 $10^4 \, M_{\odot}$ reach 
(approximate) rotational support
with  $|\vec{l}|/a_{1}$   slightly larger than  $V_{\rm c}$ 
after about 12 Myrs in simulation A. 
In simulation B  the increase in specific angular momentum 
at around 6.5 Myrs due to the merger of one of the small clumps with 
the primary clump leads to  a sharp increase in the rotation 
velocity. As we will see in more detail later the gas nevertheless 
settles into a regular rotationally supported disc despite this
rather violent dynamical evolution. Our estimate of the  rotation
velocity based on the inertia tensor exceeds the circular velocity by a factor of up to two. 
We will come back to this point in section 4.5.
\begin{figure*} \centering 
  \begin{minipage}{175mm}      \begin{center}
\centerline{\includegraphics[width=8cm]{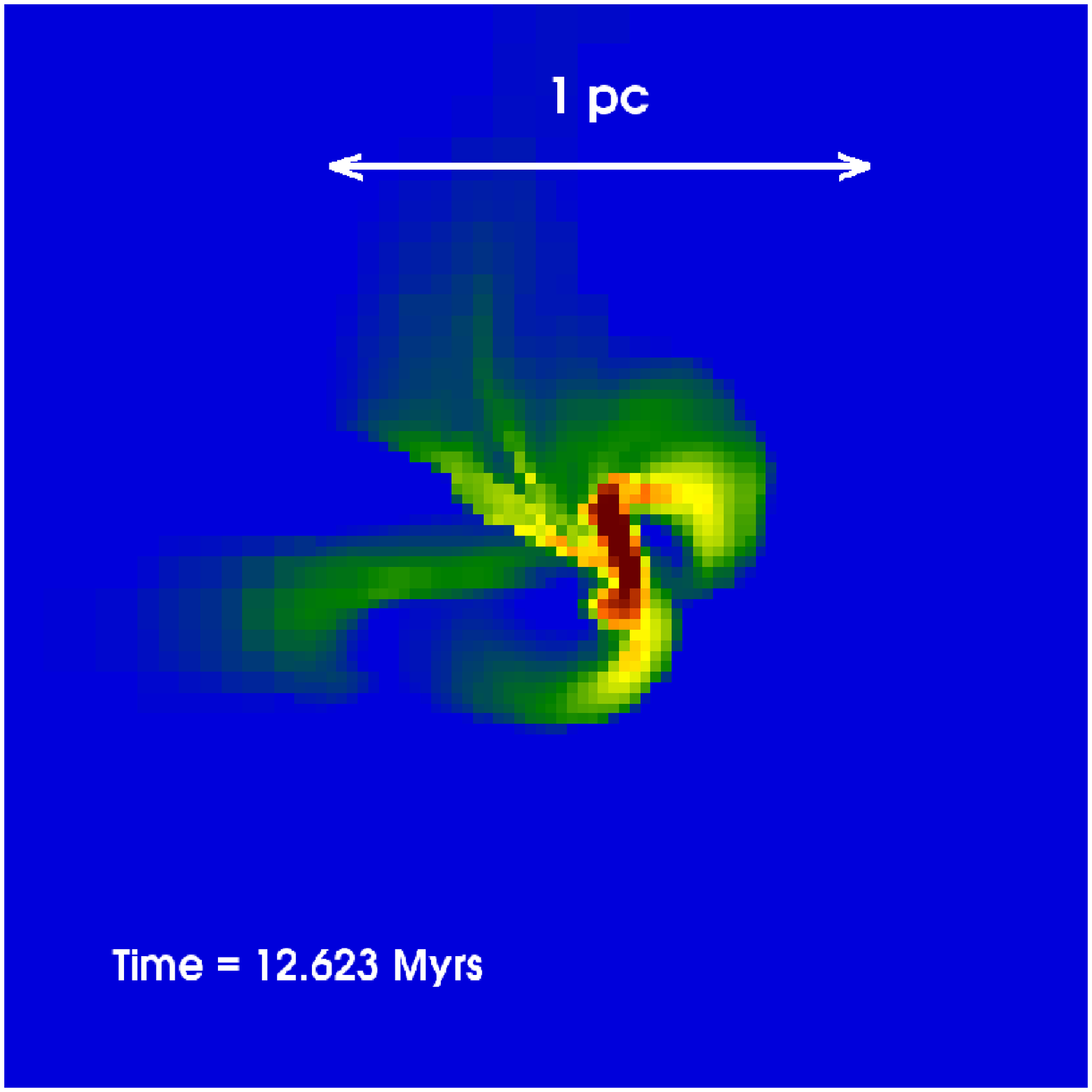}      
\includegraphics[width=8cm]{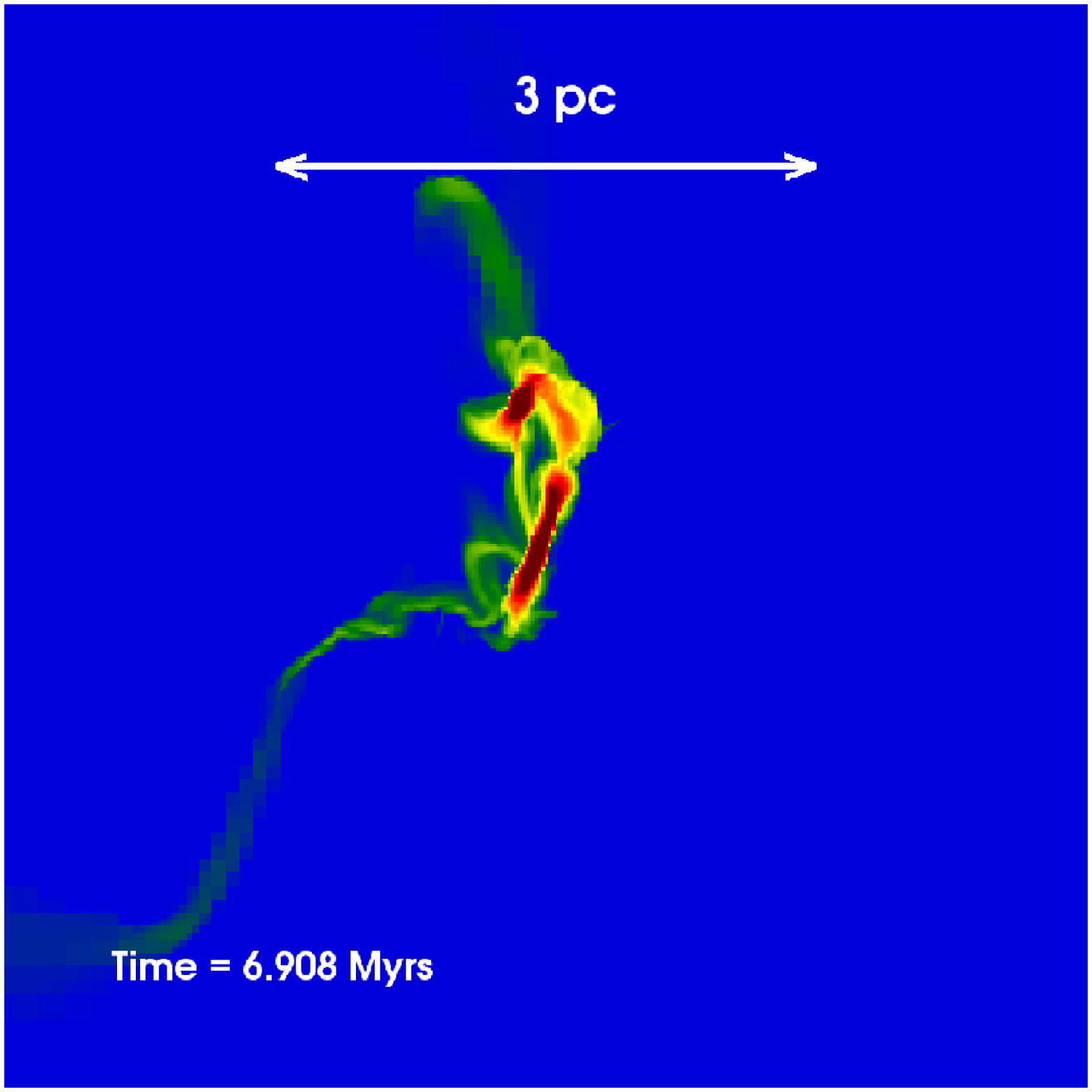}}
    \caption[\label{DensitySlices} ]{\label{DensitySlices} 
      A close up of the final panels of Figures 1 and 2 
      showing the  2D density projection of the rotationally supported 
      gas at the centre of haloes at the end of simulation A and B.} 
  \end{center} \end{minipage}
\end{figure*}
\begin{figure*} 
  \includegraphics[height=7.5cm,width=7.0cm]{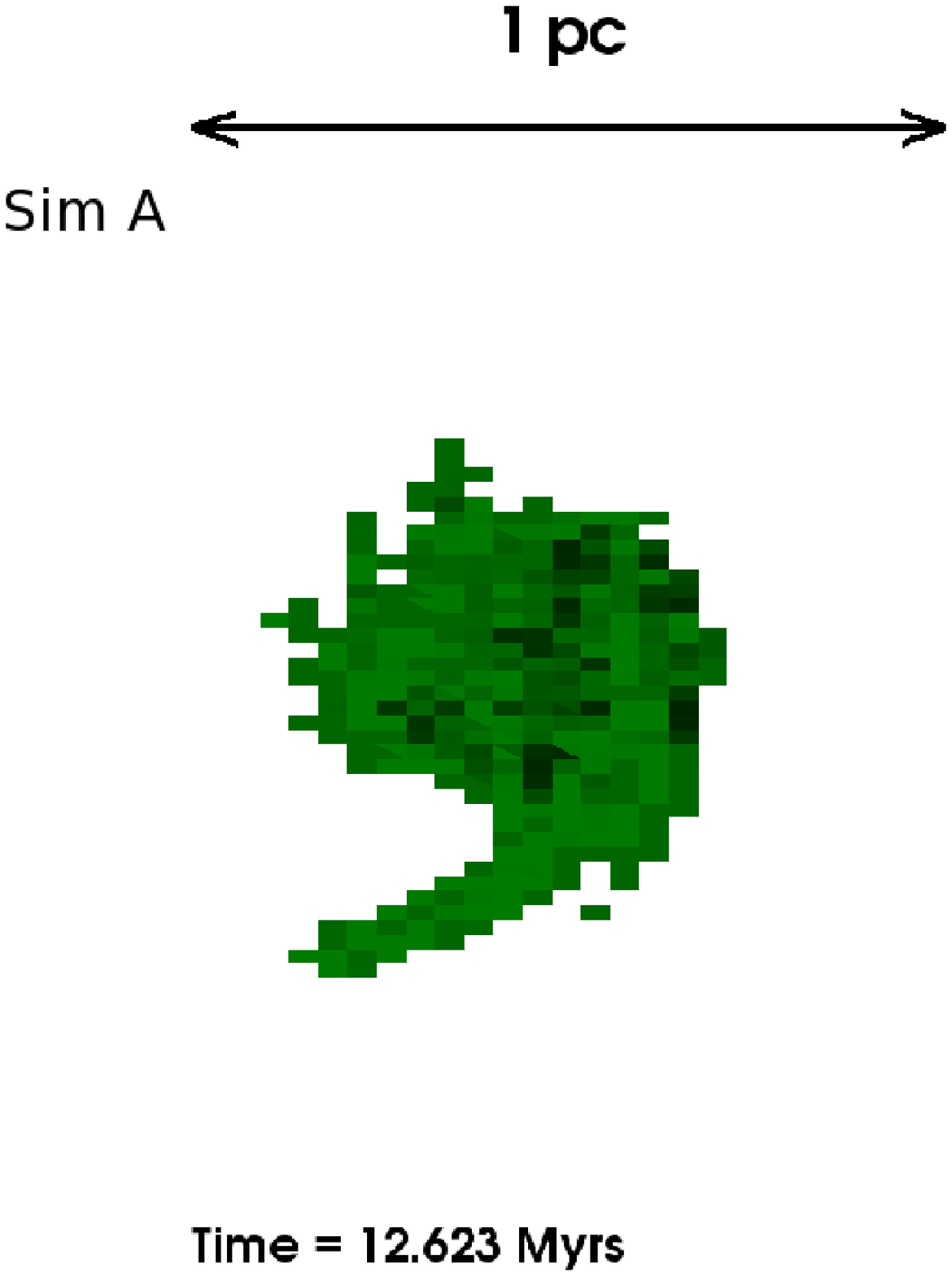} 
  \includegraphics[height=7.5cm,width=7.0cm]{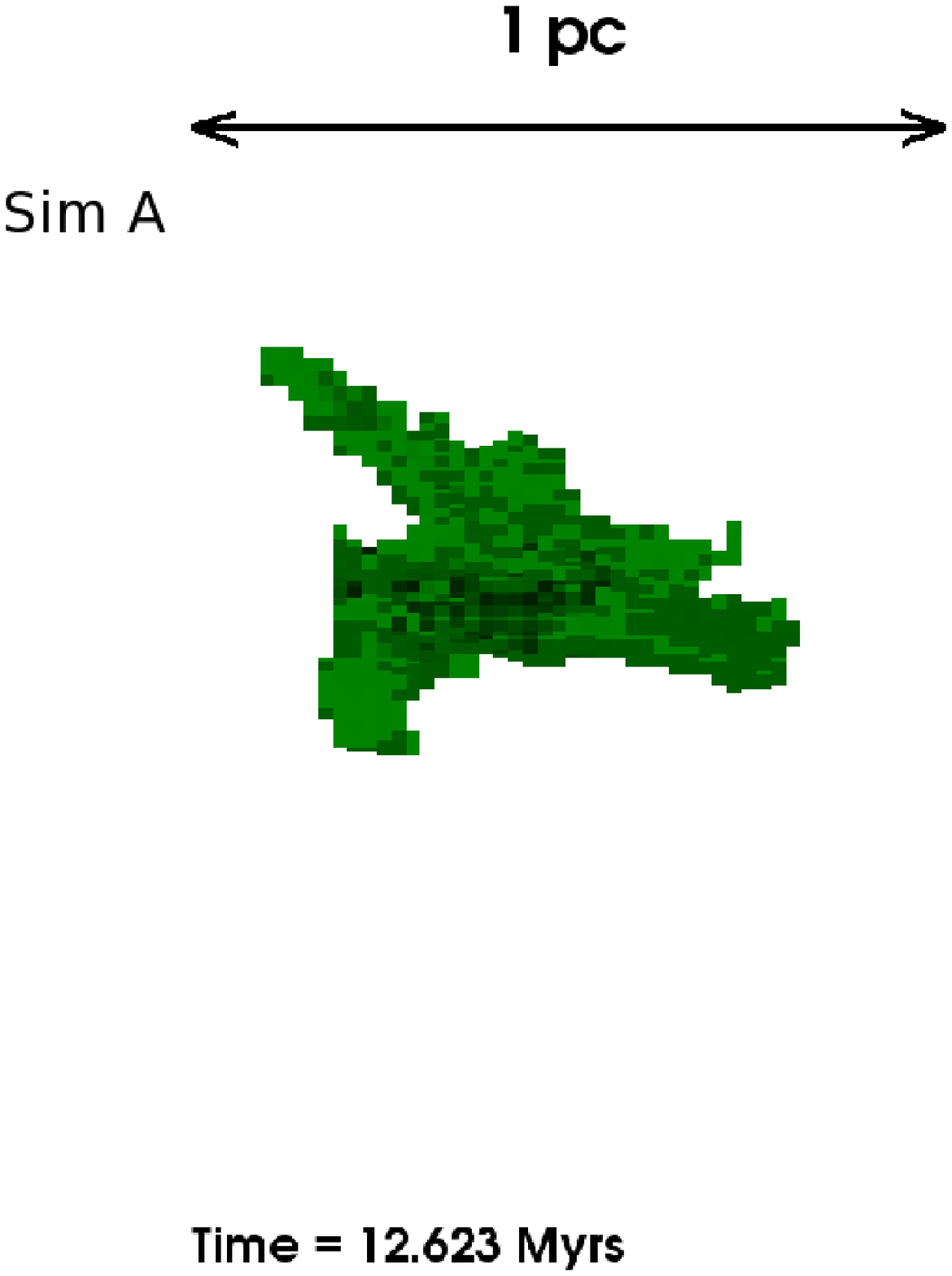}
  \includegraphics[height=7.5cm,width=7.0cm]{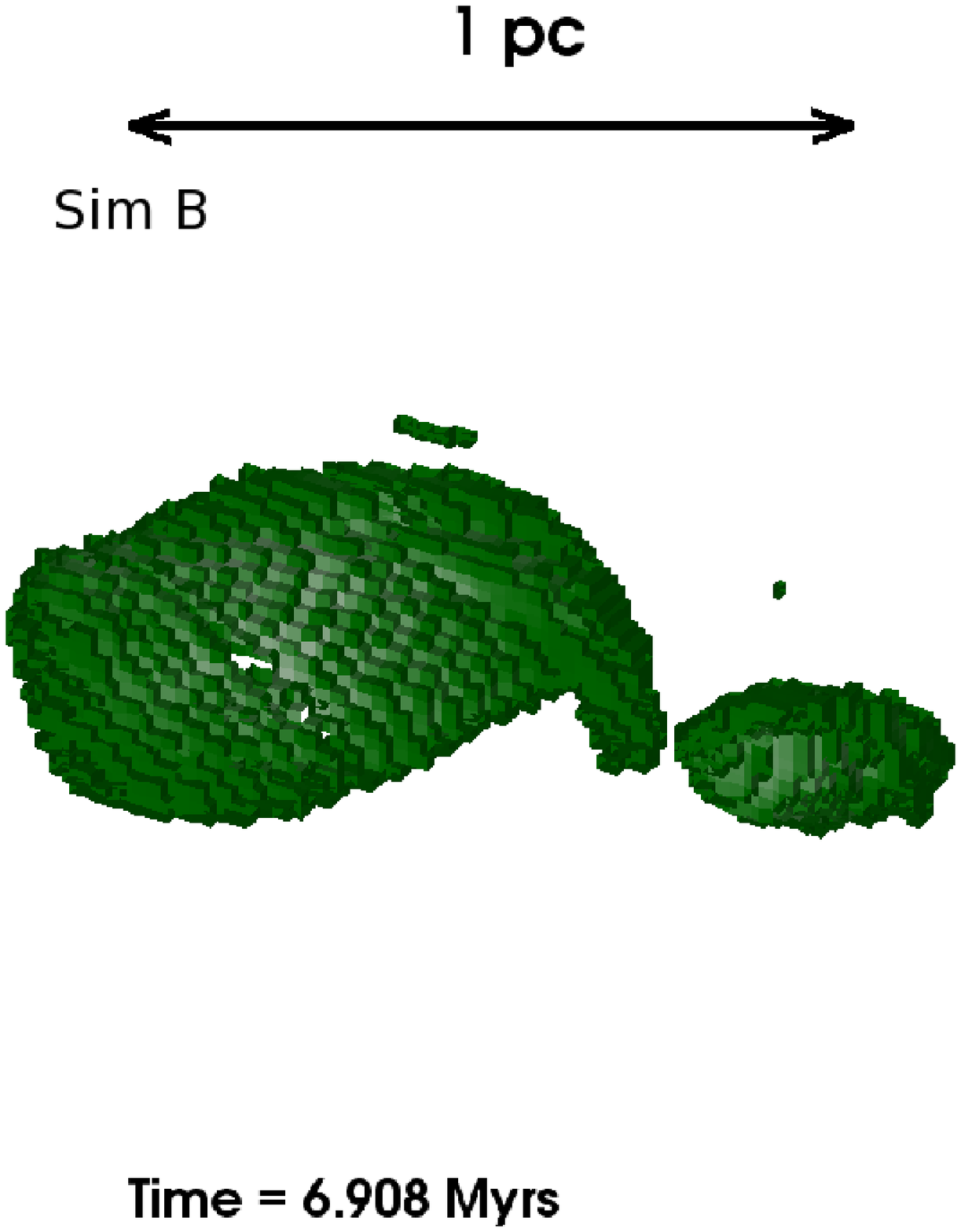}
  \includegraphics[height=7.5cm,width=7.0cm]{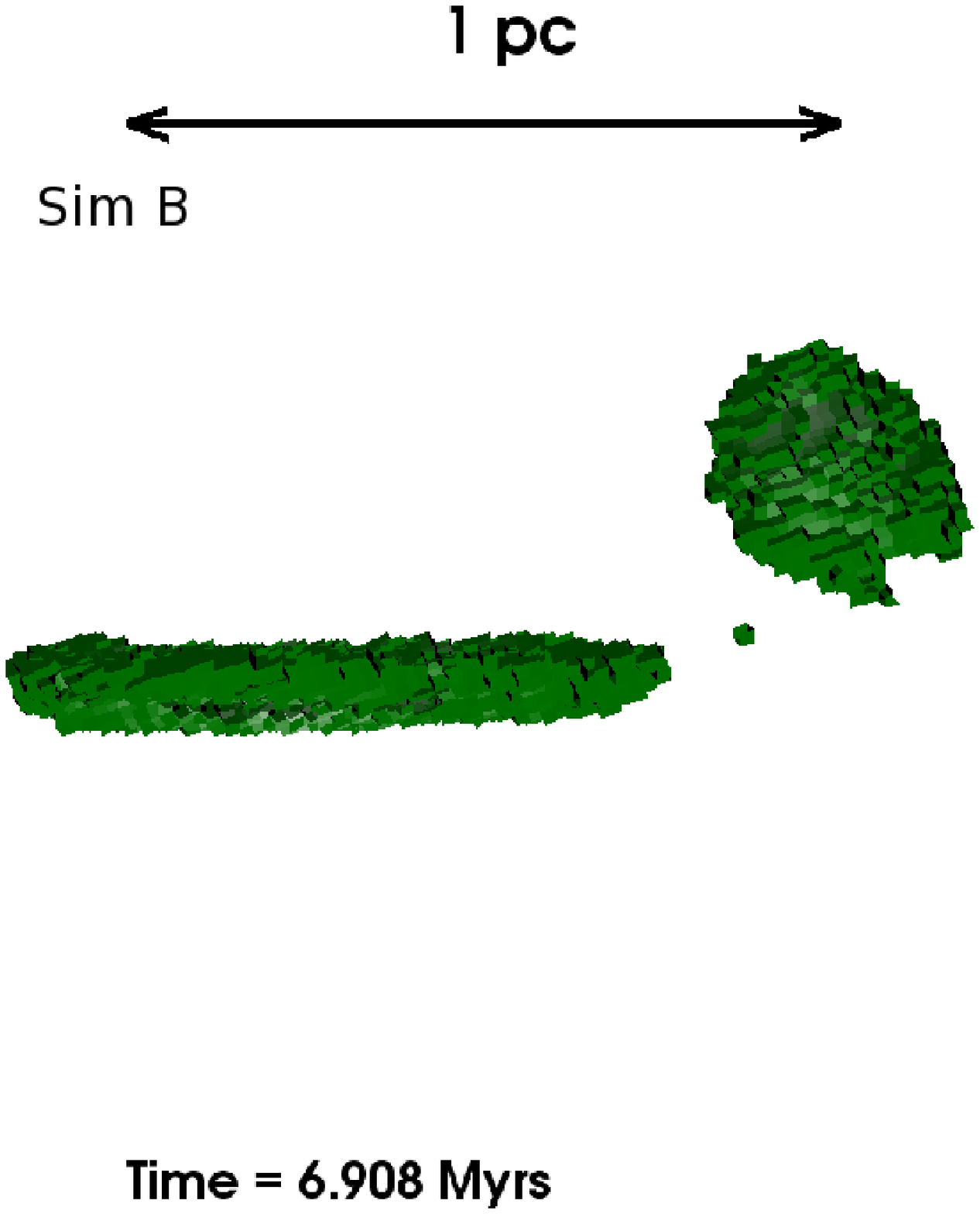} 
  \includegraphics[height=7.5cm,width=7.0cm]{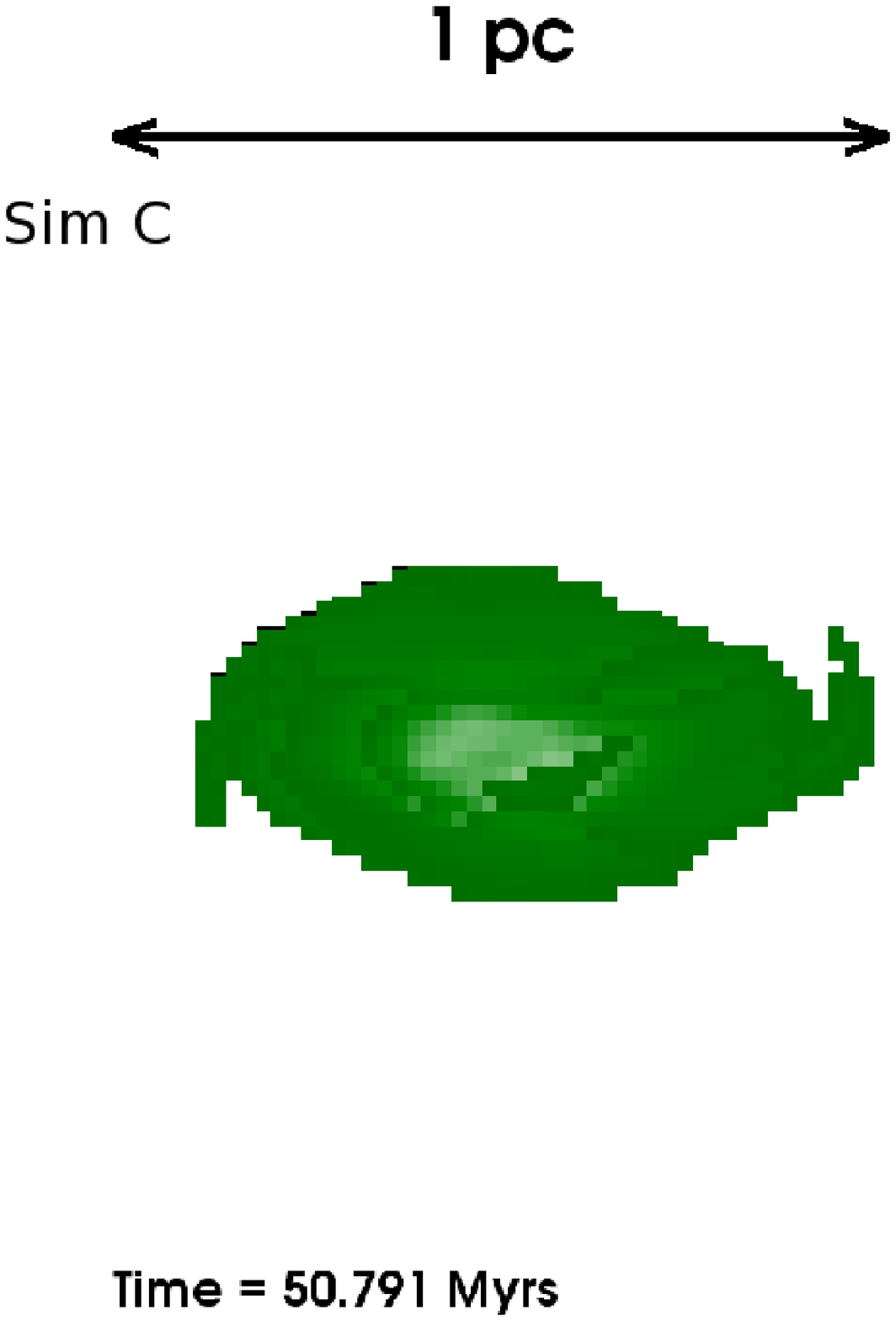}
  \includegraphics[height=7.5cm,width=7.0cm]{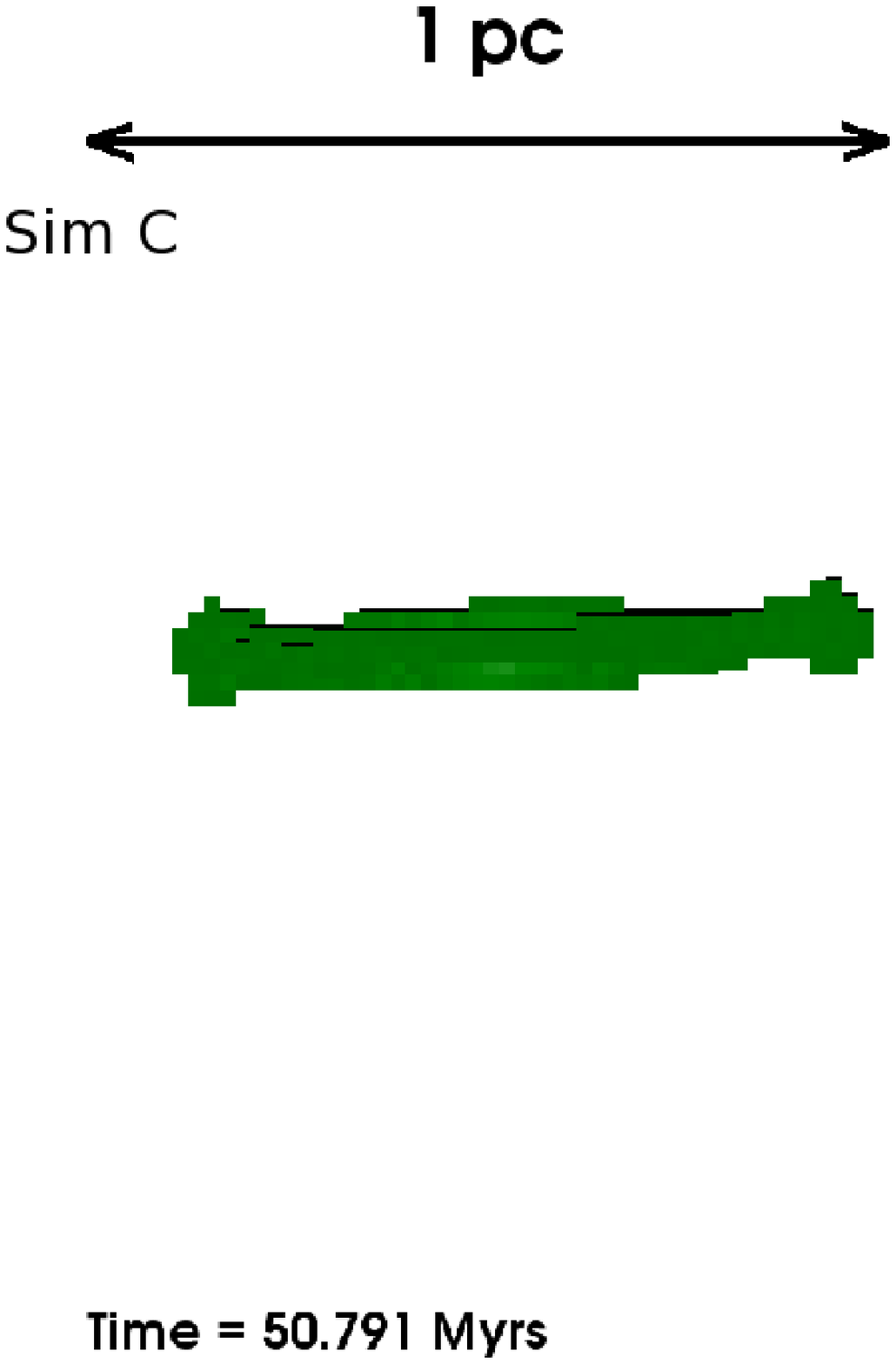} 
  \caption[\label{Disc} ]{\label{Disc}  
The density distribution of the rotationally supported fat discs from all three 
simulations shown 'face-on' (left panels)  and edge on (right panels). In all 
cases (approximate) iso-density surfaces are plotted. 
Note the prominent tidal tail(s) in simulation A  and the secondary clump in simulation B. 
The mass of the disc in simulation A is $ \approx 2 \times 10^4 \, M_{\odot}$,
while the disc in simulation B has a mass of $ \approx 1 \times 10^5 \, M_{\odot}$, with the 
smaller clump having a mass of $ \approx 5 \times 10^3 \, M_{\odot}$. The mass of the disc in
simulation C is $ \approx 1 \times 10^5 \, M_{\odot}$. }  
\end{figure*}

\begin{figure*}
    \includegraphics[width=5.77cm]{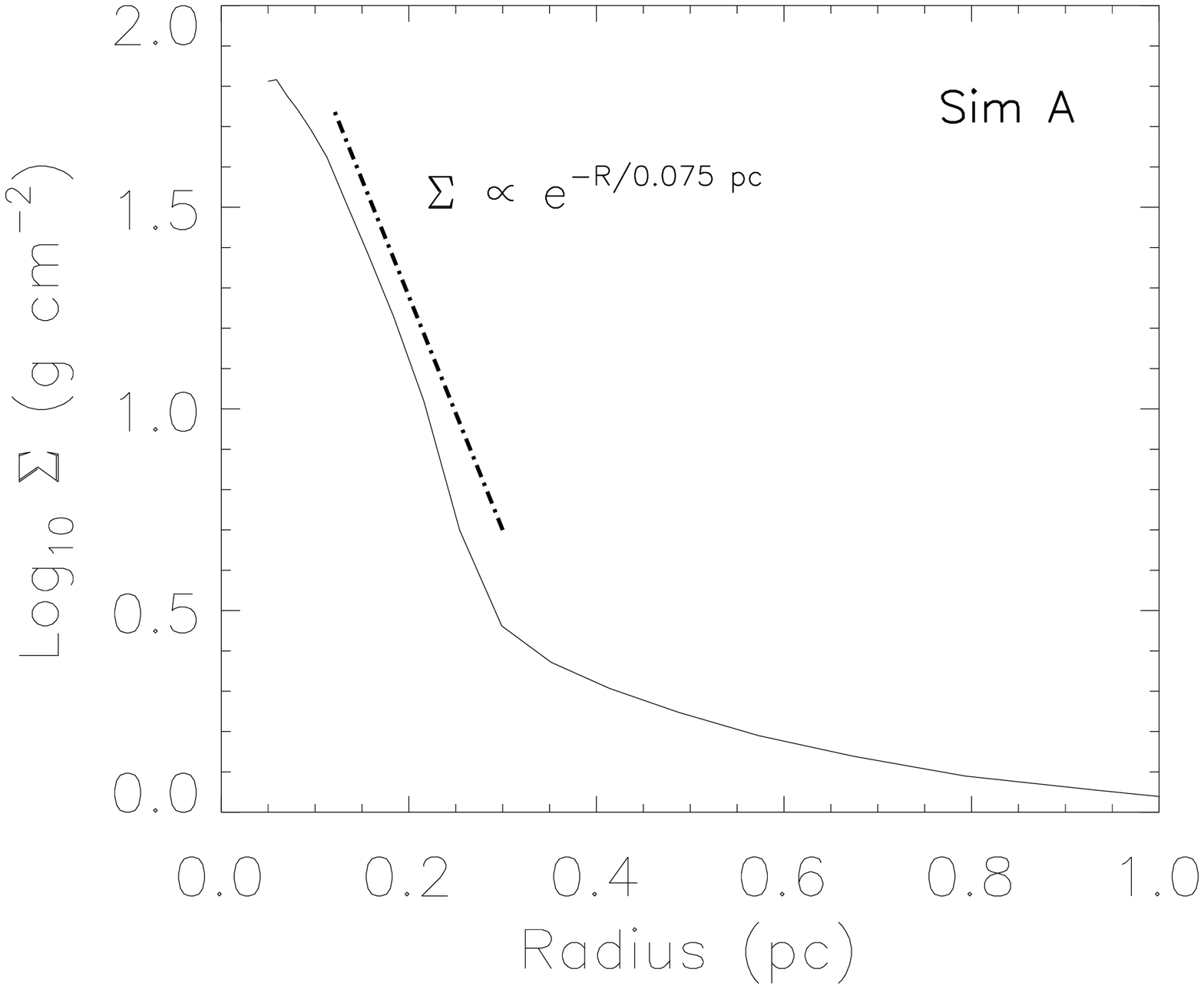}
    \includegraphics[width=5.77cm]{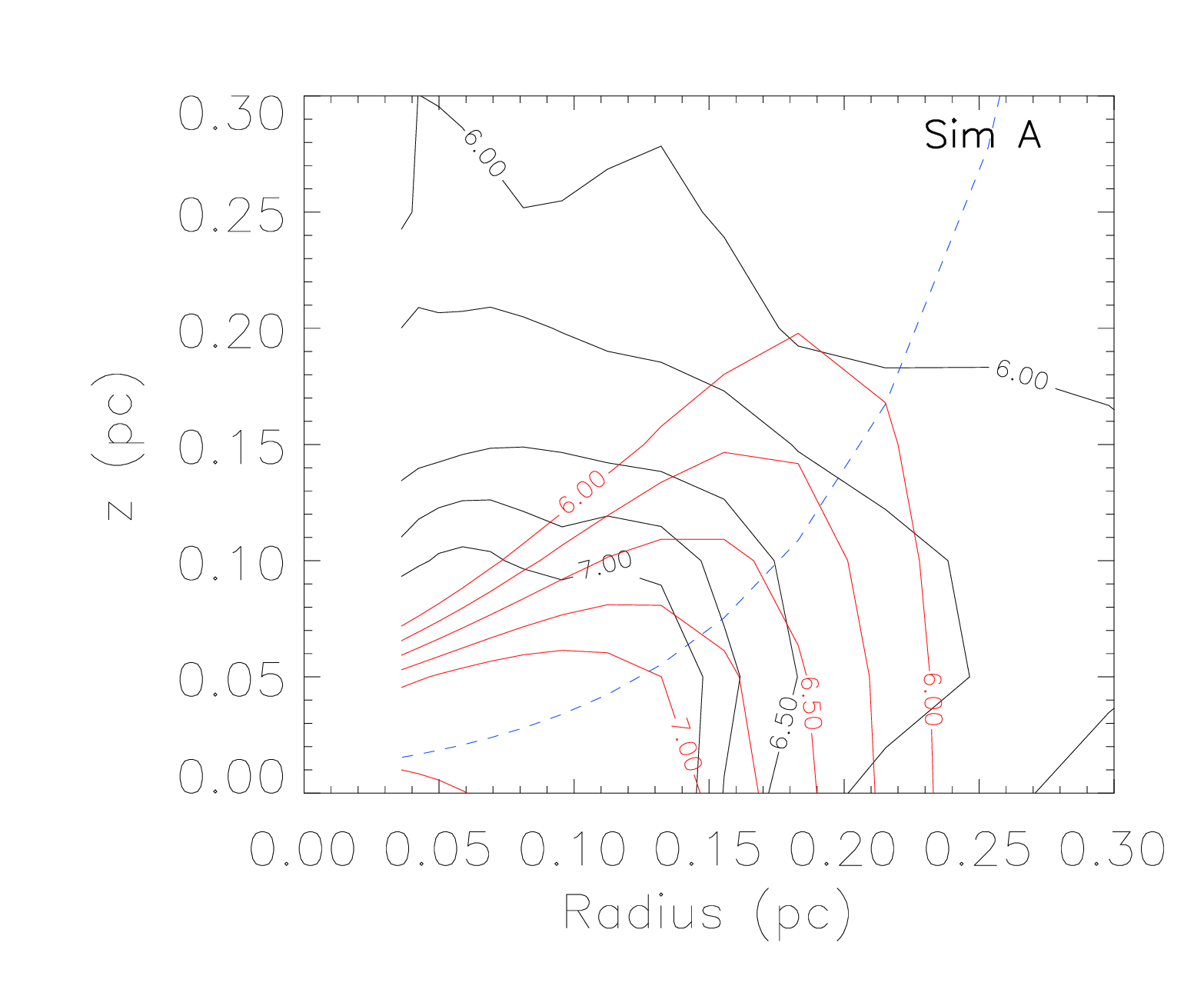}
    \includegraphics[width=5.77cm]{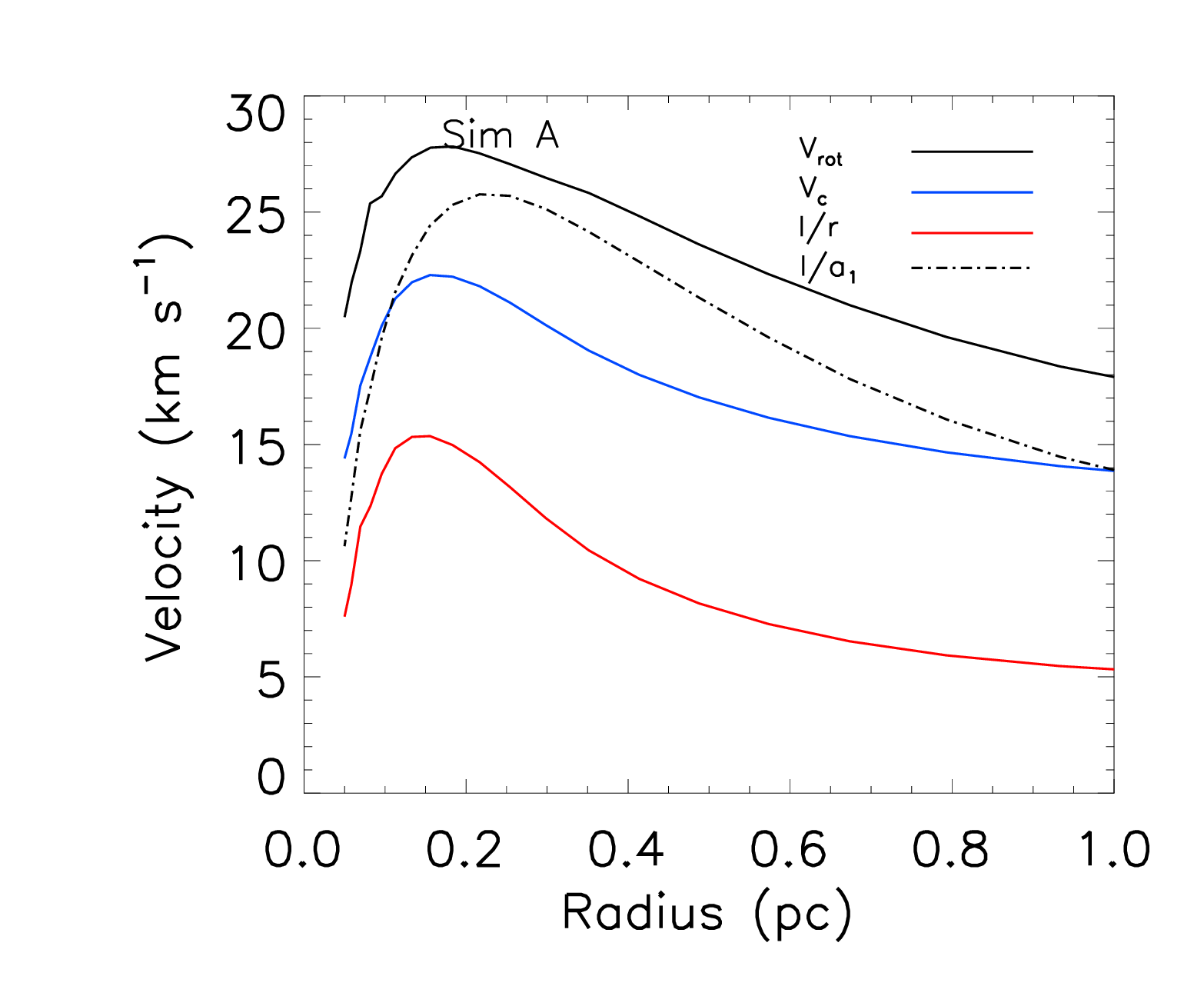}
    \includegraphics[width=5.77cm]{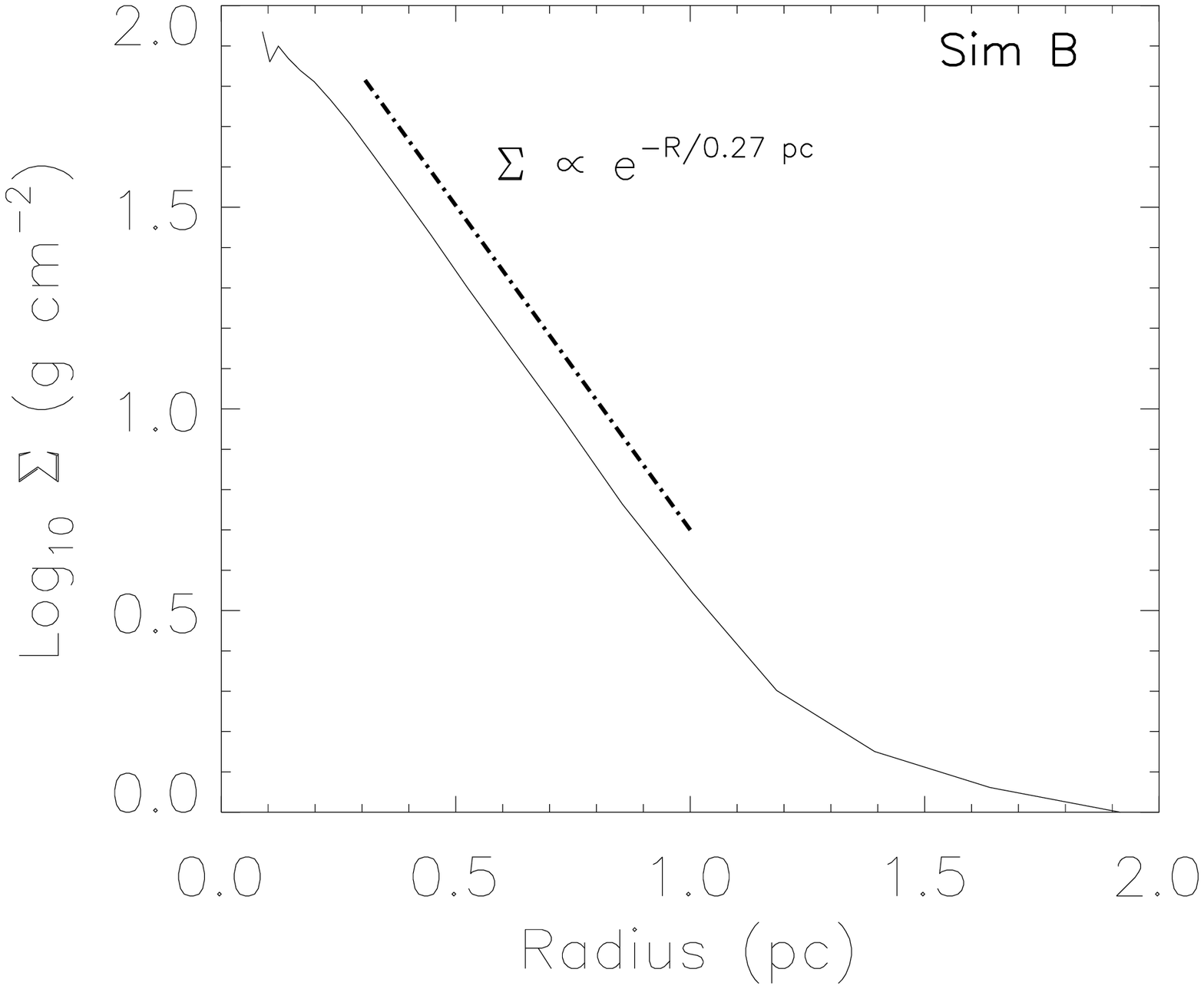}
    \includegraphics[width=5.77cm]{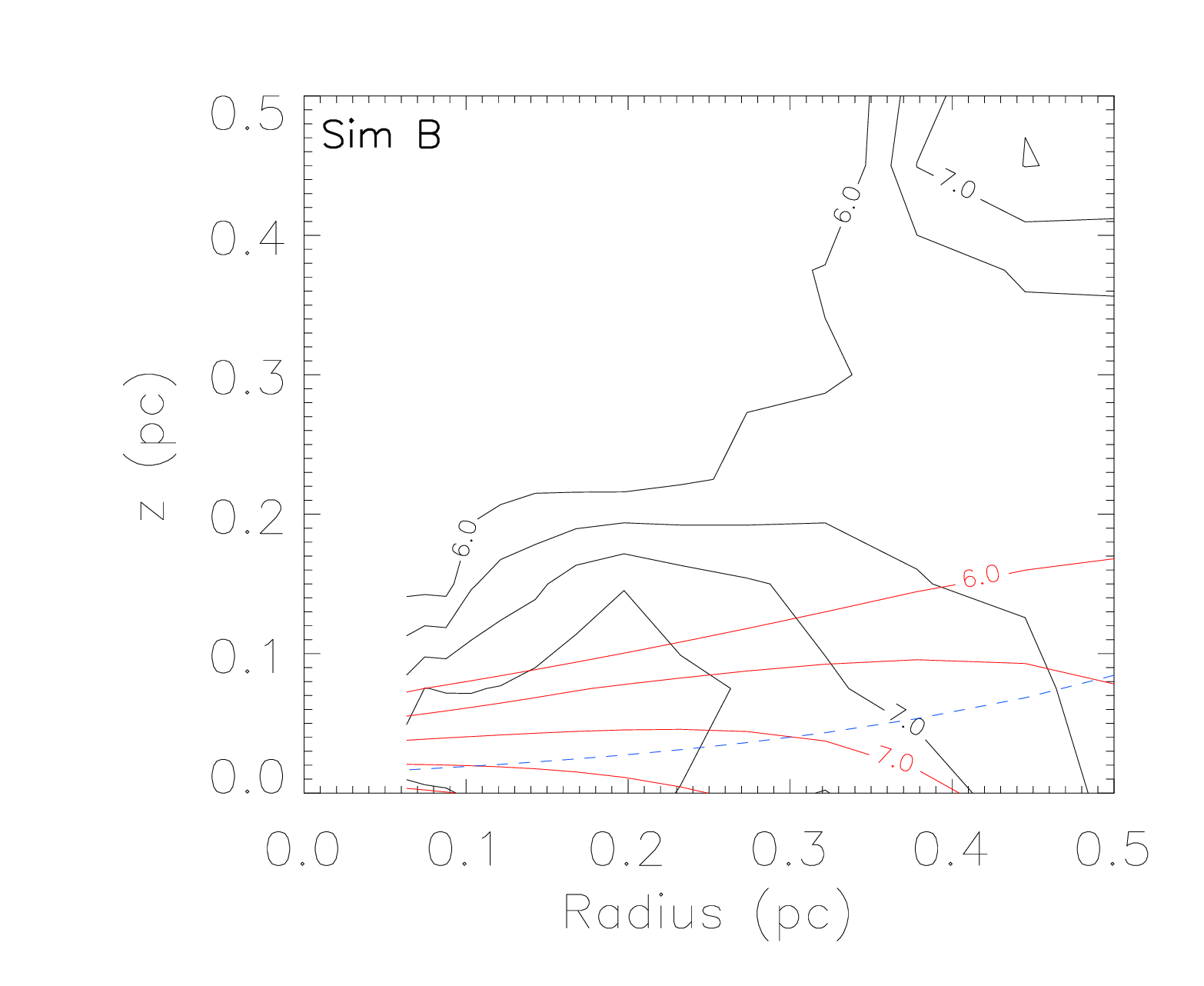}
    \includegraphics[width=5.77cm]{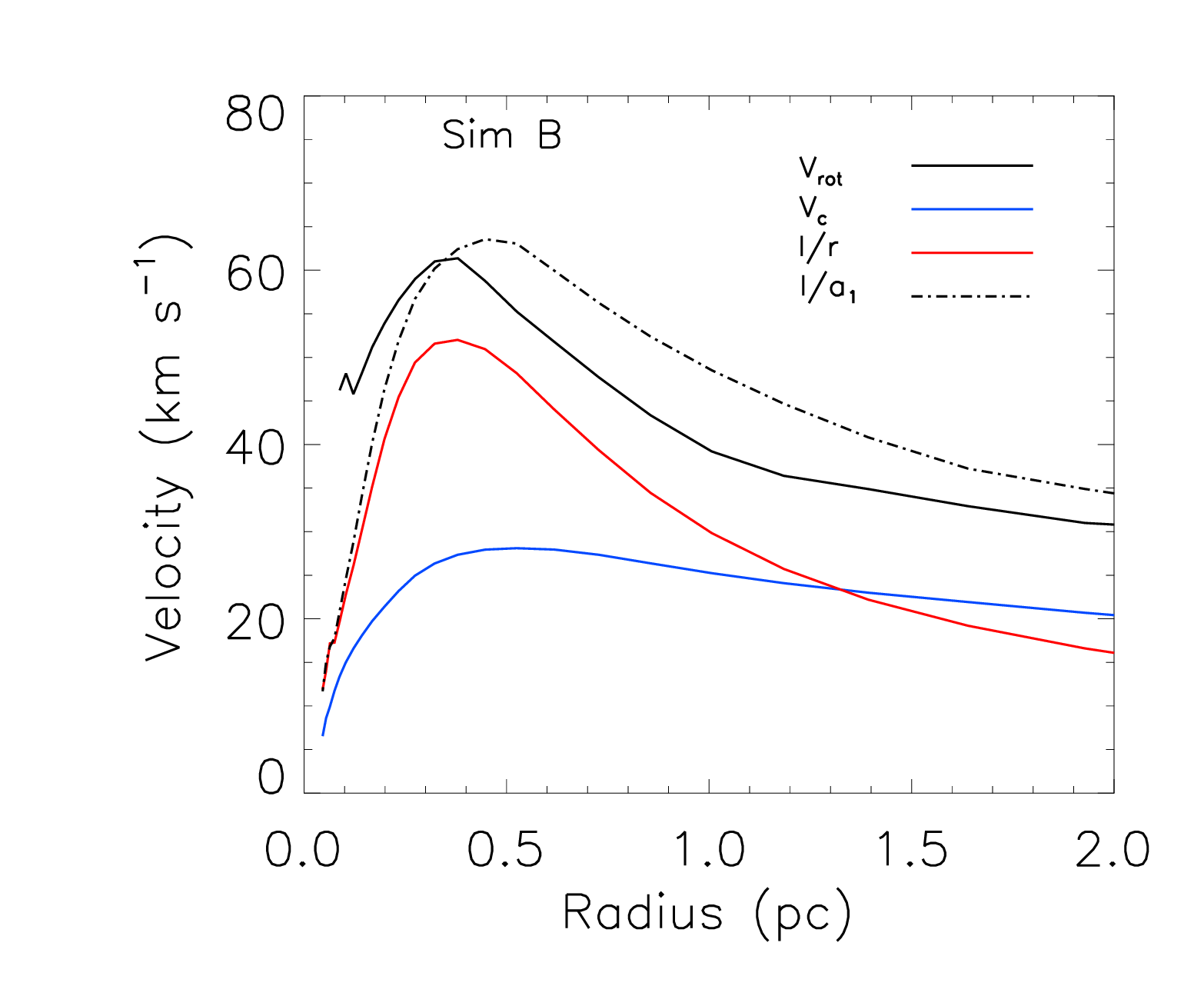}
    \includegraphics[width=5.77cm]{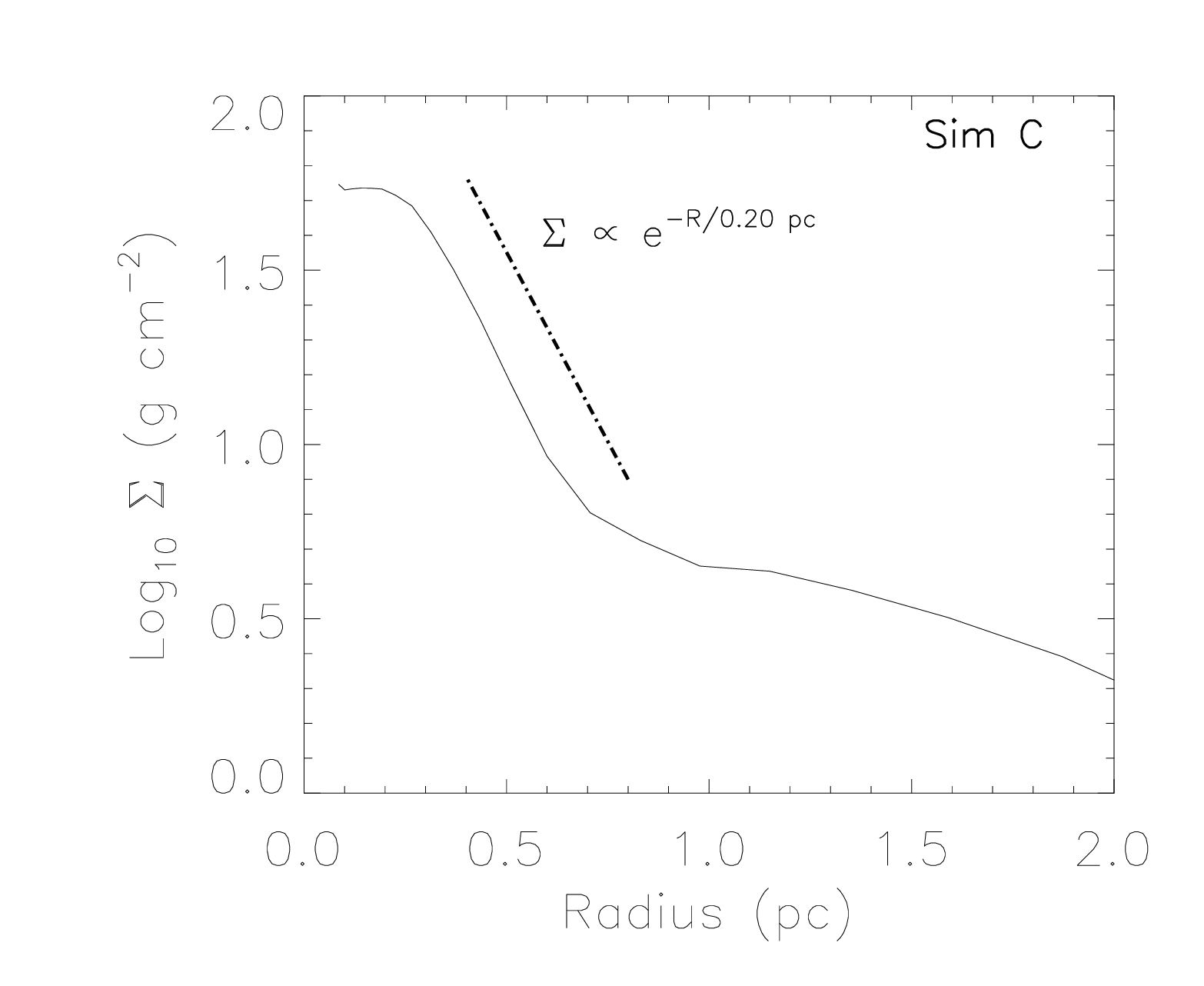}
    \includegraphics[width=5.77cm]{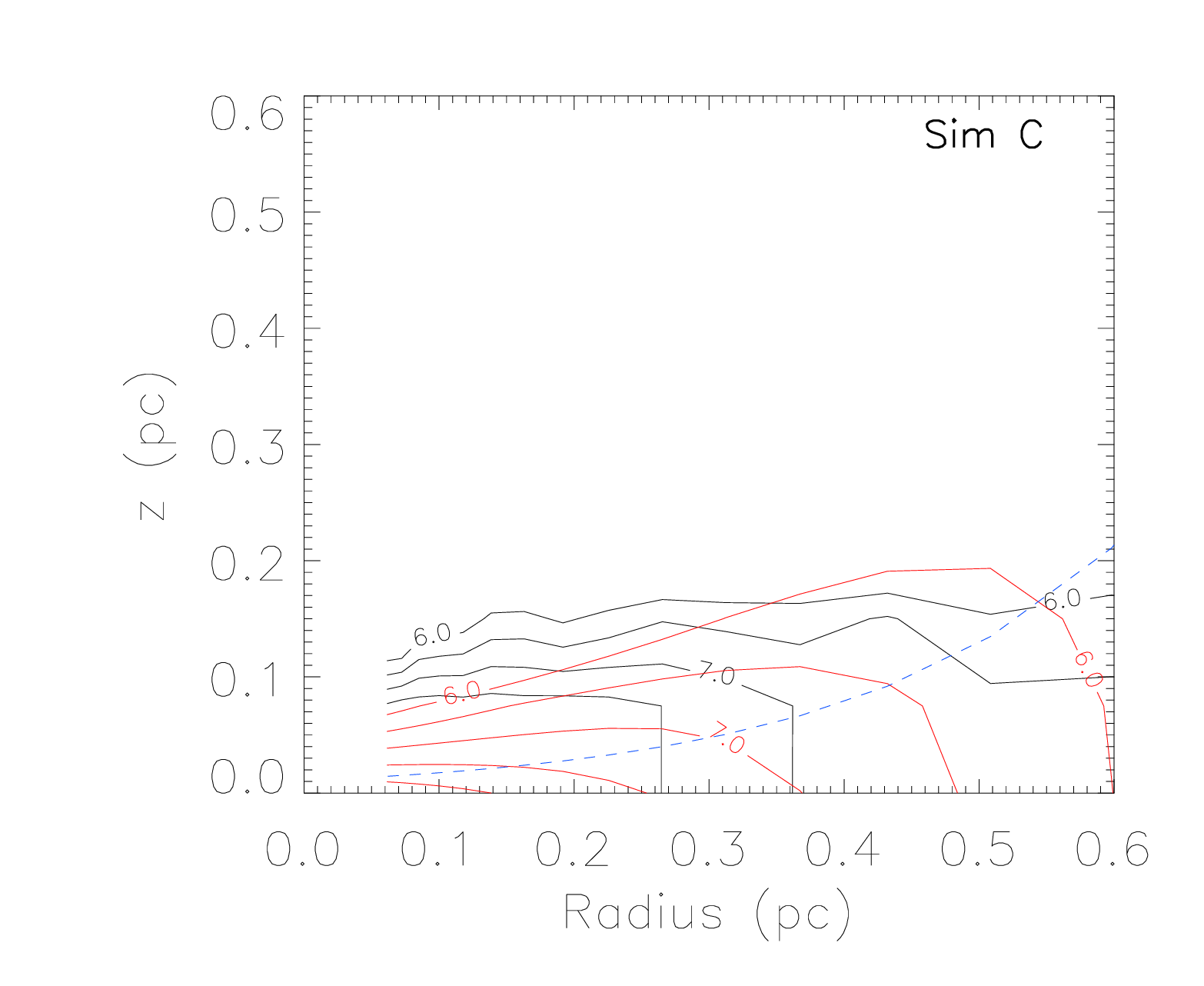}
    \includegraphics[width=5.77cm]{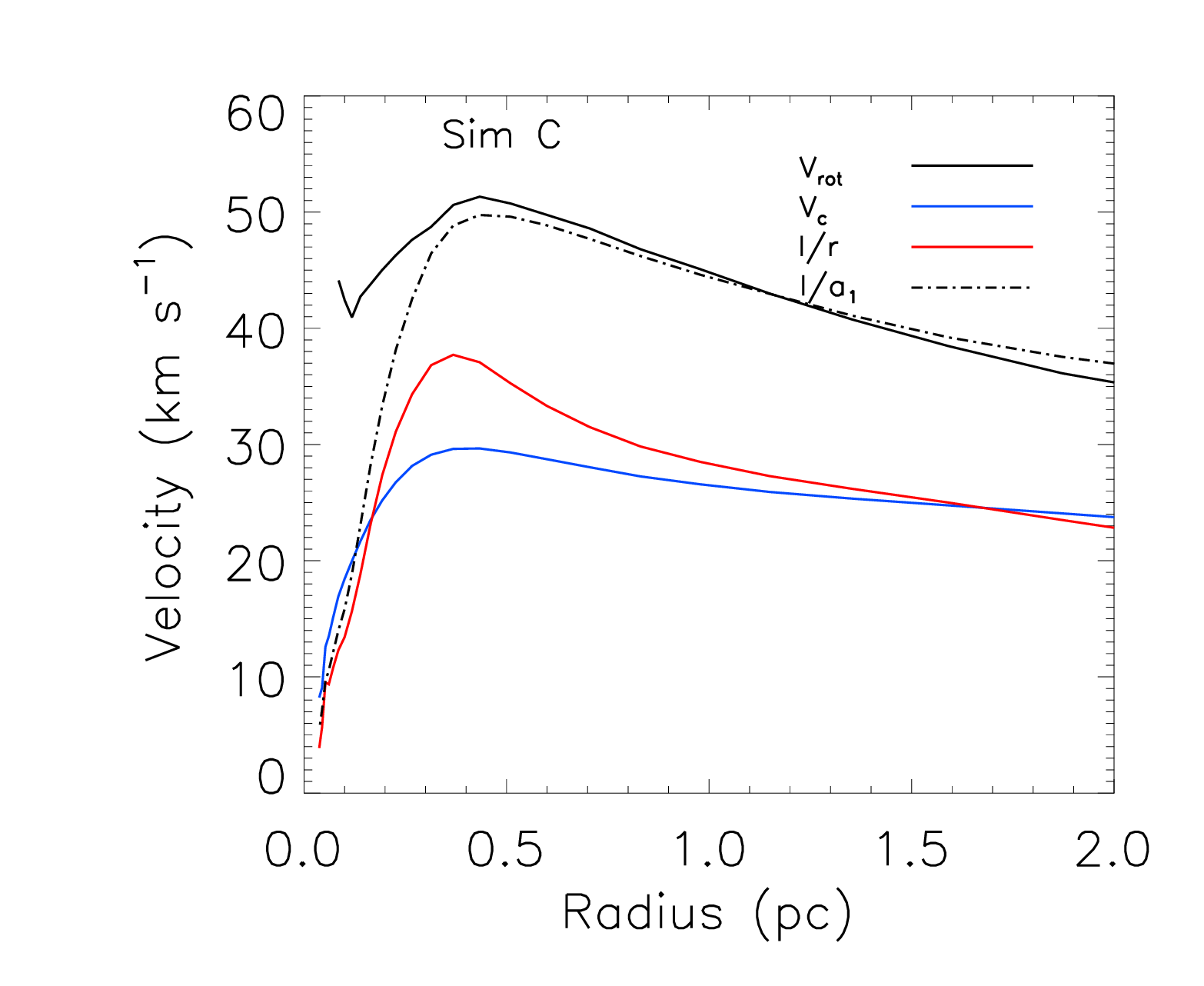}
      
    \caption[\label{rotsupport} ]{\label{rotsupport}
      {\it Left-hand Panels}: The surface mass density profile of
      the rotationally supported gas at the centre of each halo shown in figure 
      \ref{Disc}. 
      {\it Middle Panel}: The black solid lines are the contours 
      of the azimuthally averaged density in the 
      $R-z$ plane of the simulated discs. The  red contours and the blue dashed
      curve are the density and vertical scale height an
      isothermal exponential disc with the same central density, scale
      length and temperature would have.
      {\it Right Panel}: The four velocities; $V_{\rm rot}$,
      $V_{\rm c}$ ,  $|\vec{l}|/a_{1}$  and 
       $|\vec{l}|/|\vec{r}|$ as a function of radius.}
      
\end{figure*}

\indent Note that our two estimates of the  rotation velocity trace each
other,  but  that $|\vec{l}|/|\vec{r}|$  
is systematically lower 
than  $|\overrightarrow{l}| \over a_{1}$ by a factor 1.5-3.  
As the gas settles into rotational support a disc-like flattened
structure should gradually form. For a turbulent collapse like the one
considered here we found this to be most easily studied by looking at 
the   evolution of the smallest eigenvalue of the inertia tensor
which for a disc should be a good proxy for the thickness of the disc. 
In figure \ref{h/R} we show the ratio of the square root of the smallest 
eigenvalue of the inertia tensor, $a_{3}$, to the radius 
in which the inertia tensor is calculated as a function of 
the enclosed mass. The ratio reaches 
a minimum value at a few times  $10^4 \, M_{\odot}$ and a few times  $10^5
\, M_{\odot}$ for simulation A and B, respectively,  as expected for
the formation of a rotationally supported disc. The disc appears  to
fatten towards the centre and is surrounded by a more spherical
in-fall. In simulation B the minimum moves  to higher mass values 
as the merger of the initially three clumps progresses. At the end
of the simulation the minimum
occurs at $\sim 1 \times 10^5 \, M_{\odot}$.

\subsection{Formation of a  self-gravitating massive fat disc} 

About 0.1-1\% of the gas in the inner part of both haloes has settled
into a rotationally-supported self-gravitating object with rotation 
velocities of  $25-60 \kms$. We now progress to inspect
these structures in somewhat more detail. \\
\indent In figure \ref{DensitySlices}  we show the final panels of figures \ref{SimAMap}
and \ref{SimBMap} at a larger scale. The mass of the central object in
simulation A in the left hand panel is $\approx 2 \times 10^4 \,
M_{\odot}$. The radius of the central object is $ \approx 0.3$ pc with a
compact inner core with a radius of $ \approx 0.1$ pc. In the right hand panel
we show the central object in simulation B at the final output time. 
The mass of the central object (large clump) is $\approx 1 \times 10^5 \, M_{\odot}$ and
the radius is $ \approx 0.5$ pc.   The second clump on the
right hand side has a mass of  $ \approx 5 \times 10^3 \, M_{\odot}$ and a
radius of $ \approx 0.1 $ pc. The disc in simulation C (shown in figure \ref{SimCMap}) 
has a final mass of $\approx 1 \times 10^5 \, M_{\odot}$ and a radius of $ \approx 0.6$ pc.\\ 
\indent  In figure \ref{Disc}
we show the rotationally supported objects face-on and edge-on. We have plotted 
iso-density surfaces using overdensities within a relatively narrow
range (approximately an order of magnitude).
These visualisations give a clear picture of the shape of the central
objects that form in all three
simulations. Each object is quite clearly a disc.  
The feature in the edge on visualisation of simulation A (top panel) which points north-west 
is the tidal tail seen at the bottom of the face-on view.  There is
another, less prominent, 
tidal tail
at the top of the disc in simulation A which appears at the bottom left in the edge on view. 
The disc in simulation B (middle panel) is less obstructed, the smaller clump is also visible 
in both visualisations. The disc in simulation C (bottom panel) is the
cleanest example of a disc.  Like  the disc in simulation B it has a 
somewhat  ellipsoidal shape.

\subsection{Properties of the centrifugally supported disc(-like object)} 
We now have a closer look at the surface mass density profile
and the level of rotational support of  the  disc(-like) object 
for each simulation. As can be seen in the 
left panel of figure \ref{rotsupport} in all three cases the surface
mass density profile is exponential over several scale lengths. 
The scale lengths are 0.075, 0.20 and 0.27 pc in simulations A-C,
respectively. At the outer ``edge'' of the disc at about 0.3 - 1 pc the surface mass 
density profile reverts to that expected for an isothermal 
spherical density distribution in which the discs are embedded. \\
\indent In the middle  panel  of figure  \ref{rotsupport} we compare the
density structure of the discs to that of an exponential disc which is
given  by  
\begin{equation}
n(r,z) = n_0 \,\rm{exp}\Big(- {2r \over R_d} \Big) \, \rm{sech}^2 \Big({z \over \sqrt{2} z_0} \Big),
\end{equation}
with scale height 
\begin{equation}
z_0 = {c_s \over  (4 \pi G \mu m_{\rm H} n_0 e^{-2r/R_{\rm d}})^{1/2}},
\end{equation}
where $c_s$ is the sound speed, $ \mu $ is the mean
molecular weight, $n_0$ is the central density, $r$ is the distance
from the centre of mass and $R_d$ is the scale-radius of the disc \cite[]{Spitzer_1942, Oh_2002}.
We show contours of the azimuthally averaged density in the 
$R-z$ plane of the discs. The discs (especially that in simulation A) 
are somewhat fatter than expected if they were supported 
by pressure only.  This is not surprising considering
the rather unrelaxed dynamical state of the discs. 
The general agreement of the density 
structure with that of an exponential isothermal disc is
nevertheless  remarkable. In simulation B the structure in the top
right corner is due to the secondary  less massive clump.  

In the right  panel  of figure  \ref{rotsupport} we compare the actual
rotation velocities of the gas in the discs (shown by the black solid
curve) to our two estimates for the rotational velocity
$|\vec{l}|/|a_1|$ (black dot-dashed curve) and $|\vec{l}|/|\vec{r}|$
(red curve) and the  circular velocity (blue curve). We compute the actual rotational 
velocities by rotating our coordinate system into the coordinate system of the disc using 
the matrix of eigenvectors obtained from the inertia tensor (which we then checked visually).  
The rotation velocity in the plane of the disc is then easily calculated using trigonometic 
transforms. The peak of the
actual rotation velocities occurs at radii corresponding to, two to three 
times the scale radius  of the exponential discs and  range from 25 to
60 $\kms$.    The actual
rotation velocities agree with our estimate 
from the angular momentum vector and the largest eigenvalue of the inertia
tensor, $|\vec{l}|/|a_1|$ to within 10-20\%, 
despite the fact that the latter was
calculated for the enclosed mass in spheres centred on the densest cell. As already
mentioned the other estimate for the rotation velocities 
$|\vec{l}|/|\vec{r}|$ is systematically lower by a factor 
1.5-3. \\
\indent The actual rotation velocities exceed the circular velocities
by up to a factor of 2. This is probably due to a combination of reasons. 
The peak rotation velocities of a thin exponential disc  is about 15 \%
larger than that of a spherical mass configuration with the same
enclosed mass. More importantly the discs in simulation B and C
and probably also that in simulation A have an ellipsoidal shape 
which should significantly raise the 
rotation velocities compared to the circular velocity 
of a spherical mass distribution. 
Finally the discs have probably not yet reached 
centrifugal equilibrium. The gas is still settling into rotational 
support and has probably  fallen to somewhat smaller radius initially
than expected for centrifugal equilibrium and will take some time to
reach centrifugal equilibrium.

\subsection{Numerical Limitations} 

Probing the collapse of gas at the centre of dark matter haloes in
a cosmological context is an extremely challenging exercise when we
wish to follow the collapse to very high densities. In this work
we have reached radii as small as 0.01\% of the
virial radius while at the same time following the dynamical evolution of a 
substantial fraction of the gas in the halo. This presents a
considerable challenge even for an AMR code like \textsc{Enzo}. The main problem 
is that the high refinement levels necessary to achieve such a large
dynamic range  mean that the code will normally grind to a halt
following the dynamical evolution of whatever is the first high
density region to form. As discussed, in order to ameliorate  this
problem we have changed the refinement level during the simulation 
in order to allow our three haloes of choice to build up to
their full mass. To what level may this have affected 
our results? \\
\indent In order to address this we have run simulations with twice 
the initial refinement level and sixty four times the initial refinement level and found no
systematic difference in the initial value of the angular momentum of the halo. 
While the detailed dynamical evolution is obviously different especially in the 
later stages of the collapse when the initial distribution of matter in the halo  is
different the qualitative behavior of the dynamical evolution did not change. 
We would also like to point out that our results are in most
aspects similar to those of W07 who did not change the refinement
level during the simulation. This meant, however, that they were not
able to follow the gas at the centre of the halo settling into
rotational support.  We feel that this is the most interesting aspect
of our work and well worth exploring the limits of the code with some
manual intervention. 
Obviously further work is needed to address these questions in more
detail. 

\section{Conclusions}
\label{conclusions}
We have investigated the dynamical evolution of the gas in haloes
with virial temperatures of between $\sim 13000$ K and $\sim 30000$ K assuming that
cooling by atomic hydrogen and helium are the dominant cooling processes.
 The dynamical evolution of the gas in the haloes is
complex and highly turbulent with turbulent velocities approaching 
the virial velocity of the haloes. 
The gas in the inner part of our haloes collapses 
isothermally with  a temperature of 6000-7000K on a somewhat slower
time scale than the free-fall time and settles into a close to
isothermal ($\rho \propto r^{-2}$)  density profile.   
We find no signs of efficient fragmentation confirming suggestions  
that the isothermal collapse of gas in a dark matter halo 
at temperatures  close to the  virial temperature of the halo  
leads to only modest fragmentation. The inner 0.1-1 \% of the gas
in the virialised haloes loses as 
much as 95\% of its initial angular momentum during the collapse. The
gas thereby collapses by a factor of 300-1000 in radius and 
eventually settles into a very compact rotationally supported 
self-gravitating disc with peak rotation velocities of  25 - 60 $\kms$
and ``radii'' of 0.3 pc - 0.6 pc (0.05\% - 0.1\% of the virial radius 
of the host halo). \\
\indent The discs have  an exponential surface mass density profile
with scale length in the range $0.075 - 0.27$ pc which extends over several 
scale lengths. The vertical structure of the disc is somewhat more extended than expected for a
purely pressure supported isothermal axisymmetric exponential
disc. \\
\indent Massive compact self-gravitating discs such as those found in our
simulations have  been suggested  to evolve into 
massive seed black holes which later in the 
hierarchical build-up of galaxies will grow into the   supermassive black holes 
found at the centre of the bulges of present-day galaxies. 
Unfortunately we were not yet able to follow the further dynamical
evolution of the discs or the gas further out in the haloes 
in  our simulations.  However, independent of whether the gas in these
discs  will continue to efficiently lose angular momentum and
contract further or will fragment and form an ultra-compact star
cluster we most probably have identified an important intermediary  stage en
route to the formation of a massive seed black hole.

\section*{Acknowledgments}
The simulations were run on the \cosmos (SGI Altix 3700) supercomputer
at DAMTP in Cambridge and on the Cambridge High Performance Computing 
Cluster \darwin in Cambridge. \cosmos is a UK-CCC facility
which is supported  by HEFCE and PPARC. \darwin is the primary supercomputer 
of the University of Cambridge High Performance Computing Service 
(http://www.hpc.cam.ac.uk/), provided by Dell Inc. using Strategic 
Research Infrastructure Funding from the Higher Education Funding 
Council for England. We used the \textsc{Visit} (https://wci.llnl.gov/codes/visit/home.html)
visualisation software to generate the visualisation plots shown here.  
We are grateful to Brian O'Shea, John Wise, Volker Springel,  Debora Sijacki,
Tirthankar Roy Choudhury, Tom Abel, Mitch Begelman, Giuseppe Lodato, 
Priya Natarajan, Jim Pringle and  Martin Rees 
for useful discussions. We are also grateful to Greg Bryan, Mike Norman, Brian O'Shea and the 
rest of the \enzo team for making the \enzo code publicly available.

\bibliographystyle{mn2e}

\end{document}